\documentclass[twocolumn,aps,superscriptaddress,floatfix,prb,nobibnotes,longbibliography]{revtex4}
\usepackage[normalem]{ulem}

\usepackage[T1]{fontenc}
\usepackage{graphicx}
\usepackage{amsmath}
\usepackage{amssymb}
\usepackage{bbm}
\usepackage{xspace}
\usepackage{color}
\usepackage{footnote}
\usepackage{comment}
\usepackage{hyperref}
\usepackage{subfigure}
\usepackage{bbold}
\usepackage{ulem}
\usepackage{braket}
\usepackage{multirow}
\usepackage{enumerate}

\definecolor{grey}{RGB}{100,100,100}

\definecolor{orange}{RGB}{252,77,6}
\definecolor{brown}{RGB}{200,127,50}
\definecolor{blue}{RGB}{00,000,100}
\definecolor{blue2}{RGB}{00,000,250}
\definecolor{green1}{RGB}{00,100,00}
\definecolor{green2}{RGB}{00,150,00}
\definecolor{green3}{RGB}{00,200,00}
\definecolor{green4}{RGB}{00,250,00}
\definecolor{dgreen}{rgb}{0.0,0.5,0.0}

\begin{document}

\title{Spin-Orbit coupling in the Kagome lattice with flux and time-reversal symmetry}

\author{Irakli Titvinidze}
\email{titvinidze@itp.uni-frankfurt.de}
\affiliation{Institut f\"ur Theoretische Physik, Goethe-Universit\"at, 60438 Frankfurt am Main, Germany}

\author{Julian Legendre}
\affiliation{CPHT, CNRS, Institut Polytechnique de Paris, Route de Saclay, 91128 Palaiseau, France}

\author{Maarten Grothus}
\affiliation{Institut f\"ur Theoretische Physik, Goethe-Universit\"at, 60438 Frankfurt am Main, Germany}

\author{Bernhard Irsigler}
\affiliation{Institut f\"ur Theoretische Physik, Goethe-Universit\"at, 60438 Frankfurt am Main, Germany}

\author{Karyn Le Hur}
\affiliation{CPHT, CNRS, Institut Polytechnique de Paris, Route de Saclay, 91128 Palaiseau, France}

\author{Walter Hofstetter}
\affiliation{Institut f\"ur Theoretische Physik, Goethe-Universit\"at, 60438 Frankfurt am Main, Germany}

\date{\today}

\begin{abstract} 
We study the topological properties of a spin-orbit coupled tight-binding model  with flux on the Kagome lattice. The model is time-reversal invariant and realizes a $\mathbb{Z}_2$ topological insulator as a result of artificial gauge fields. We develop topological arguments to describe this system showing three inequivalent sites in a unit cell and a flat band in its energy spectrum in addition to the topological dispersive energy bands. We show the stability of the topological phase towards spin-flip processes and different types  of on-site potentials. In particular, we also address the situation where on-site energies may differ inside a unit cell.  Moreover, a staggered potential on the lattice may realize topological phases for the half-filled situation. Another interesting result is the occurrence of a topological phase for large on-site energies. To describe topological properties of the system we use a numerical approach based on the twisted boundary conditions and we develop a mathematical approach, related to smooth fields. 
\end{abstract}


\maketitle
%

\begin{figure}[t]
\includegraphics[width=8cm]{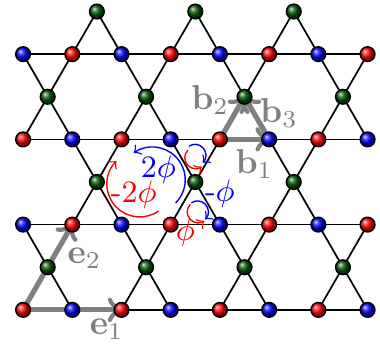}
\caption{
The schematic representation of the Kagome lattice. The lattice contains three sites  per unit cell, which we depict by red ($R$), blue ($B$), and green ($G$). ${\bf e}_1$ and ${\bf e}_2$ are the displacement vectors between the neighboring unit cells which form the triangular lattice. ${\bf b}_1$, ${\bf b}_2$, and ${\bf b}_3$ are the displacement vectors within the unit cell between $R$ and $B$, $R$ and $G$, and $B$ and $G$ sites, respectively. }
\label{schematicp}
\end{figure}

\section{Introduction}

The discovery of topological insulators has revolutionized the field of condensed matter physics in recent decades\cite{ka.me.05a, ka.me.05b, ko.wi.07, ha.ka.10, qi.zh.11, be.hu.13, gr.ab.14, co.da.19, be.li.16, go.bu.16, zh.zh.18, zh.ho.18}. 
Topological insulators have been observed not only in solid-state materials\cite{ge.be.13, yi.ma.20}, but also in the artificial systems created by ultracold atoms loaded in optical lattices\cite{ai.at.13, ai.lo.15, mi.hi.13, jo.me.14, fl.re.16, ma.pa.15, st.lu.15} and photonic systems\cite{kita.06, mi.na.18}. 
In particular, the Azbel-Harper-Hofstadter\cite{ai.at.13, mi.hi.13, ai.lo.15} and the Haldane\cite{jo.me.14, fl.re.16} models were experimentally realized. Robust edge states, indicating topological insulators, were also observed with ultracold atoms in synthetic dimensions\cite{ma.pa.15, st.lu.15}.
A common approach to achieve topological phases in cold atomic gas experiments is to employ artificial gauge fields\cite{ge.da.10, da.ge.11, go.da.14, go.ju.14, go.bu.16, ga.ju.19, ga.sp.13}. 
Realization of artificial spin-orbit coupling (SOC) was discussed in Ref. \onlinecite{ga.sp.13}. 
SOC has already been realized experimentally in ultracold atoms in the absence of optical lattices\cite{li.ji.11, wa.yu.12, ch.so.12,hu.me.16} and there are proposals how to realize it in the presence of the optical lattices\cite{du.di.04, gr.li.17}.

Topological insulators behave as insulators in the bulk, while they are conducting at their boundary. Despite that, their topological properties are classified according to the topological invariants which are determined based on their bulk properties.  Time-reversal symmetric nonmagnetic insulators are characterized by a $\mathbb{Z}_2$ invariant $\nu$,  
i.e., they are divided into two categories: a topological insulator with $\mathbb{Z}_2$ number $\nu=1$ and a trivial band insulator with $\nu=0$. The latter is adiabatically connected to the trivial state, while the former cannot be connected to the trivial state without closing a bulk gap.\cite{ka.me.05a, ka.me.05b}

There are a number of works that have already investigated topological  properties of the tight-binding model on the Kagome lattice\cite{oh.mu.00, ko.ho.10, gr.sa.10, zhan.11, pe.ho.12, xu.li.15, li.su.18, gu.ve.20, le.hu.20, gu.fr.09, li.zh.09,  wa.zh.10, li.zh.10, ta.me.11,  li.ch.12, li.ch.13, ch.ch.14, du.ch.18, bo.na.19, ku.yo.19, wa.we.19}, related to spin-orbit coupling\cite{gu.fr.09,ta.me.11, li.zh.10, wa.zh.10,li.zh.09}, staggered fluxes\cite{li.ch.13},  Hofstadter butterflies\cite{wa.we.19,du.ch.18}, breathing Kagome lattice\cite{bo.na.19},
flat band phases\cite{li.ch.12,ch.ch.14}, chiral edge modes\cite{pe.ho.12}, higher order topological Mott insulators\cite{ku.yo.19},
quantum anomalous Hall \cite{pe.ho.12,zhan.11, oh.mu.00, ko.ho.10, gr.sa.10, xu.li.15, li.su.18, gu.ve.20, le.hu.20}, fractional quantum Hall\cite{ta.me.11}, and spin Hall\cite{li.zh.10,wa.zh.10} effects.

Hereafter, we study topological properties of the spin-orbit coupled 
tight-binding model  with flux on the Kagome lattice, which is  non-Bravais lattice and contains three sites per unit cell.
In addition to the SOC and the flux produced by the artificial gauge fields imprinted as spin-dependent Peierls phases, we also take into account the effect of site-dependent on-site energies, which can be realized in experiments with ultracold atoms. 
We consider the cases where on-site energies can be different within the lattice unit cell but are the same for all lattice unit cells, and we also consider cases with staggered on-site energies between different lattice unit cells.

We show that fluxes and SOC, generated by artificial gauge fields, induce topological phases. We show that these phases are stable after applying on-site energies. 
Interestingly, a staggered potential may also induce a topological phase for the half-filled situation. A remarkable result we obtained is the existence of the topological phase for divergingly large on-site energies.

In this work, we also develop an analytical approach to describe the topological properties of the phases which allows us to understand the variations of the $\mathbb{Z}_2$ number.

The paper is organized as follows. In the next section (Sec.~\ref{Model}), we introduce the model Hamiltonian and rewrite it in momentum space. In Sec.~\ref{Method}, we first give an overview of the numerical method to compute the $\mathbb{Z}_2$ number based on twisted boundary conditions (Sec.~\ref{twisted-boundary-condition}). Afterwards, in Sec.~\ref{Z2-method} we present the analytical method which brings a different view on the topological property of the phases and allows us to understand the variations of the $\mathbb{Z}_2$ number. We present our results in Secs.~\ref{Results_without} and \ref{section_staggered_potential}. In Sec.~\ref{Results_without} we consider on-site energies that may differ within the unit cell but are the same for all unit cells, while in Sec.~\ref{section_staggered_potential} we consider a setup with the staggered potential on the lattice. 
In Sec.~\ref{Conclusions} we elaborate conclusive remarks. The paper also contains three appendices. In Appendix~\ref{appendixA}, 
we review and compare two other methods introduced in Refs.~\onlinecite{fu.ka.07}~and~\onlinecite{pe.ho.12}, to compute the $\mathbb{Z}_2$ number for our model. 
In Appendix~\ref{Conservation_along_E2} we derive expressions for conservation of the average current, and finally in Appendix~\ref{Details_Effective Hamiltonian}, we derive an effective Hamiltonian for large on-site energies.

\section{Model}\label{Model}

\subsection{Hamiltonian in real-space}
\label{rela-space-H}

We investigate the tight-binding model  with flux on the Kagome lattice. Experimentally, this lattice has been realized using ultracold atoms by superimposing two triangular optical lattices with different wavelengths.\cite{jo.gu.12} The resulting structure has three sites per unit cell, which we denote by $R$, $B$, and $G$ (see Fig.~\ref{schematicp}). The unit cells are arranged on a triangular lattice. The displacement vectors between the neighboring unit cells are 
\begin{eqnarray}
\label{Lattice_vectors}
{\bf e}_{1}=a\left(1,0\right) \quad {\rm and} \quad  {\bf e}_{2}=a\left(\frac{1}{2},\frac{\sqrt{3}}{2}\right) \, .
\end{eqnarray}
Here $a$ is the lattice constant. Here we have defined the displacement vector ${\bf e}_{3}={\bf e}_{2}-{\bf e}_{1}$.
We also introduce ${\bf b}_1=\frac{1}{2}{\bf e}_{1}$, ${\bf b}_2=\frac{1}{2}{\bf e}_{2}$, and ${\bf b}_3=\frac{1}{2}{\bf e}_{3}$
within a unit cell between  $R$ and $B$, $R$ and $G$, and $B$ and $G$, respectively.

The Hamiltonian in real-space representation reads
\begin{eqnarray}
\label{Hamiltonian}
{\cal H}&=&-t\sum_{{\bf r}}\Biggl[
c_{R,{\bf r}}^{\dagger}\mathbb{1}c_{B,{\bf r}}^{\phantom\dagger}
+c_{R,{\bf r}+{\bf e}_1}^{\dagger}\mathbb{1}c_{B,{\bf r}}^{\phantom\dagger} 
\nonumber\\
&+&c_{R,{\bf r}}^{\dagger}e^{-i2\pi\gamma\sigma^x}c_{G,{\bf r}}^{\phantom\dagger}
+c_{G,{\bf r}}^{\dagger}e^{-i2\pi\gamma\sigma^x}c_{R,{\bf r}+{\bf e}_2}^{\phantom\dagger}
\nonumber\\
&+&c_{B,{\bf r}}^{\dagger}e^{i\phi \sigma^z}c_{G,{\bf r}}^{\phantom\dagger}+c_{B,{\bf r}+{\bf e}_3}^{\dagger}e^{i\phi\sigma^z}c_{G,{\bf r}}^{\phantom\dagger} 
+ H.c.\Biggl] 
\nonumber \\
&+&\sum_{{\bf r}}\sum_{\alpha=R,B,G} V_{\alpha,{\bf r}} n_{\alpha,{\bf r}} \,.
\end{eqnarray}
Here, $c_{\alpha,{\bf r}}^{\dagger}=\left(c_{\alpha,{\bf r},\uparrow}^{\dagger},c_{\alpha, {\bf r},\downarrow}^{\dagger}\right)$ creates a fermion at site $\bf r$ for $\alpha=R$, at site ${\bf r}+{\bf b}_1$ for $\alpha=B$, and at site ${\bf r}+{\bf b}_2$ for $\alpha=G$, respectively. 
$n_{\alpha,{\bf r},\sigma}=c_{\alpha,{\bf r},\sigma}^{\dagger}c_{\alpha,{\bf r},\sigma}^{\phantom\dagger}$ is the fermion number operator for spin $\sigma$ on the corresponding site and $n_{\alpha,{\bf r}}=n_{\alpha,{\bf r},\uparrow}+n_{\alpha,{\bf r},\downarrow}$. 
We define the filling $n$ as $n=\frac{1}{3N_1N_2}\sum\limits_{\alpha,{\bf r}}\langle n_{\alpha,{\bf r}}\rangle$. Here $N_1$ and $N_2$ are the numbers of unit cells along ${\bf e}_1$ and ${\bf e}_2$, respectively.
$\sigma^x$ and $\sigma^z$ are the $x$ and $z$ Pauli matrices acting in spin space, while $\mathbb{1}$ is unit matrix. $t$ is the hopping amplitude of fermions  between  neighboring lattice sites. The third and the fourth terms are the Rashba-type  spin-orbit  coupling  terms\cite{by.ra.84}  of  strength $\gamma$,  which  determine the coupling between the two spin species via $e^{-i2\pi\gamma\sigma^x}=\mathbb{1}\cos(2\pi\gamma)-i\sigma^x\sin(2\pi\gamma)$. 
This form allows us to study linear effects associated to spin-flip terms and also non-linear processes.
Here, $\phi$ introduces a phase which acquires opposite signs for $\sigma=\uparrow$ and $\downarrow$ particles, such that the Hamiltonian preserves time-reversal symmetry. For small values of $\phi$, this term is similar to a Kane-Mele spin-orbit coupling introduced here between nearest-neighbor sites\cite{ka.me.05a, ka.me.05b}.
 
Finally $V_{\alpha,{\bf r}}$ is the on-site energy on the $\alpha$ sublattice and in general, it depends on the unit cell coordinate ${\bf r}$.  
On the one hand, we consider cases where the on-site energies are independent of $\bf r$ but can be different from each other inside the unit cell, i.e.,  $V_{\alpha,{\bf r}}=\lambda_\alpha$. In particular, we consider four different setups: 
\begin{enumerate}[(i)]
\item all on-site energies are zero ($\lambda_R=\lambda_B=\lambda_G=0$), 
\item $\lambda_B=-\lambda_R=\lambda$ and $\lambda_G=0$, 
\item $\lambda_R=\lambda$ and $\lambda_B=\lambda_G=0$,  
\item $\lambda_B=\lambda$ and $\lambda_R=\lambda_G=0$. 
\end{enumerate}
Here we note that the case when $\lambda_G=\lambda$ and $\lambda_R=\lambda_B=0$ is equivalent to the case where the on-site energies are non-zero for $R$ sublattice sites, i.e., case (ii) due to the symmetry discussed below (see Eq.~\eqref{R-G}). 

On the other hand, we also consider a staggered potential. For the latter, we have  $V_{\alpha,{\bf r}}=\lambda_{\alpha,1}$ for ${\bf r}=2n_1{\bf e}_1+n_2{\bf e}_2$ and  
$V_{\alpha,{\bf r}}=\lambda_{\alpha,2}$ for ${\bf r}=(2n_1+1){\bf e}_1+n_2{\bf e}_2$. Here $1 \leq n_1 \leq N_1 $ and $1\leq  n_2 \leq N_2 $
are integer numbers. Therefore, the size of the unit cell of the model is twice as large as the size of the unit cell of the lattice. Also in this case the unit cells are arranged on a triangular lattice, but with the displacement vectors between the neighboring unit cells ${\tilde{\bf e}_1=2{\bf e}_1}$ and ${\tilde{\bf e}_2={\bf e}_2}$.  We consider three cases  
\begin{enumerate}[(a)]
\item  $\lambda_{R,1}=\lambda_{B,1}=\lambda_{G,1}=-\lambda_{R,2}=-\lambda_{B,2}=-\lambda_{G,2}=\lambda$ for $n=2/3$ filling, 
\item  $\lambda_{R,1}=\lambda_{B,1}=\lambda_{G,1}=-\lambda_{R,2}=-\lambda_{B,2}=-\lambda_{G,2}=\lambda$ for half-filling ($n=1$),
\item $\lambda_{R,1}=-\lambda_{R,2}=\lambda$ and $\lambda_{B,s=1,2}=\lambda_{G,s=1,2}=0$ for $n=2/3$ filling.
\end{enumerate}

In contrast to our model, Harper\cite{harp.55} and Hofstadter\cite{hofs.76} in their original works considered a Peierls phase $\phi$ that depends on the site coordinate. As it was shown in Ref.~\onlinecite{hofs.76} such a Peierls phase produces a fractal spectrum known as the ``Hofstadter butterfly''.

\subsection{Hamiltonian in momentum space}
\label{momentum-space-H}

\subsubsection{Without staggered potential: $V_{\alpha,{\bf r}}=\lambda_{\alpha}$}

First we consider the system without the staggered potential, i.e. for $V_{\alpha,{\bf r}}=\lambda_{\alpha}$. We consider periodic boundary conditions along the ${\bf e}_1$ and the ${\bf e}_2$ directions, i.e. ${\bf r}+N_1{\bf e}_1={\bf r}$ and ${\bf r}+N_2{\bf e}_2={\bf r}$. 
We perform a Fourier transform
\begin{equation}
\label{Hk_matrix}
{\cal H}=\sum_{{\bf k}}\psi_{{\bf k}}^{\dagger}\left(
\begin{array}{cc}
{\cal H}_{\uparrow}({\bf k}) & {\cal H}_{RSO}({\bf k}) \\
{\cal H}_{RSO}^{\dagger}({\bf k}) & {\cal H}_{\downarrow}({\bf k})\\
\end{array}
\right)\psi_{{\bf k}}^{\phantom\dagger} \\
\end{equation}
with
\begin{eqnarray}
\label{Hsigmak_matrix}
&&\hspace{-0.75cm}{\cal H}_{\sigma}({\bf k})\hspace{-0.1cm}=\hspace{-0.1cm}\left(
\hspace{-0.1cm}
\begin{array}{ccc}
\lambda_R & \varepsilon_1({\bf k}) & \cos(2\pi\gamma)\varepsilon_2({\bf k}) \\
\varepsilon_1({\bf k}) & \lambda_B & e^{i s_z\phi}\varepsilon_3({\bf k}) \\
\cos(2\pi\gamma)\varepsilon_2({\bf k}) &e^{-i s_z\phi}\varepsilon_3({\bf k})& \lambda_G\\
\end{array}
\hspace{-0.1cm}
\right) \\
\label{HSO_matrix}
&&\hspace{-0.75cm}{\cal H}_{RSO}({\bf k})=\left(
\begin{array}{ccc}
0& 0 & \sin(2\pi\gamma)\xi({\bf k})\\
0 & 0 & 0\\
\sin(2\pi\gamma)\xi({\bf k}) & 0 & 0\\
\end{array}
\right) \,.
\end{eqnarray}
Here we have defined 
\begin{eqnarray}
\label{Psik}
\psi_{{\bf k}}^{\dagger}=\left(c_{R,{\bf k},\uparrow}^{\dagger},c_{B,{\bf k},\uparrow}^{\dagger},c_{G,{\bf k},\uparrow}^{\dagger},c_{R,{\bf k},\downarrow}^{\dagger},c_{B,{\bf k},\downarrow}^{\dagger},c_{G,{\bf k},\downarrow}^{\dagger}  \right) \,,
\end{eqnarray}
as well as $\varepsilon_\alpha=-2t\cos({\bf k}\cdot{\bf b}_\alpha)$ and $\xi({\bf k})=2t\sin({\bf k}\cdot{\bf b}_2)$.
Note that $s_z=1\,(-1)$ for $\sigma=\uparrow(\downarrow)$. 
The reciprocal lattice of the triangular lattice has the following basis vectors: 
\begin{eqnarray}
&&{\bf g}_1=2\pi\frac{R{\bf e}_2}{{\bf e}_1 \cdot R{\bf e}_2}=\frac{4\pi}{a\sqrt{3}}
\left(\frac{\sqrt{3}}{2},-\frac{1}{2}\right) \\
&&{\bf g}_2=2\pi\frac{R{\bf e}_1}{{\bf e}_2 \cdot R{\bf e}_1}=\frac{4\pi}{a\sqrt{3}}(0, 1) \, .
\end{eqnarray}
Here 
$$
R=\left(
\begin{array}{cc}
0 & -1\\
1 & 0
\end{array}
\right)
$$ represents a $\pi/2$ rotation matrix. Therefore 
\begin{equation}
{\bf k}=\frac{n_1}{N_1}{\bf g}_1 + \frac{n_2}{N_2}{\bf g}_2 
\end{equation}
We obtain ${{\bf k}\cdot {\bf b}_{\alpha} =\frac{\pi n_\alpha}{N_\alpha}}= k_\alpha$ for $\alpha=1,2$ and ${{\bf k}\cdot {\bf b}_{3}=\frac{\pi n_2}{N_2} - \frac{\pi n_1}{N_1}}=k_2-k_1$.

The Hamiltonian \eqref{Hk_matrix} has $6$ eigenvalues for each value of $\bf k$. Due to time-reversal symmetry the spectrum possesses non-movable band crossings which are known as Kramers degeneracies. This leaves $3$ bands potentially non-overlapping for any value of $\bf k$.  Based on that, gaps may appear between the second and third bands, when two from six bands are filled, as well as between the fourth and fifth bands when four from six bands are filled. Therefore, a gap may appear for the fillings $n=2/3$ and $n=4/3$. Just reminder when all bands are filled $n=2$.

\subsubsection{Staggered potential}

Now we consider the case when a staggered potential is applied, i.e. $V_{\alpha,{\bf r}}=\lambda_{\alpha,1}$  for ${\bf r}=2n_1{\bf e}_1+n_2{\bf e}_2$ and $V_{\alpha,{\bf r}}=\lambda_{\alpha,2}$ for
${\bf r}=(2n_1+1){\bf e}_1+n_2{\bf e}_2$. In this case, as it was mentioned above, the displacement vectors between the neighboring unit cells are $\tilde{\bf e}_1=2{\bf e}_1$ and $\tilde{\bf e}_2={\bf e}_2$. We again consider periodic boundary conditions. 
Taking into account the size of the unit cell which is twice as large we have
${\bf r}+\frac{N_1}{2}\tilde{\bf e}_1={\bf r}$ and  ${\bf r}+N_2\tilde{\bf e}_2={\bf r}$. In this case, the number of unit cells along the ${\bf e}_1$ direction is $N_1/2$. After the Fourier transform we obtain
\begin{eqnarray}
\label{Hk_matrix_staggered}
&&\hspace{-0.5cm}{\cal H}=\sum_{{\bf k}}\tilde\psi_{{\bf k}}^{\dagger}\hspace{-0.1cm}\left(
\hspace{-0.15cm}
\begin{array}{cccc}
{\cal H}_{1,\uparrow}({\bf k}) & {\cal H}_{t,\uparrow}({\bf k}) & {\cal H}_{RSO}({\bf k}) & 0\\
{\cal H}_{t,\uparrow}^\dagger({\bf k}) & {\cal H}_{2,\uparrow}({\bf k}) &  0 & {\cal H}_{RSO}({\bf k})\\
{\cal H}_{RSO}^{\dagger}({\bf k}) & 0 & {\cal H}_{1,\downarrow}({\bf k}) & {\cal H}_{t,\downarrow}({\bf k})\\
0 & {\cal H}_{RSO}^{\dagger}({\bf k}) & {\cal H}_{t,\downarrow}^\dagger({\bf k}) & 
{\cal H}_{2,\downarrow}^\dagger({\bf k})
\end{array}
\hspace{-0.15cm}
\right)\hspace{-0.1cm}\tilde\psi_{{\bf k}}^{\phantom\dagger} \nonumber\\
\\
&&\hspace{-0.5cm}\mbox{with:} \nonumber\\
\label{Hisigmak_matrix_staggered}
&&\hspace{-0.5cm}{\cal H}_{i,\sigma}({\bf k})\hspace{-0.1cm}=\hspace{-0.1cm}\left(
\hspace{-0.1cm}
\begin{array}{ccc}
\lambda_{R,i} & -te^{-i{\bf k} \cdot {\bf b}_1}
& \cos(2\pi\gamma)\varepsilon_2({\bf k}) \\
-te^{i{\bf k} \cdot {\bf b}_1} & \lambda_{B,i} & -te^{i\sigma\phi}e^{-i{\bf k} \cdot {\bf b}_3}  \\
\cos(2\pi\gamma)\varepsilon_2({\bf k}) &-te^{-i\sigma\phi}e^{i{\bf k} \cdot {\bf b}_3} & \lambda_{G,i}
\end{array}
\hspace{-0.1cm}
\right) \nonumber\\
\\
\label{Htsigmak_matrix_staggered}
&&\hspace{-0.5cm}{\cal H}_{t,\sigma}({\bf k})\hspace{-0.1cm}=\hspace{-0.1cm}\left(
\hspace{-0.1cm}
\begin{array}{ccc}
0 & -te^{i{\bf k} \cdot {\bf b}_1}
& 0 \\
-te^{-i{\bf k} \cdot {\bf b}_1} & 0 & -te^{i\sigma\phi}e^{i{\bf k} \cdot {\bf b}_3} \\
0 &-te^{-i\sigma\phi}e^{-i{\bf k} \cdot {\bf b}_3} & 0
\end{array}
\hspace{-0.1cm}
\right)\,. 
\end{eqnarray}
Here
\begin{eqnarray}
\label{Psiktilde}
&&\hspace{-0.75cm}\tilde\psi_{{\bf k}}^{\dagger}=\left(c_{R1,{\bf k},\uparrow}^{\dagger},c_{B1,{\bf k},\uparrow}^{\dagger},c_{G1,{\bf k},\uparrow}^{\dagger},c_{R2,{\bf k},\uparrow}^{\dagger},c_{B2,{\bf k},\uparrow}^{\dagger},c_{G2,{\bf k},\uparrow}^{\dagger},\right.
\nonumber\\ 
&&\hspace{-0.1cm}\left.c_{R1,{\bf k},\downarrow}^{\dagger},c_{B1,{\bf k},\downarrow}^{\dagger},c_{G1,{\bf k},\downarrow}^{\dagger},c_{R2,{\bf k},\downarrow}^{\dagger},c_{B2,{\bf k},\downarrow}^{\dagger},c_{G2,{\bf k},\downarrow}^{\dagger}  \right) \,.
\end{eqnarray}

The reciprocal lattice of the triangular lattice with the extended unit cell has the following basis vectors: 
\begin{eqnarray}
&&\tilde {\bf g}_1=2\pi\frac{R\tilde{\bf e}_2}{\tilde{\bf e}_1 \cdot R\tilde{\bf e}_2}=\frac{2\pi}{a\sqrt{3}}
\left(\frac{\sqrt{3}}{2},-\frac{1}{2}\right) \\
&&\tilde{\bf g}_2=2\pi\frac{R\tilde{\bf e}_1}{\tilde{\bf e}_2 \cdot R\tilde{\bf e}_1}=\frac{4\pi}{a\sqrt{3}}(0, 1) \, ,
\end{eqnarray}
and taking into account that the number of unit cells along ${\bf e}_1$ is $N_1/2$ we have
\begin{equation}
{\bf k}=\frac{n_1}{N_1/2}\tilde {\bf g}_1 + \frac{n_2}{N_2}\tilde {\bf g}_2   \,.
\end{equation}
Here, $-\frac{N_1}{4} < n_1 \leq \frac{N_1}{4} $ and $-\frac{N_2}{2} < n_2 \leq \frac{N_2}{2} $ are integer numbers.
We obtain ${{\bf k}\cdot {\bf b}_{1}=\frac{\pi n_1}{N_1}}= \frac{1}{2} \tilde k_1 $, ${{\bf k}\cdot {\bf b}_{2} =\frac{\pi n_2}{N_2}}= \tilde k_2$
and ${{\bf k}\cdot {\bf b}_{3}=\frac{\pi n_2}{N_2} - \frac{\pi n_1}{N_1}}=\tilde k_2-\frac{1}{2}\tilde k_1$.

The Hamiltonian \eqref{Hk_matrix_staggered} has $12$ eigenvalues for each value of $\bf k$. Due to the Kramers degeneracy, discussed above, there are $6$ bands potentially non-overlapping for any value of $\bf k$.  They are between the $2m$-th and $(2m+1)$th bands ($m=1,\ldots,6$). Therefore gaps may appear for the fillings $n=1/3$, $n=2/3$, $n=1$, $n=4/3$, and $n=5/3$.

\subsection{Symmetries}
\label{subsectin:symmetry}

In this subsection, we discuss the symmetries of the Hamiltonian \eqref{Hamiltonian}. To detect them we consider different gauge transformations as well as the particle-hole symmetry. 

We start with noting that there is no physical difference between $R$ and $G$ sites in the unit cell. Interchanging them is equivalent to the following gauge transformation:
\begin{equation}
\label{R-G}
c_{B,{\bf r}} \leftrightarrow c_{B,{\bf r}}e^{i\phi\sigma^z} \quad
R \leftrightarrow G \,.
\end{equation}

One can easily show that due to the gauge transformation
\begin{eqnarray}
c_{\alpha,{\bf r}}^{\dagger}  \leftrightarrow  c_{\alpha,{\bf r}}^{\dagger}  e^{i\frac{\pi}{2}\sigma^x}  
= i c_{\alpha,{\bf r}}^{\dagger} \sigma^x \quad\quad\quad \alpha=R\,,\,B\,,\,G\,,
\end{eqnarray}
the following relations hold
\begin{eqnarray}
\label{symmetry_sigma_x} 
{\cal H}(\gamma,\phi,V_{\alpha,{\bf r}}) \leftrightarrow
{\cal H}(\gamma,2\pi-\phi, V_{\alpha,{\bf r}}) \,.
\end{eqnarray}
The gauge transformation
\begin{eqnarray}
c_{\alpha,{\bf r}}^{\dagger}  \leftrightarrow  c_{\alpha,{\bf r}}^{\dagger}  e^{i\frac{\pi}{2}\sigma^y}  
=i c_{\alpha,{\bf r}}^{\dagger} \sigma^y    \quad\quad\quad \alpha=R\,,\,B\,,\,G\,.
\end{eqnarray}
gives us 
\begin{eqnarray}
\label{symmetry_sigma_y} 
{\cal H}(\gamma,\phi,V_{\alpha,{\bf r}}) \leftrightarrow
{\cal H}(1-\gamma,2\pi-\phi, V_{\alpha,{\bf r}}) \,.
\end{eqnarray}
While after the gauge transformation
\begin{eqnarray}
c_{\alpha,{\bf r}}^{\dagger}  \leftrightarrow  c_{\alpha,{\bf r}}^{\dagger}  e^{i\frac{\pi}{2}\sigma^z}  =
i c_{\alpha,{\bf r}}^{\dagger} \sigma^z  \quad\quad\quad \alpha=R\,,\,B\,,\,G\,.
\end{eqnarray}
we obtain
\begin{eqnarray}
\label{symmetry_sigma_z} 
{\cal H}(\gamma,\phi,V_{\alpha,{\bf r}}) \leftrightarrow
{\cal H}(1-\gamma,\phi, V_{\alpha,{\bf r}}) \,.
\end{eqnarray}
Finally, the gauge transformation
\begin{eqnarray}
\label{gauge_pi_RB}
c_{R,{\bf r}}^{\dagger}  \leftrightarrow  c_{R,{\bf r}}^{\dagger}  e^{i \pi}  
=- c_{R,{\bf r}}^{\dagger}\,,\,\,
c_{B,{\bf r}}^{\dagger}  \leftrightarrow  c_{B,{\bf r}}^{\dagger}  e^{i \pi}  
=- c_{B,{\bf r}}^{\dagger}  \,,
\end{eqnarray}
gives us
\begin{eqnarray}
\label{symmetry_pi_RB} 
{\cal H}(\gamma,\phi,V_{\alpha,{\bf r}}) \leftrightarrow
{\cal H}(\gamma-\frac{1}{2},\pi-\phi, V_{\alpha,{\bf r}}) \,.
\end{eqnarray}

Due to the fact that $\gamma$ and $\phi$ are in the exponent and are periodic variables it is enough to consider only 
\begin{equation}
0 \leq \gamma < 1 
\quad\quad \mbox{and} \quad\quad 
0 \leq \phi < 2\pi \,.
\end{equation}
Based on the Eqs. \eqref{symmetry_sigma_x}, \eqref{symmetry_sigma_y}, \eqref{symmetry_sigma_z}, and \eqref{symmetry_pi_RB}, we can focus on the region  
\begin{eqnarray}
\label{Symmetry_1}
0\leq \gamma \leq 0.25
\quad\quad \mbox{and} \quad\quad
0 \leq \phi  \leq \pi \,,
\end{eqnarray}
but alternatively one can also consider
\begin{eqnarray}
\label{Symmetry_2}
0\leq \gamma \leq 0.5 
\quad\quad \mbox{and} \quad\quad
0 \leq \phi \leq \frac{\pi}{2}  \,.
\end{eqnarray}

To check further symmetries of the model, we consider the particle-hole transformation. In general, the Hamiltonian under consideration (Eq. \eqref{Hamiltonian}) is not symmetric under the particle-hole symmetry, but it  maps different parameter sets to each other.

We perform the following particle-hole transformation:
\begin{equation}
\label{Particle-hole}
c_{R,{\bf r}}^{\dagger}  \leftrightarrow  \sigma^z c_{R,{\bf r}}^{\phantom\dagger}\,,\,
c_{B,{\bf r}}^{\dagger}  \leftrightarrow - \sigma^z c_{B,{\bf r}}^{\phantom\dagger}\,,\,
c_{G,{\bf r}}^{\dagger}  \leftrightarrow - \sigma^z c_{G,{\bf r}}^{\phantom\dagger}\,.
\end{equation}
After simple calculations, we obtain that the Hamiltonian under consideration has the following symmetry
\begin{eqnarray}
\label{symmetry} 
{\cal H}(\gamma,\phi,V_{\bf r})\,\mbox{ for }\,n \leftrightarrow
{\cal H}(\gamma,\pi-\phi, -V_{\bf r})\, \mbox{ for }\, 2-n  \,. 
\end{eqnarray}
Here $n$ is the average filling.

Eventually, we consider the time-reversal symmetry. We write the time-reversal operator $\Theta=i \sigma^y K$ with $\sigma^y$ the $y$ Pauli matrix acting in spin space and $K$ denoting complex conjugation. The Hamiltonian~\eqref{Hamiltonian} is invariant under time-reversal symmetry: $\Theta {\cal H} \Theta^{-1} = {\cal H}$. We can check it explicitly by rewriting ${e^{-i2\pi\gamma\sigma^x} = \mathbb{1} \cos 2\pi\gamma - i \sigma^x \sin 2\pi\gamma }$ and ${e^{i\phi\sigma^z} = \mathbb{1} \cos \phi  + i \sigma^z \sin \phi }$ and remembering that ${\sigma^y \sigma^x \sigma^{y*} = \sigma^x}$ and ${\sigma^y \sigma^z \sigma^{y*} = \sigma^z}$. Then we write ${\cal H}=\sum_{{\bf k}}\psi_{{\bf k}}^{\dagger} {\cal H}({\bf k}) \psi_{{\bf k}}^{\phantom\dagger}$, with ${\cal H}({\bf k})$ a generic notation for the momentum space Hamiltonian matrix which appear in Eq.~\eqref{Hk_matrix} or in Eq.~\eqref{Hk_matrix_staggered}, depending on the situation we consider. We have $\Theta {\cal H}({\bf k}) \Theta^{-1}= {\cal H}({-\bf k})$. Indeed we notice that ${\cal H}_{\sigma}({-\bf k})={\cal H}_{\overline{\sigma}}({\bf k})^*$, ${\cal H}_{RSO}({-\bf k})=-\left({\cal H}_{RSO}({\bf k})^{\dagger}\right)^*$, ${\cal H}_{i,\sigma}({-\bf k})={\cal H}_{i,\overline{\sigma}}({\bf k})^*$ and ${\cal H}_{t,\sigma}({-\bf k})={\cal H}_{t,\overline{\sigma}}({\bf k})^*$. Here $\overline{\uparrow}=\downarrow$ and $\overline{\downarrow}=\uparrow$. 

\section{Methods}
\label{Method}

In this Section, we review the different methods which allow us to calculate the $\mathbb{Z}_2$ number and which are useful to characterize the topological phases. One of them is numerical and based on the twisted boundary conditions, while the other two are analytical and also introduce geo\-metrical arguments specific to the Kagome lattice (see Appendix \ref{appendixA}). 
Another way to calculate the $\mathbb{Z}_2$ number is to use the Wilson loop.\cite{he.so.20, yu.qi.11, gr.ab.14, al.da.14, bo.bl.19, ir.zh.20}

\subsection{Calculation of the $\mathbb{Z}_2$ number using twisted boundary conditions}
\label{twisted-boundary-condition}

First, we review the approach employing twisted boundary conditions\cite{fu.ha.05, fu.ha.07, ku.me.16, ir.zh.20} following closely the definitions of Ref.~\onlinecite{ku.me.16}. We consider spin-dependent twisted boundary conditions along the ${\bf e}_1$ direction and spin-independent twisted boundary along the ${\bf e}_2$ direction. So we have
\begin{equation}
\label{twsited_boundary_condition}
c_{{\bf r}+L_1{\bf e}_1,\alpha}=e^{i\sigma_z \theta_{1}}c_{{\bf r},\alpha}
\quad{\rm and}\quad
c_{{\bf r}+L_2{\bf e}_2,\alpha}=e^{i \mathbb{1}\theta_{2}}c_{{\bf r},\alpha}\,.
\end{equation}
Here, $L_1 \times L_2$ is the 2D sample area while $\boldsymbol{\theta}=(\theta_1,\theta_2)$  is the vector of the two twist angles. Here, it is worth mentioning that the periodic boundary condition is the special case of the twisted boundary condition \eqref{twsited_boundary_condition} with $\theta_1=\theta_2=0$. 

We perform calculations for a relatively small real-space sample $L_1 \times L_2$ for the different values of $\boldsymbol{\theta}$. Therefore we introduce another grid of size  $N_{\theta_1}$ and  $N_{\theta_2}$, such that $\theta_{\kappa=1,2}=2\pi n_{\kappa}/N_{\theta_\kappa}$, where $-N_{\theta_\kappa}/2 \leq n_{\kappa} <N_{\theta_\kappa}/2$.

Furthermore we define the $U(1)$ link variables
\begin{eqnarray}
U_{\kappa=1,2}(\boldsymbol{\theta})=\frac{\det g_\kappa}{|\det g_\kappa|}\,,
\end{eqnarray}
which are functions of the twist angle $\boldsymbol{\theta}$. Here
\begin{eqnarray}
[g_\kappa((\boldsymbol{\theta})]_{ab}=\langle \psi_a(\boldsymbol{\theta})|\psi_b(\boldsymbol{\theta}+\boldsymbol{\mu}_\kappa)\rangle 
\end{eqnarray}
are matrices with dimension equal to the number of occupied eigenstates of the Hamiltonian ${|\psi_a(\boldsymbol{\theta})\rangle}$ for a given twist angle $\boldsymbol{\theta}$. 
$\boldsymbol{\mu}_{1}=\left(2\pi/N_{\theta_1},0\right)$ and 
$\boldsymbol{\mu}_{2}=\left(0,2\pi/N_{\theta_2}\right)$ 
are unit vectors in the respective directions. 

Now we can define 
the Berry curvature
\begin{eqnarray}
\label{Bery_curvature}
\Omega(\boldsymbol{\theta})
=\log \left[U_1(\boldsymbol{\theta})
U_2(\boldsymbol{\theta}+\boldsymbol{\mu}_{1})
U_1(\boldsymbol{\theta}+\boldsymbol{\mu}_{2})^{-1}
U_2(\boldsymbol{\theta})^{-1}
\right] \,.
\end{eqnarray}

For time-reversal invariant systems the $\mathbb{Z}_2$ number can be calculated by summing over all $\boldsymbol{\theta}$
\begin{equation}
\label{nu_number_definition} 
\nu \equiv \left[\frac{1}{4\pi i} \sum_{\boldsymbol{\theta}}\Omega(\boldsymbol{\theta})\right]  \mod \,\, 2 \,. 
\end{equation}
One can show that Eq. \eqref{nu_number_definition} always produces an integer for the gapped phase, where it is well-defined. This number quickly converges to the actual $\mathbb{Z}_2$ number when increasing the numerical accuracy.
The advantage of this method is that  Eq. \eqref{Bery_curvature} is gauge invariant due to the fact that the phases of a $U(1)$ gauge transformation will always cancel out.

\subsection{Analytical computation of the $\mathbb{Z}_2$ number}
\label{Z2-method}

Here we consider the case $V_{\alpha,{\bf r}}=\lambda_{\alpha}$. We describe an analytical computation for the $\mathbb{Z}_2$ number at $n=2/3$ filling. This computation is first performed at $\gamma=0$ 
for all the values of the flux $\phi$ and for arbitrary on-site energy $\lambda$ and it is consistent with both approaches described in appendix~\ref{appendixA}.  Then the results are extended to the $\gamma \neq 0$ case, either by adiabatic evolution of the Hamiltonian or by considering the evolution of the current operator average value when turning on the $\gamma$ term. This analytical approach is complementary to the previous approach. It allows to understand (mathematically speaking) the variations of the $\mathbb{Z}_2$ number.

\begin{figure}[t!]
\includegraphics[width=8cm]{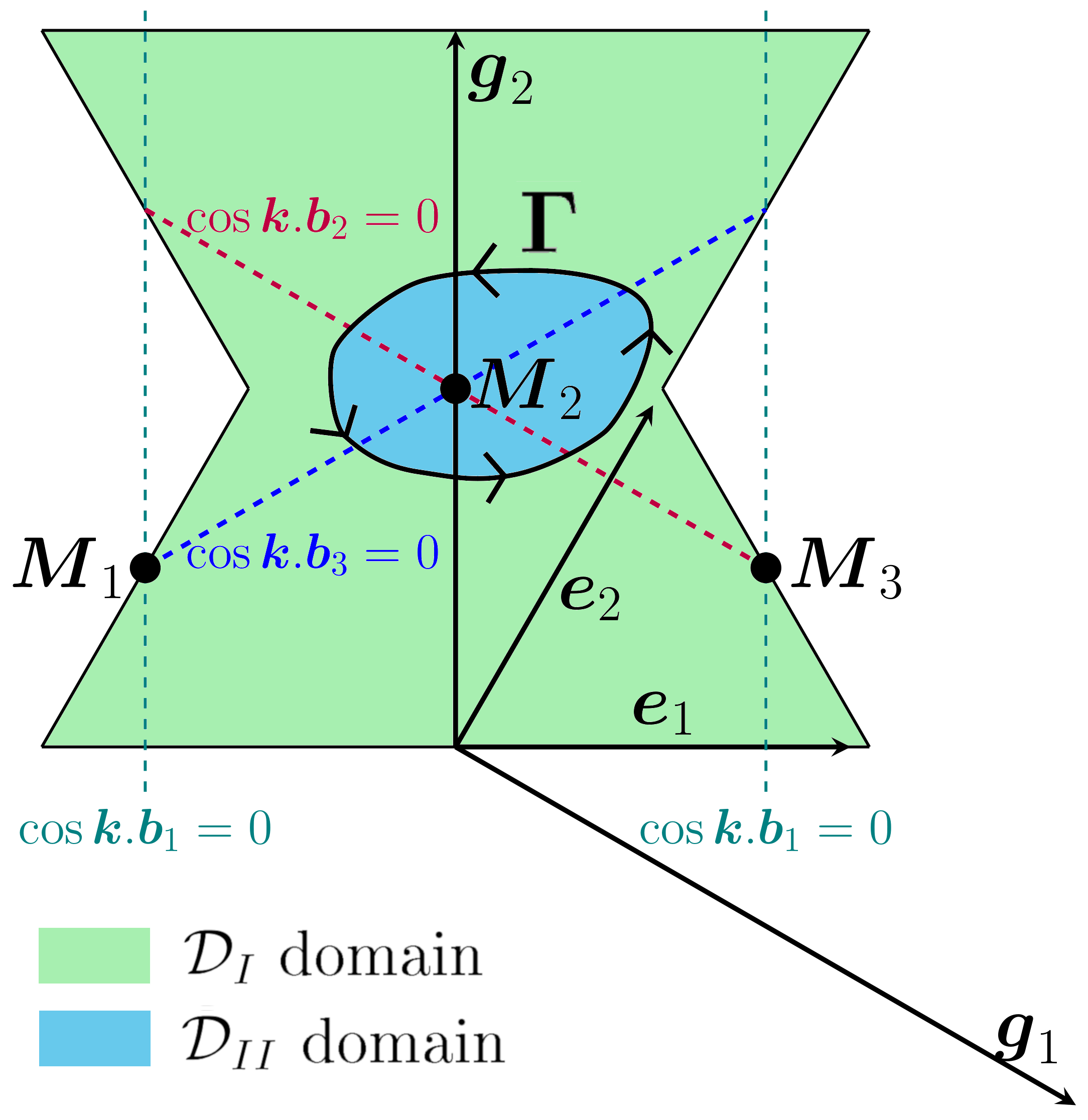}
\caption{Definition of the Brillouin zone into two domains (see Sec. \ref{Z2-method}),
each associated to a different gauge choice for the eigenvectors.}
\label{domains}
\end{figure}

\subsubsection{The $\gamma =0$ case}

When $\gamma =0$, the Hamiltonian (see expression in Eq.~\ref{Hk_matrix}) decouples into two independent parts 
for $\sigma=\uparrow$ and $\sigma=\downarrow$
%
It gives 3 energy bands associated to spin up states, which are degenerate with the 3 energy bands associated to the spin down states. 
Due to this decoupling, we have the following relation for the $\mathbb{Z}_2$ topological number\cite{fu.ka.06, sh.we.06, kane.13}
\begin{equation}
\label{nuequaldifferenceC}
\nu=\frac{1}{2}(C_{\uparrow}-C_{\downarrow})\,\,\, \mbox{mod}\,\,\, 2 =
C_{\uparrow}\,\,\, \mbox{mod}\,\,\, 2 \,,
\end{equation}
where
\begin{equation}
	C_\sigma = \dfrac{1}{2 i \pi} \int d^2k \left[ \boldsymbol{\nabla}_{{\bf k}} \times \boldsymbol{A}_{\sigma,{\bf k}} \right]\cdot\boldsymbol{e}_z ,
\end{equation}
is the spin Chern number and
$\boldsymbol{e}_z$ is a unit vector perpendicular to the Kagome lattice and the so-called Berry gauge field ${\boldsymbol{A}_{\sigma,{\bf k}} = \int d^2r u_{\sigma,{\bf k}}^{*}(\boldsymbol{r}) \boldsymbol{\nabla}_{{\bf k}} u_{\sigma,{\bf k}}(\boldsymbol{r}) = \bra{u_{\sigma,{\bf k}}}\boldsymbol{\nabla}_{{\bf k}} \ket{u_{\sigma,{\bf k}}} }$, associated to the lowest energy band with spin $\sigma$. $u_{\sigma,{\bf k}}(\boldsymbol{r}) = u_{\sigma,{\bf k}}(\boldsymbol{r}+\boldsymbol{R})$, with $\boldsymbol{R}$ the (magnetic) Bravais lattice vector, is the periodic (in real space) part of the Bloch eigenvector $\Psi_{{\bf k}}(\boldsymbol{r}) = \textrm{e}^{i{\bf k}\cdot\boldsymbol{r}} u_{\sigma,{\bf k}}(\boldsymbol{r})$. The integration is performed over the whole Brillouin zone (BZ). Because the BZ is a torus, if we can find a gauge choice such that $\boldsymbol{A}_{\sigma,{\bf k}}$ is uniquely and smoothly defined over all the BZ, then using the Stokes' theorem, we find that $C_\sigma$ is vanishing. Therefore, a non-trivial topology comes from the impossibility of finding such a gauge choice that makes $\boldsymbol{A}_{\sigma,{\bf k}}$ uniquely and smoothly defined. 
Under a gauge transformation $u_{\sigma,{\bf k}}(\boldsymbol{r}) \rightarrow u_{\sigma,{\bf k}}(\boldsymbol{r})\textrm{e}^{if_\sigma({\bf k})}$, with $f_\sigma$ a smooth function of ${\bf k}$ and independent of $\boldsymbol{r}$, we have $\boldsymbol{A}_{\sigma,{\bf k}} \rightarrow \boldsymbol{A}_{\sigma,{\bf k}} + i \boldsymbol{\nabla}_{{\bf k}} f_\sigma({\bf k})$, so $C_\sigma \rightarrow C_\sigma$. Now, suppose that the BZ is divided into 2 domains denoted $\mathcal{D}_{I}$ and $\mathcal{D}_{II}$, where the respective gauge choices $\ket{u_{\sigma,{\bf k},I}}$ and $\ket{u_{\sigma,{\bf k},II}}$ are unique and smooth (for illustration see Fig. \ref{domains}). We define $\varphi_\sigma({\bf k})$ a smooth function of ${\bf k}$ such that $\ket{u_{\sigma,{\bf k},I}} =\textrm{e}^{i\varphi_\sigma({\bf k})}\ket{u_{\sigma,{\bf k},II}}$ and we define ${\boldsymbol{A}_{\sigma,{\bf k},I} = \bra{u_{\sigma,{\bf k},I}}\boldsymbol{\nabla}_{{\bf k}} \ket{u_{\sigma,{\bf k},I}}}$ and ${\boldsymbol{A}_{\sigma,{\bf k},II} = \bra{u_{\sigma,{\bf k},II}}\boldsymbol{\nabla}_{{\bf k}} \ket{u_{\sigma,{\bf k},II}}}$ which are both uniquely and smoothly defined fields respectively inside $\mathcal{D}_{I}$ and $\mathcal{D}_{II}$. 
Then after using Stokes' theorem we obtain
\begin{equation}
\label{Stokes_theorem}
	C_\sigma = \dfrac{1}{2 i \pi} \left( \oint_{\Gamma} d{\bf k}  \cdot \boldsymbol{A}_{\sigma,{\bf k},I} - \oint_{\Gamma} d{\bf k} \cdot \boldsymbol{A}_{\sigma,{\bf k},II}\right).
\end{equation}
Here $\Gamma$ is the boundary between $\mathcal{D}_{I}$ and $\mathcal{D}_{II}$. 
Using $\boldsymbol{A}_{\sigma,{\bf k},I} = \boldsymbol{A}_{\sigma,{\bf k},II} + i \boldsymbol{\nabla}_{{\bf k}} \varphi_\sigma({\bf k})$, we find
\begin{equation} \label{chern_number}
	C_\sigma = \dfrac{1}{2 \pi} \oint_{\Gamma} d{\bf k} \cdot \boldsymbol{\nabla}_{{\bf k}} \varphi_\sigma ({\bf k}).
\end{equation}
In the results of Sec.~\ref{Results_without}, we detail this computation with convenient gauge choices depending on the value of the parameters.
We should notice that the phase $\oint_{\Gamma} d{\bf k} \cdot \boldsymbol{\nabla}_{{\bf k}} \varphi_\sigma ({\bf k})$ accumulated here is different from a Berry phase accumulated by the wave packet $\ket{u_{\sigma,{\bf k}}}$ along the $\Gamma$ path\cite{pe.ho.12}. The latter can be written $\oint_{\Gamma} d{\bf k}  \cdot \boldsymbol{A}_{\sigma,{\bf k}}$ and it is the contribution to the Chern number of the Berry curvature integrated over the domain delimited by $\Gamma$. 
The former is the Berry curvature integrated over all the BZ and is then related to the difference of Berry phases in Eq. \eqref{Stokes_theorem}. When the Berry curvature takes non-negligible values only around $L$ points that we denote $\boldsymbol{K}_l, \, l=1, \dots, L$, the Chern number is well approximated by $\sum_{l} \oint_{\Gamma_{\boldsymbol{K}_l}} d{\bf k}  \cdot \boldsymbol{A}_{\sigma,{\bf k}}$ where $\Gamma_{\boldsymbol{K}_l}$ is the boundary delimiting the domain around the $\boldsymbol{K}_l$ point where the Berry curvature takes non-negligible values. Usually, the contributions $\oint_{\Gamma_{\boldsymbol{K}_l}} d{\bf k}  \cdot \boldsymbol{A}_{\sigma,{\bf k}}$ are numerically evaluated.
This is what is done in Ref.~\onlinecite{pe.ho.12} and it provides a way to access the quantity in Eq.\eqref{Stokes_theorem}.
\\

\begin{figure*}[t!]
\subfigure[$\gamma=0$ and $\phi=\pi/2$]{
\label{Fig:Spectrum_gamma0_phi0p5pi}
\centering \includegraphics[width=7cm]{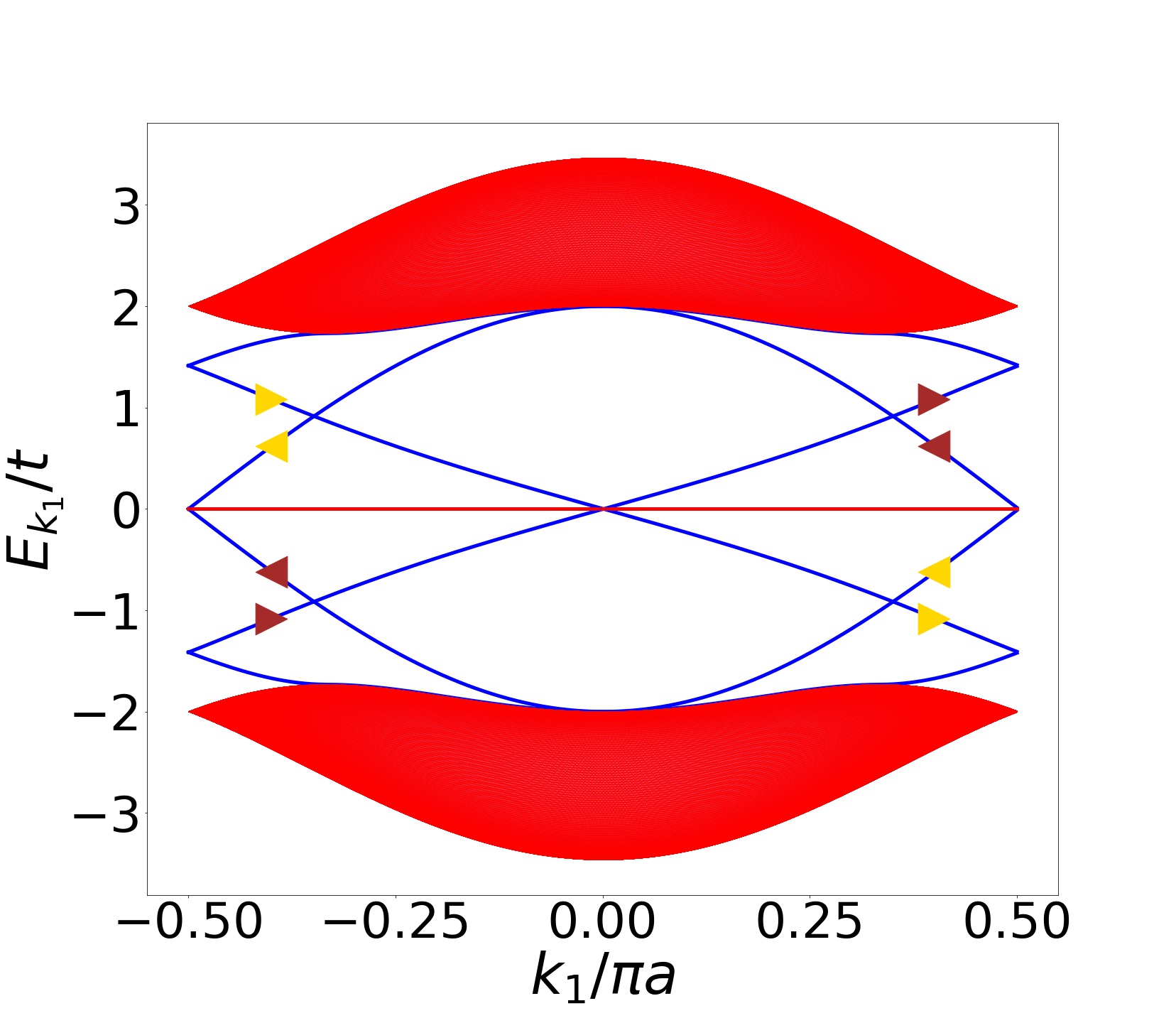} }
\subfigure[$\gamma=0.05$ and $\phi=\pi/2$]{
\label{Fig:Spectrum_gamma0p05_phi0p5pi}
\centering \includegraphics[width=7cm]{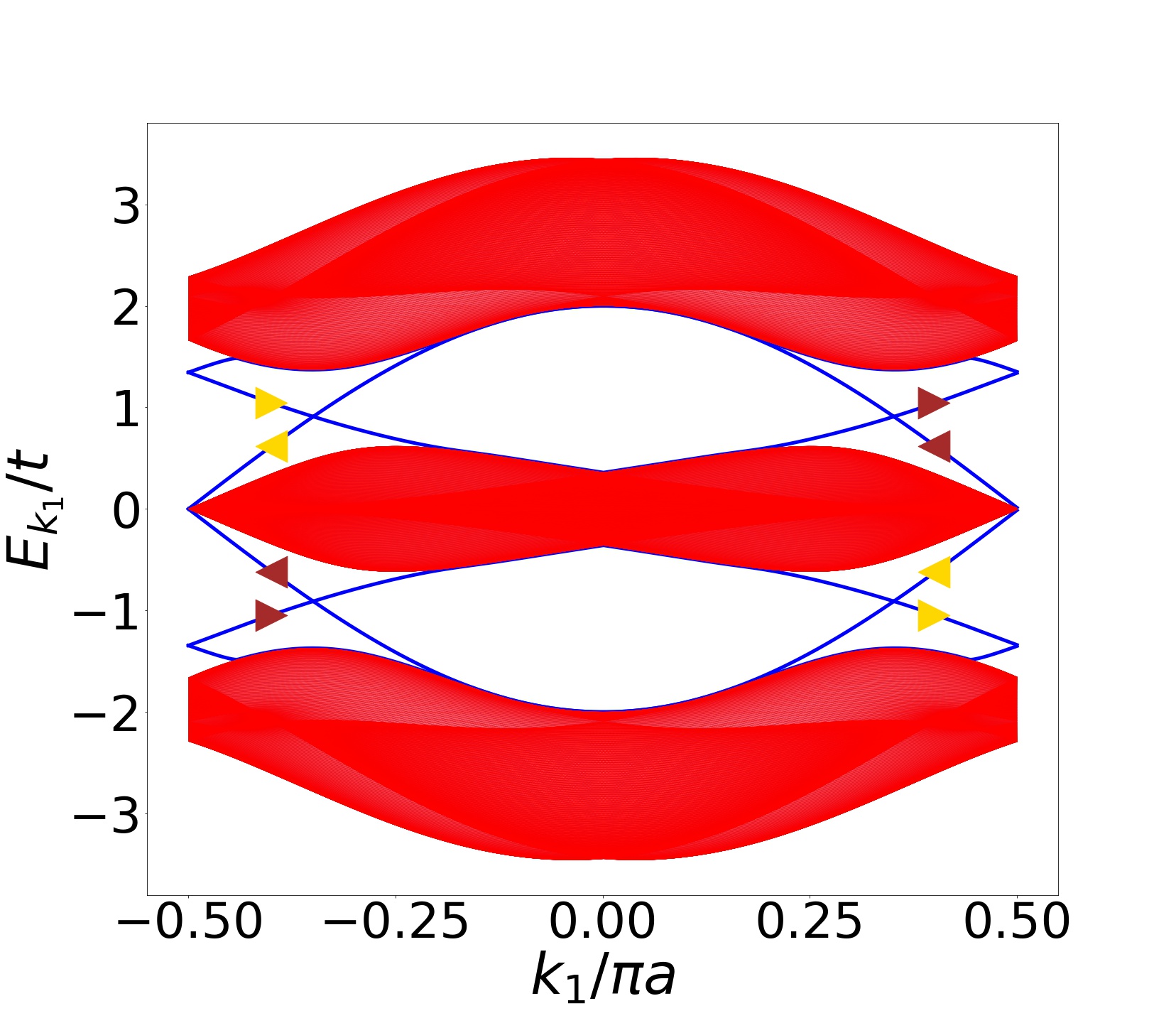} }\\
\subfigure[$\gamma=0.15$ and $\phi=\pi/2$]{
\label{Fig:Spectrum_gamma0p15_phi0p5pi}
\centering \includegraphics[width=7cm]{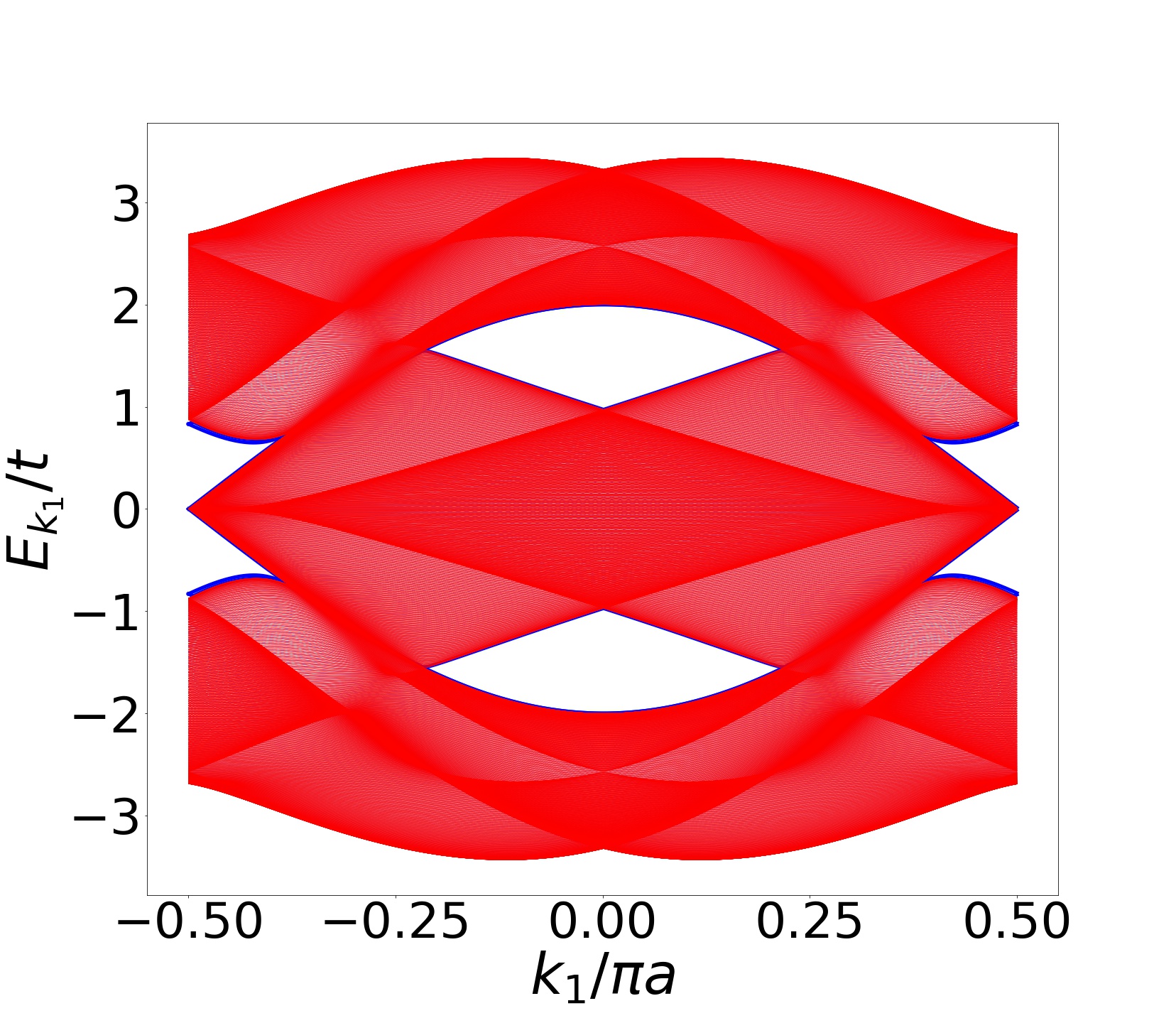} }
\subfigure[$\gamma=0.1$ and $\phi=\pi/4$]{
\label{Fig:Spectrum_gamma0p1_phi0p25pi}
\centering \includegraphics[width=7cm]{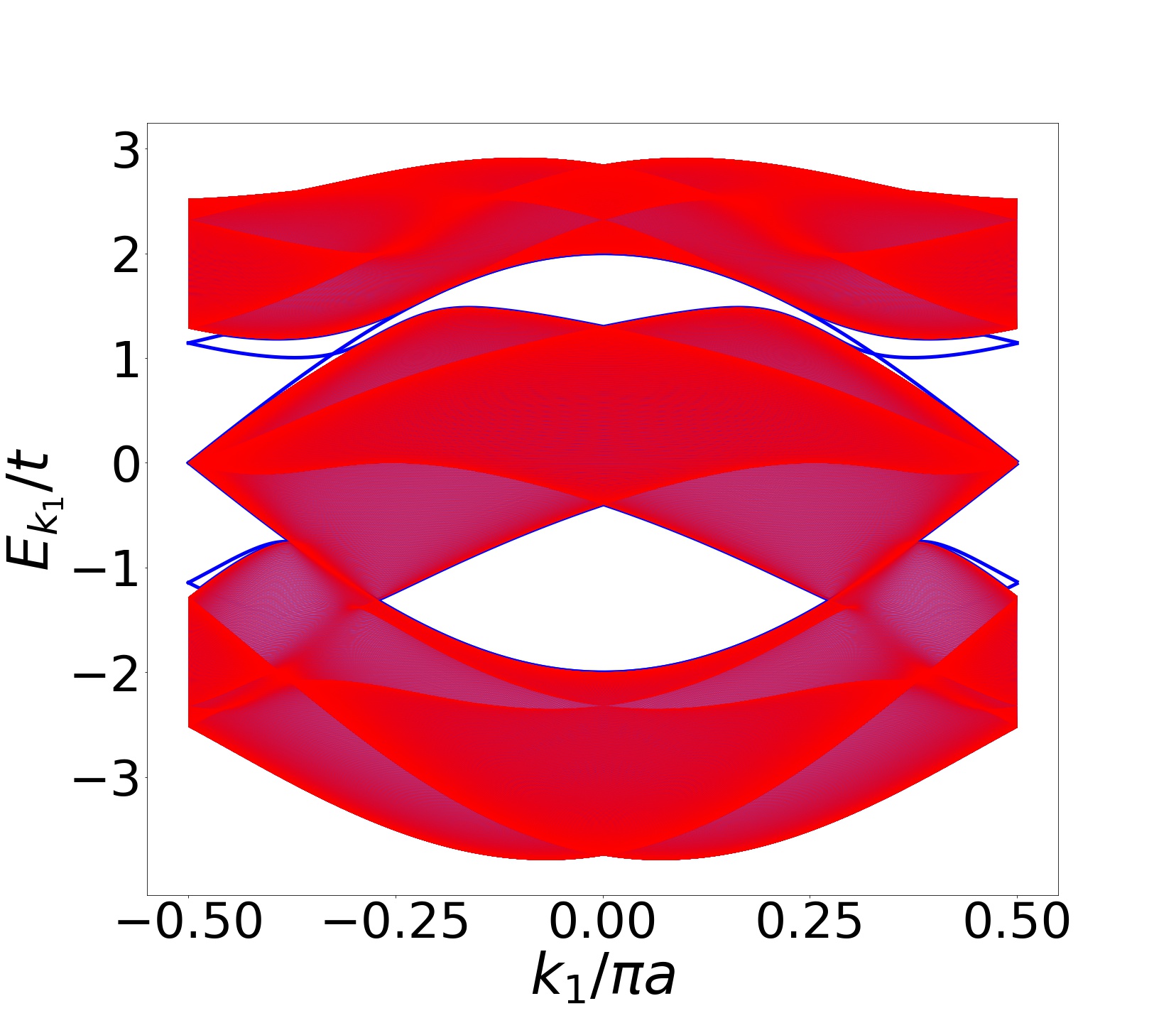} }
\caption{
The spectra $E_{k_1}$ for vanishing on-site energies ($\lambda_R=\lambda_B=\lambda_G=0$) and for different values of $\gamma$ and $\phi$ .  Blue lines correspond to the edge states obtained by calculations where we consider periodic boundary conditions in the ${\bf e}_1$ direction and  open boundary conditions in the ${\bf e}_2$ direction. Triangles right (left) indicate that the corresponding edge state is localized at the $x_2=N_2$ ($x_2=1$) edge of the system. Brown (gold)  triangles indicate a state which dominantly contains down (up) spin fermions.     
}
\label{Fig:Spectrum_hom}
\end{figure*}

\begin{figure*}[t!]
\subfigure[$\gamma=0$ and $\phi=\pi/2$, $k_1=-0.4\pi$, $l=N_2/2-1$]{
\label{Fig:Wavefunction_gamma0_phi0p5pi_k1n0p4_l199}
\centering \includegraphics[width=4cm]{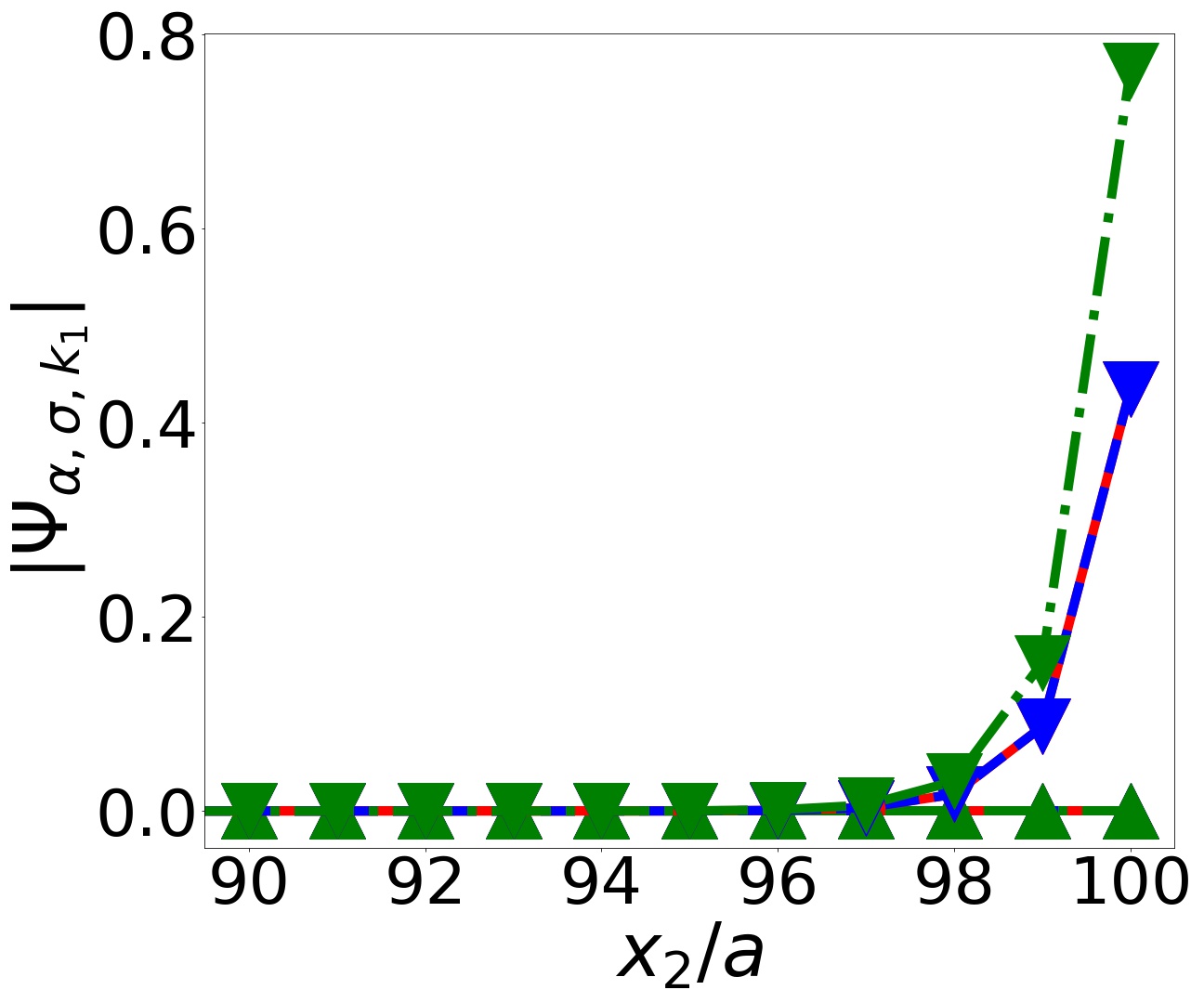} }
\subfigure[$\gamma=0$ and $\phi=\pi/2$, $k_1=0.4\pi$, $l=2N_2-1$]{
\label{Fig:Wavefunction_gamma0_phi0p5pi_k10p4_l199}
\centering \includegraphics[width=4cm]{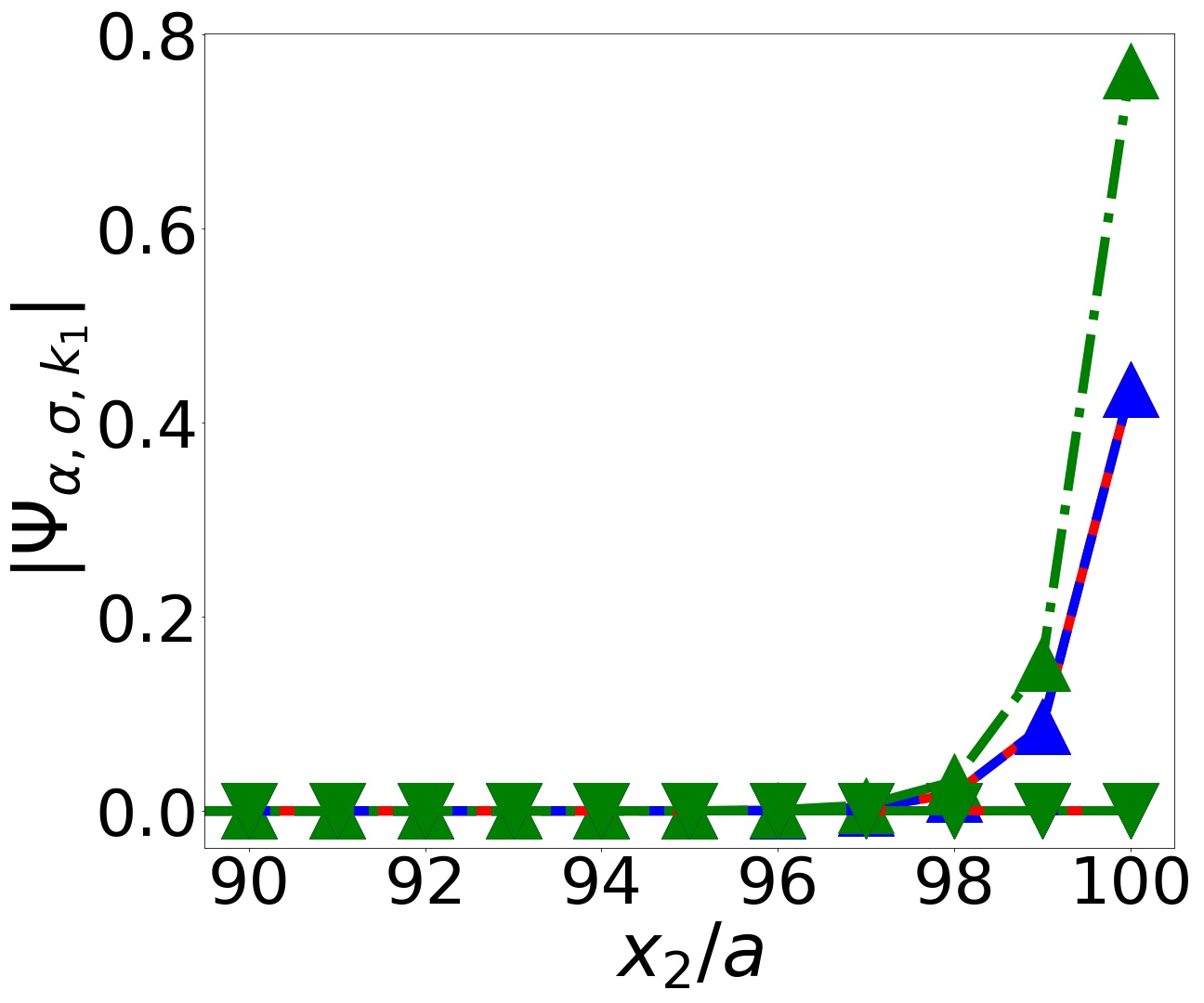} }
\subfigure[$\gamma=0$ and $\phi=\pi/2$, $k_1=-0.4\pi$, $l=2N_2$]{
\label{Fig:Wavefunction_gamma0_phi0p5pi_k1n0p4_l200}
\centering \includegraphics[width=4cm]{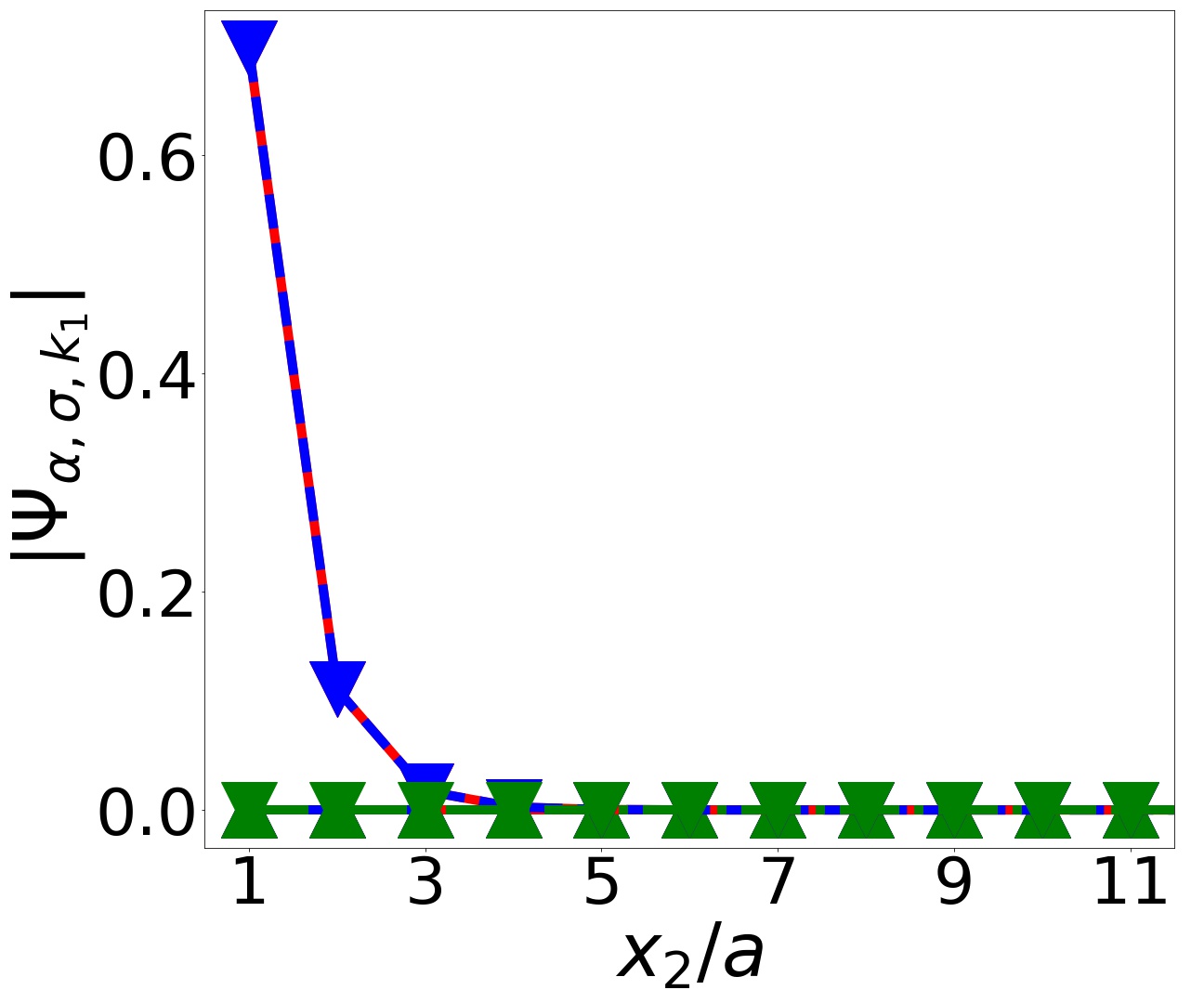} }
\subfigure[$\gamma=0$ and $\phi=\pi/2$, $k_1=0.4\pi$, $l=2N_2$]{
\label{Fig:Wavefunction_gamma0_phi0p5pi_k10p4_l200}
\centering \includegraphics[width=4cm]{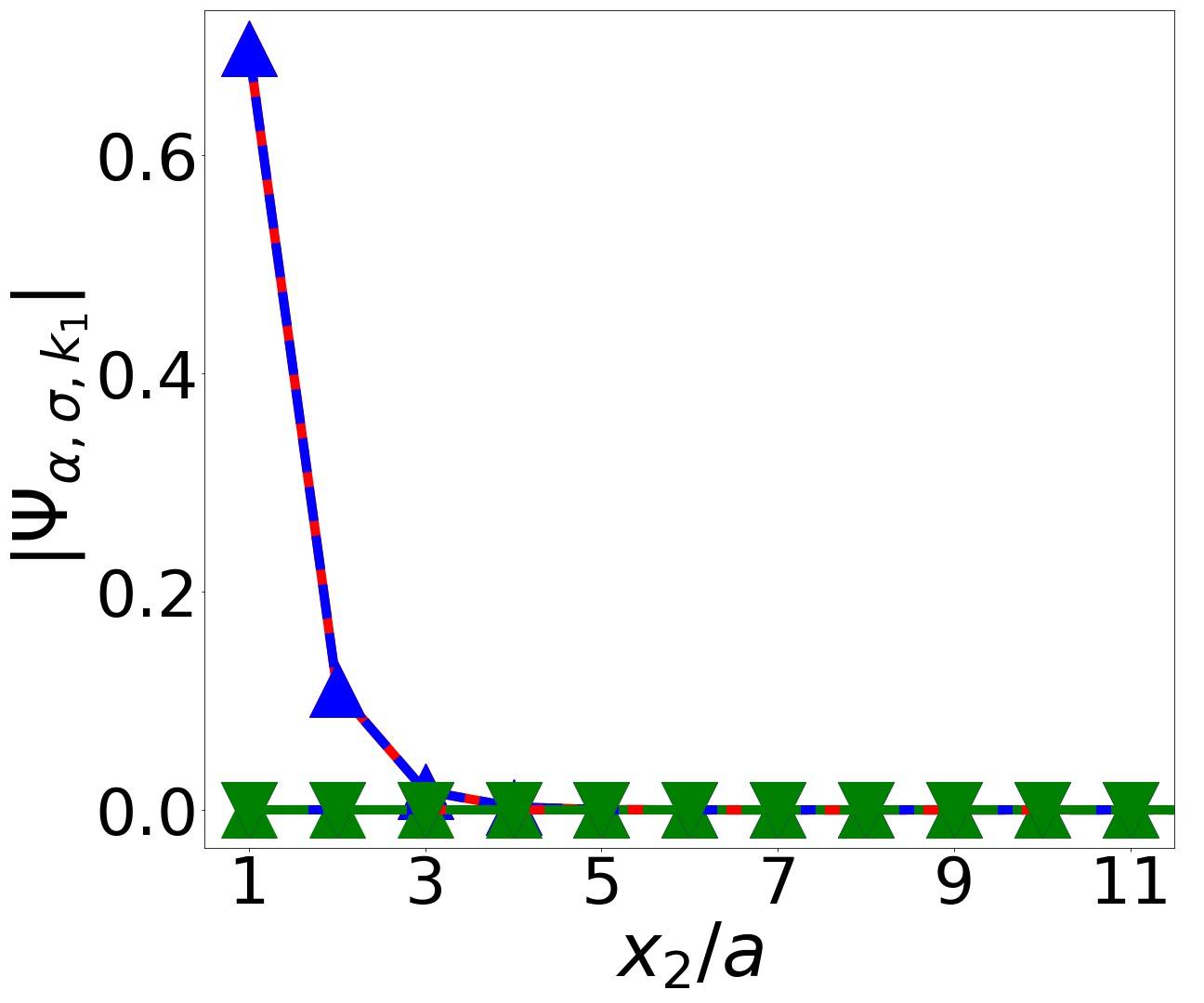} } \\
\subfigure[$\gamma=0.05$ and $\phi=\pi/2$, $k_1=-0.4\pi$, $l=N_2/2-1$]{
\label{Fig:Wavefunction_gamma0p05_phi0p5pi_k1n0p4_l199}
\centering \includegraphics[width=4cm]{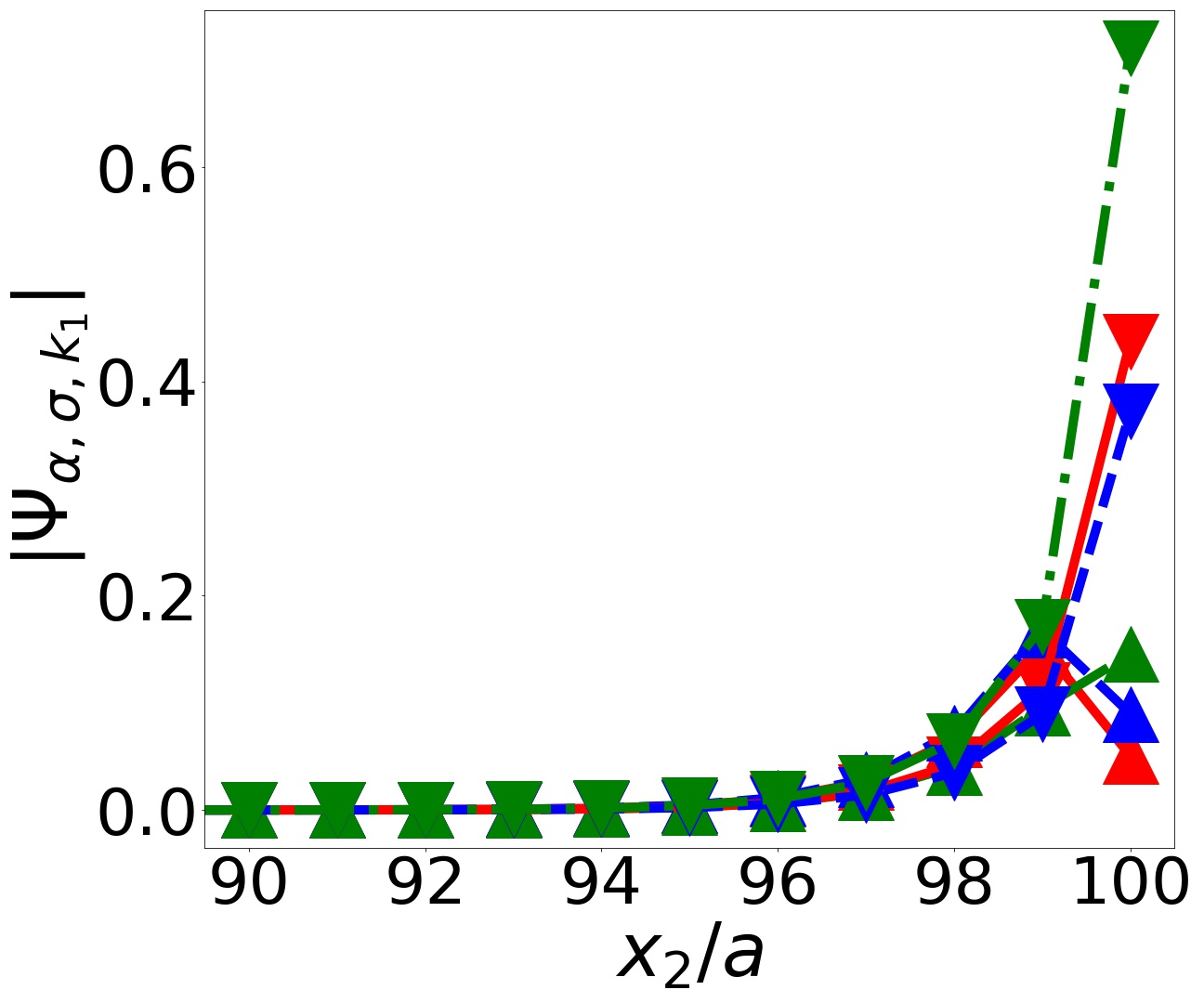} }
\subfigure[$\gamma=0.05$ and $\phi=\pi/2$, $k_1=0.4\pi$, $l=2N_2-1$]{
\label{Fig:Wavefunction_gamma0p05_phi0p5pi_k10p4_l199}
\centering \includegraphics[width=4cm]{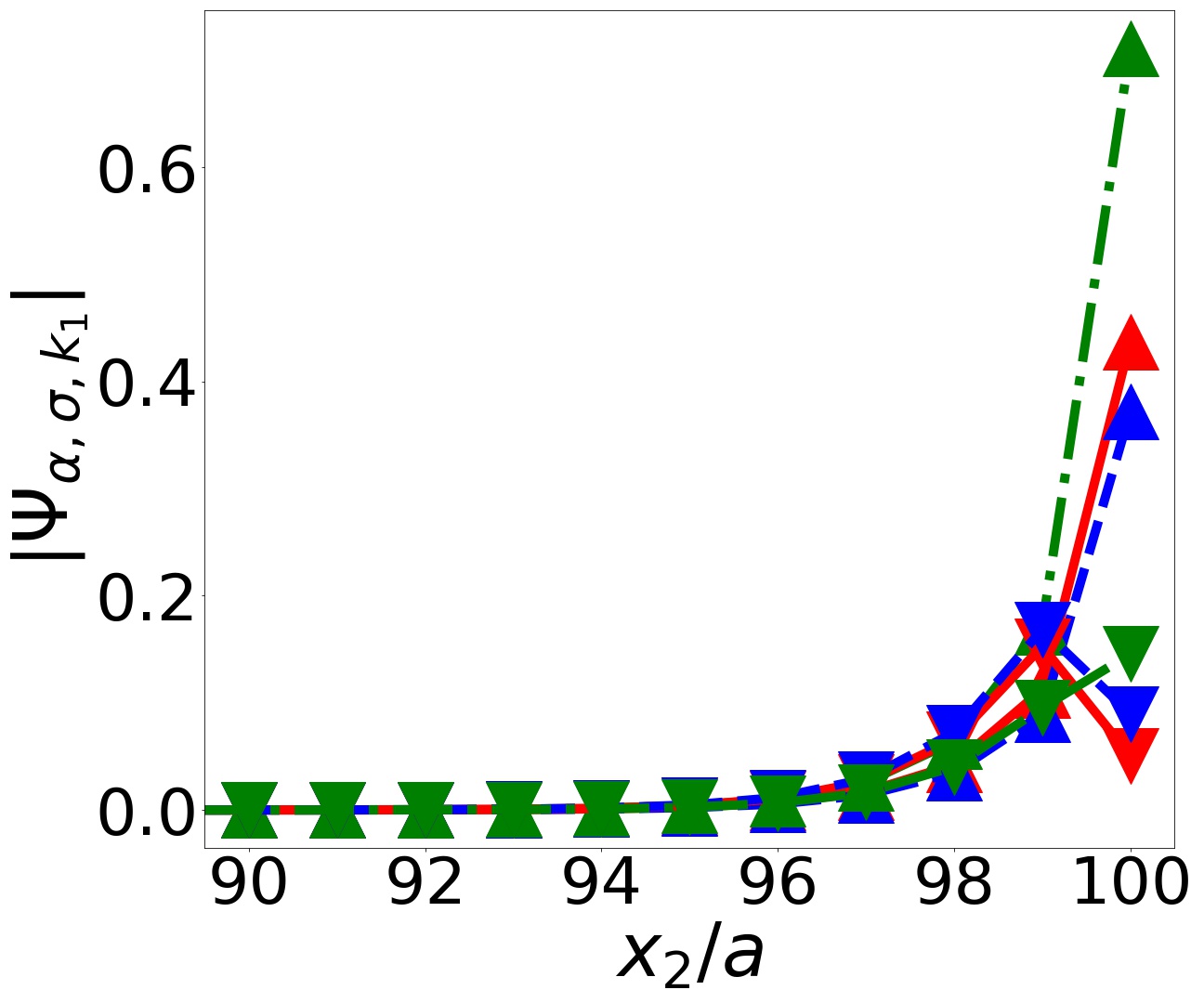} }
\subfigure[$\gamma=0.05$ and $\phi=\pi/2$, $k_1=-0.4\pi$, $l=2N_2$]{
\label{Fig:Wavefunction_gamma0p05_phi0p5pi_k1n0p4_l200}
\centering \includegraphics[width=4cm]{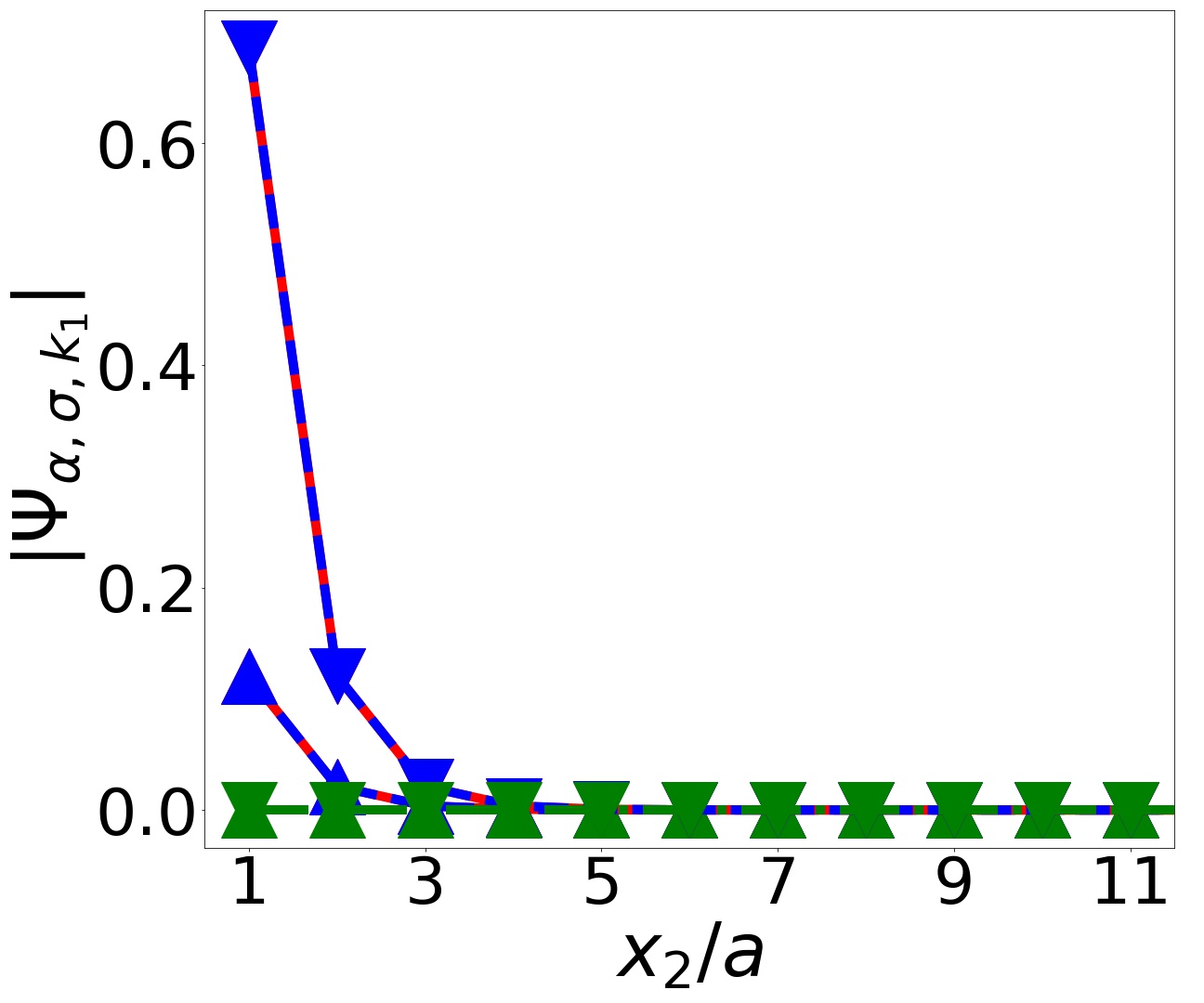} }
\subfigure[$\gamma=0.05$ and $\phi=\pi/2$, $k_1=0.4\pi$, $l=2N_2$]{
\label{Fig:Wavefunction_gamma0p05_phi0p5pi_k10p4_l200}
\centering \includegraphics[width=4cm]{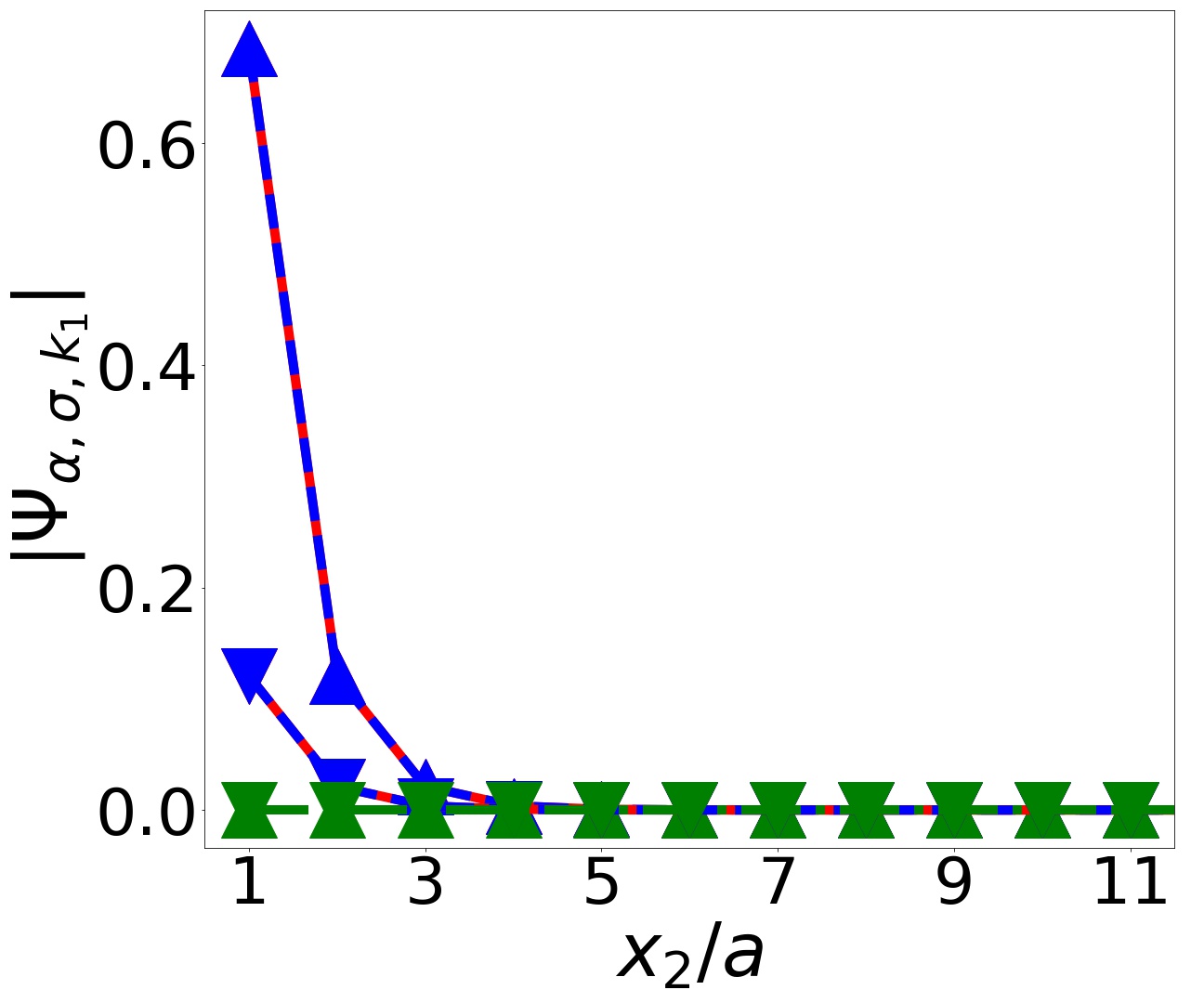} } 
%
%
\caption{
The edge states for $\gamma=0$ and $\phi=\pi/2$ (a-d) and for $\gamma=0.05$ and $\phi=\pi/2$ (e-h). Triangles pointing up correspond to spin up fermions and triangles pointing down correspond to spin down fermions. Red (solid), blue (dashed), green (dashed-dotted) lines correspond to $R$, $B$ and $G$ sites, respectively. $l$ corresponds to the  energy level. The number of unit cells along ${\bf e}_1$ and ${\bf e}_2$  directions are $N_1=500$ and $N_2=100$ respectively. Other parameters are $\lambda_R=\lambda_B=\lambda_G=0$.  
}
\label{Fig:Wavefunctions}
\end{figure*}

\subsubsection{Extending the results to the $\gamma \neq 0$ case}

We can generalize the results of the $\gamma=0$ study by considering adiabatic deformations of the Hamiltonian. Because the $\mathbb{Z}_2$ number is an integer, it is unchanged under adiabatic deformations of the Hamiltonian, \textit{i.e.} as long as there is no gap closing. Therefore, we expect that a $\gamma \neq 0$ point is characterized by the same $\mathbb{Z}_2$ number as a $\gamma=0$ point which lies in the same phase (no gap closure between both points).

We can use another argument, maybe more intuitive, from the evolution of the conductivity for a two-dimensional open geometry, when we turn the $\gamma$ term on. When $\gamma=0$, we assume that the system at $n=2/3$ is in a topological insulating phase. It means that we can observe chiral edge modes\cite{hats.93,pe.ho.12}, each of which is associated to a finite chiral current average value, while the total current vanishes. If we can show that the chiral current associated to each edge mode is conserved when we turn $\gamma$ on,  as well as the insulating feature of the bulk, then we prove that the $\mathbb{Z}_2$ number is also conserved. 

We introduce a (real-)time $\tau$ dependence in the $\gamma$ parameter, say a simple linear one $2 \pi \gamma = \tau$. 
We want to evaluate the electronic current operator ${\bf j}_{(\alpha,{\bf r}),(\alpha',{\bf r}')}(\tau)$ between 2 nearest neighbor points $(\alpha,{\bf r})$ and $(\alpha',{\bf r}')$ of the lattice. First we combine the continuity equation for the charge density $n_{\alpha,{\bf r}} = c_{\alpha,{\bf r}}^\dagger c_{\alpha,{\bf r}}$ with the (Heisenberg picture) evolution equation for $n_{\alpha,{\bf r}}$ to get
\begin{equation}
\boldsymbol{\nabla} \cdot \boldsymbol{j}_{\alpha,{\bf r}}(\tau)
=- i\left[{\cal H}(\tau),n_{\alpha,{\bf r}}\right],
\end{equation} 
with ${\cal H}$ corresponding to the form in Eq. \eqref{Hamiltonian} and with
\begin{equation}
\label{curren_alphar}
\boldsymbol{j}_{\alpha,{\bf r}}(\tau) 
= \sum_{\alpha',{\bf r}'} {\bf j}_{(\alpha,{\bf r}),(\alpha',{\bf r}')}(\tau) 
\end{equation} 
the electronic current operator at the $(\alpha,{\bf r})$ point of the lattice. In Eq. \eqref{curren_alphar} the summation with $(\alpha',{\bf r}')$ runs over all neighboring sites of $(\alpha,{\bf r})$.

Now, we show that when $\gamma$ goes from $0$ to a finite value, the chiral currents associated to the chiral edge modes are conserved as well as the insulating feature of the bulk. To this end, we  write the time derivative of the average value of the current operator, at whatever time $\tau$. 
We denote $\ket{\Psi(\tau)}$ the state of the system described by the Hamiltonian ${\cal H}(\tau)$. We have 
\widetext
\begin{equation}
\dfrac{d }{d\tau} \bra{\Psi(\tau)} {\bf j}_{(\alpha,{\bf r}),(\alpha',{\bf r}')}(\tau) \ket{\Psi(\tau)}= \bra{\Psi(\tau)} \left( \dfrac{i}{\hbar} \left[{\cal H}(\tau),{\bf j}_{(\alpha,{\bf r}),(\alpha',{\bf r}')}(\tau)\right] + \dfrac{d}{d\tau} {\bf j}_{(\alpha,{\bf r}),(\alpha',{\bf r}')}(\tau) \right)\ket{\Psi(\tau)}.
\end{equation}
\endwidetext

Relying on the symmetries of the Hamiltonian, the sum of the average current along one directional line in the system is conserved (for details see in  Appendix~\ref{Conservation_along_E2}). What we call a directional line is one straight line of atoms in the system along one of the ${\bf e}_1$, ${\bf e}_2$ or ${\bf e}_3$ directions. 

In the Appendix \ref{Conservation_along_E2} we give an example of this computation for one directional line, that we denote ${\cal E}_2$, along the ${\bf e}_2$ direction. 
We consider an inversion symmetric lattice with line-shaped boundaries. 
For each unit cell at position ${\bf r} \in {\cal E}_2$, 
we need to consider the currents along ${\cal E}_2$, which are ${\bf j}_{(G,{\bf r}),(R,{\bf r})}(\tau)$ and ${\bf j}_{(R,{\bf r}+{\bf e}_2),(G,{\bf r})}(\tau)$. We show (see Appendix~\ref{Conservation_along_E2}) that  
\begin{equation}
\sum_{{\bf r} \in {\cal E}_2} \dfrac{d }{d\tau} \bra{\Psi(\tau)}  {\bf j}_{(G,{\bf r}),(R,{\bf r})}(\tau) + {\bf j}_{(R,{\bf r}+{\bf e}_2),(G,{\bf r})}(\tau) \ket{\Psi(\tau)} = 0.
\end{equation}

For one line along the ${\bf e}_2$ direction, the currents, associated to an eigenstate, are conserved when we vary $\gamma$. The proof can also be done for the currents associated to the ${\bf e}_1$ and ${\bf e}_3$ directions, but is not shown here because it is similar to ${\bf e}_2$ the proof. 

The currents that flow in the system are those associated to the occupied eigenstates. At $n=2/3$, as long as the gap between the second and the third band does not close, all the eigenstates associated to the first and the second band remain occupied. Therefore, when we turn on the $\gamma$ term, as long as the gap (between the second and the third band) does not close, the chiral currents associated to the chiral edge modes are conserved, as well as the insulating feature of the bulk, which shows that the $\mathbb{Z}_2$ number is conserved.

\section{Results without staggered potential: $V_{\alpha, {\bf r}}=\lambda_\alpha$}
\label{Results_without}

\subsection{Without on-site energies: $\lambda_R=\lambda_B=\lambda_G =0$}
\label{Without_on-site_energies}

We start to present our results for $V_{\alpha, {\bf r}}=\lambda_\alpha$.  First, we consider a setup for which all the on-site energies are zero, i.e., $\lambda_R=\lambda_B=\lambda_G =0$. 
We present spectra $E_{k_1}$ for different values of $\gamma$ and $\phi$ in the Fig.~\ref{Fig:Spectrum_hom}. 
For $\gamma=0$, the spin up and spin down fermions are decoupled from each other and we obtain three bands, each of them are doubly degenerate corresponding to spin up and spin down fermions. We obtain flat bands for $\phi=0$ (upper band), $\phi=\pi/2$ (middle band), and $\phi=\pi$ (lower band). Spectra for  $\phi=\pi/2$ is presented in Fig.~\ref{Fig:Spectrum_gamma0_phi0p5pi}.
For finite $\gamma$, the spin up and spin down fermions are mixed. Degeneracy is partially removed, but bands are still pairwise partially degenerate. For small values of $\gamma$, the system is still gapped for both $n=2/3$ and $n=4/3$ fillings (see Fig.~\ref{Fig:Spectrum_gamma0p05_phi0p5pi}), but with increasing $\gamma$ the gaps are closing (see Figs.~\ref{Fig:Spectrum_gamma0p15_phi0p5pi} and \ref{Fig:Spectrum_gamma0p1_phi0p25pi}).

\begin{figure}[t!]
\includegraphics[width=8cm]{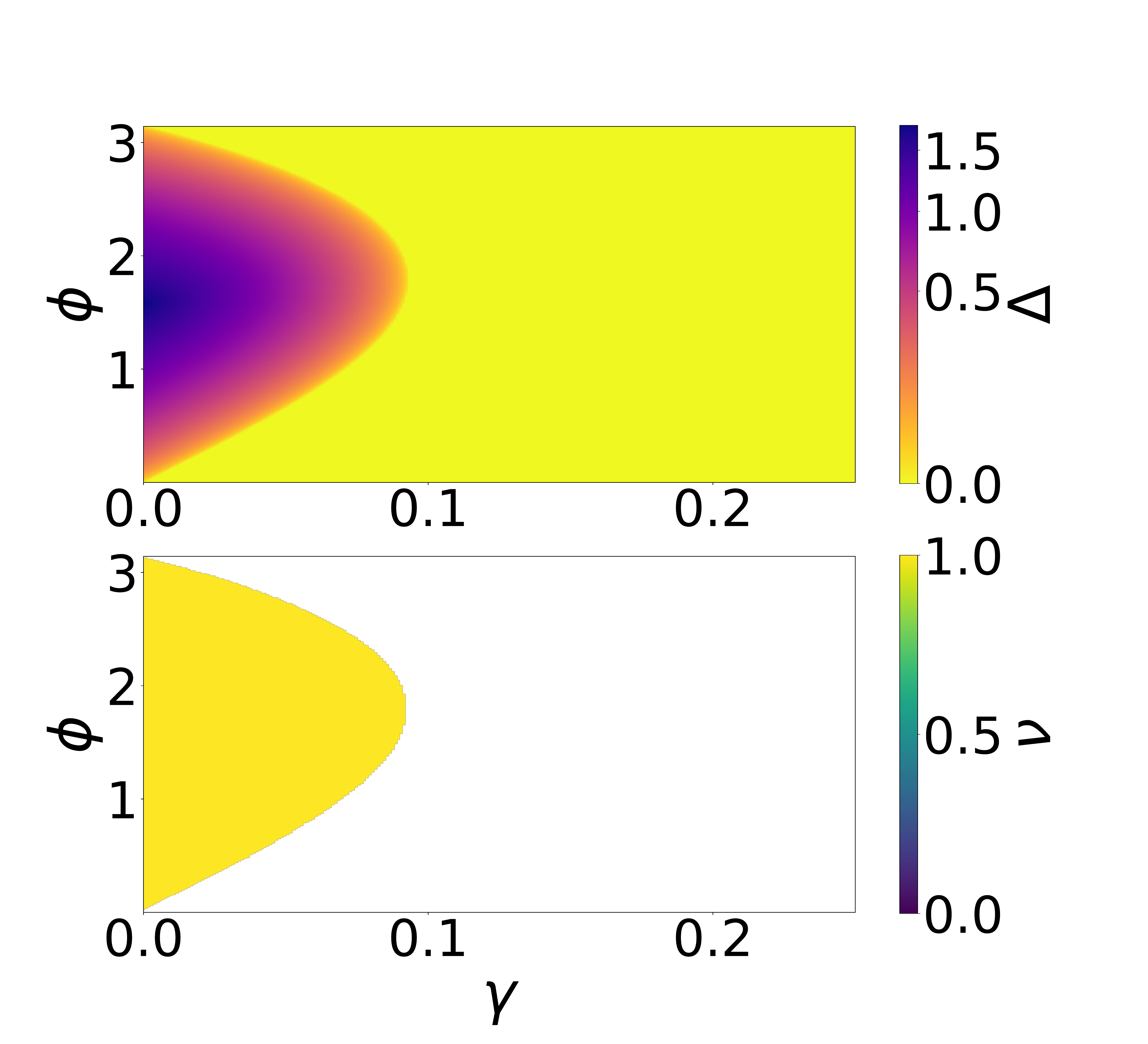}
\caption{The phase diagram for $n=2/3$ filling and for $\lambda_R=\lambda_B=\lambda_G=0$. The upper panel shows the size of the gap $\Delta$, while the lower panel shows the $\mathbb{Z}_2$ number $\nu$. White region in the lower panel corresponds to the parameter set when the gap is closed.
}
\label{Fig:Phase-Diagram_hom_n2d3}
\end{figure}

As mentioned above, the gap may appear for $n=2/3$ and $n=4/3$ fillings and the results obtained for these fillings are related to each other by the mapping in Eq.  \eqref{symmetry}. Therefore, we present our results for $n=2/3$ and the results for $n=4/3$ can be obtained by replacing $\phi$ by $\pi-\phi$. First, we consider the gap. Our results are presented in the upper panel of Fig.~\ref{Fig:Phase-Diagram_hom_n2d3}. The next step is to find out whether we have a trivial band insulator or a topological band insulator. For this purpose we calculate the $\mathbb{Z}_2$ number using twisted boundary conditions. Our calculations show that for all values of $(\gamma,\phi)$ for which the gap is finite, the  $\mathbb{Z}_2$ number is $\nu=1$ (see lower panel of Fig.~\ref{Fig:Phase-Diagram_hom_n2d3}), i.e., the gapped phase is a topological insulator. 
To conclude, for large values of $\gamma$, the system is in the metallic phase, while for small values of $\gamma$, the system is a topological insulator.

\begin{figure}[t!]
\includegraphics[width=8cm]{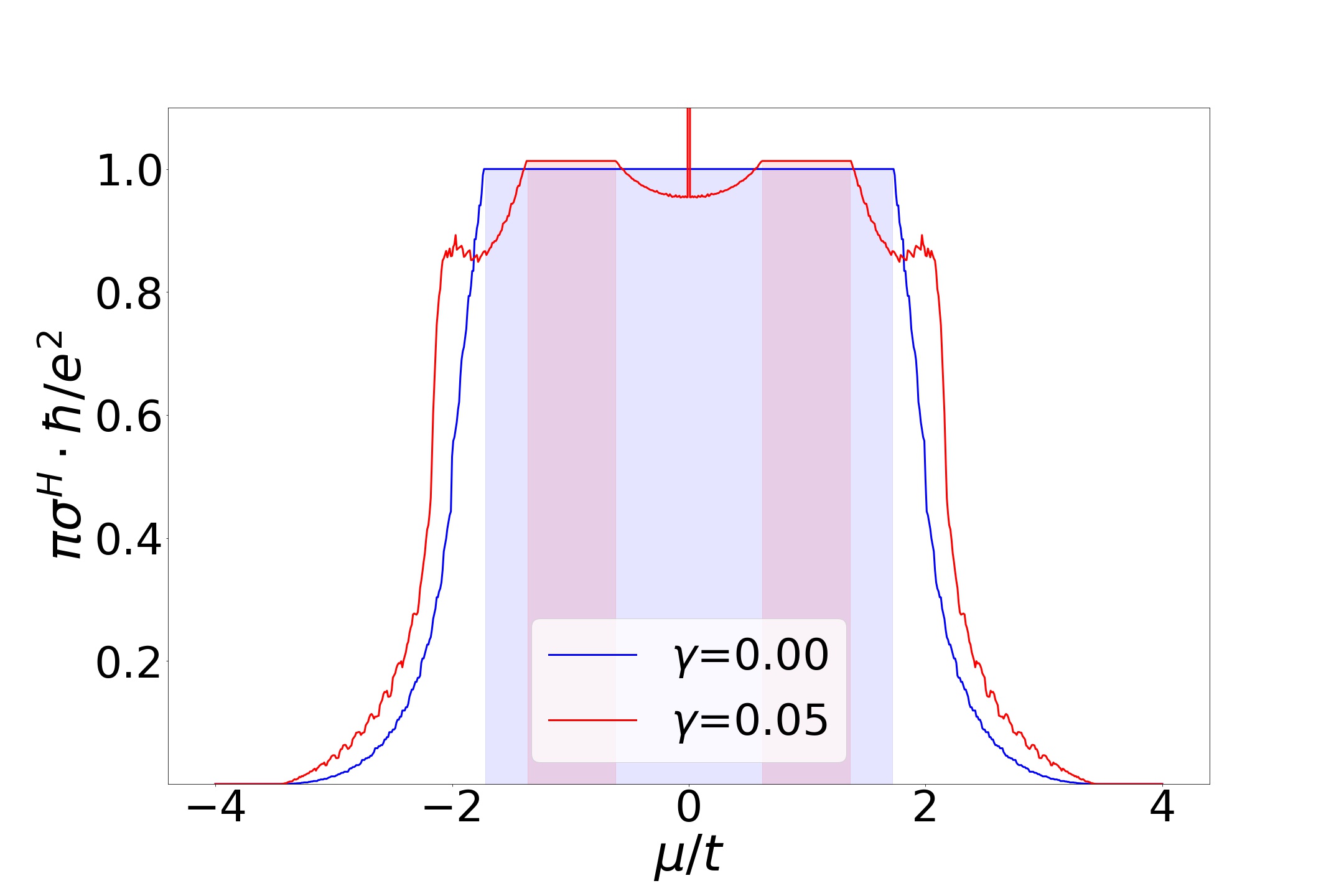}
\caption{
Spin Hall conductivity $\sigma^H(\mu)$ as a function of chemical potential $\mu$ for $\gamma=0$ and $\gamma=0.05$. The shaded region corresponds to the values of $\mu$ where the chemical potential is inside the gap (blue for $\gamma=0$ and red for $\gamma=0.05$). Other parameters are $\phi=\pi/2$ and $\lambda_R=\lambda_B=\lambda_G=0$.  
}
\label{Fig:sigmaH}
\end{figure}

As mentioned above,  topological insulators behave as insulators in the bulk, while they are conducting at their boundary. Indeed as one can see from Figs.~\ref{Fig:Spectrum_gamma0_phi0p5pi} and ~\ref{Fig:Spectrum_gamma0p05_phi0p5pi} considering only the bulk bands  the system is an insulator. With edge states, however, the system becomes gapless at the boundary (blue curves inside the gap).

To investigate the edge states in more detail we perform calculations where we consider periodic boundary conditions in the ${\bf e}_1$ direction and  open boundary conditions in the ${\bf e}_2$ direction. We obtain four edge states. Two edge states are localized at the $x_2=1$ edge of the system (see left triangles in Figs. \ref{Fig:Spectrum_gamma0_phi0p5pi} and \ref{Fig:Spectrum_gamma0p05_phi0p5pi} and also Figs. \ref{Fig:Wavefunction_gamma0_phi0p5pi_k1n0p4_l200}, \ref{Fig:Wavefunction_gamma0_phi0p5pi_k10p4_l200}, \ref{Fig:Wavefunction_gamma0p05_phi0p5pi_k1n0p4_l200}, and \ref{Fig:Wavefunction_gamma0p05_phi0p5pi_k10p4_l200}) and the other two are localized at the $x_2=N_2$ edge of the system (see right triangles in Figs. \ref{Fig:Spectrum_gamma0_phi0p5pi} and \ref{Fig:Spectrum_gamma0p05_phi0p5pi} and also Figs.  \ref{Fig:Wavefunction_gamma0_phi0p5pi_k1n0p4_l199}, \ref{Fig:Wavefunction_gamma0_phi0p5pi_k10p4_l199}, \ref{Fig:Wavefunction_gamma0p05_phi0p5pi_k1n0p4_l199}, and \ref{Fig:Wavefunction_gamma0p05_phi0p5pi_k10p4_l199}). As one can see from Fig. \ref{Fig:Wavefunctions} the penetration length of the edge states into the bulk is $2$-$3$ unit cells. 

For $\gamma=0$ the spin up fermions and the spin down fermions are decoupled from each other. In that case we have for the spin up  fermions one edge state localized at the $x_2=1$ edge of the system and one edge state localized at the $x_2=N_2$ edge of the system. Similar for the spin down fermions.  Each of these edge states can be presented by wave functions $\Psi_{\alpha,\sigma,k_1}(x_2)$ describing fermions in $R$, $B$, and $G$ sublattices with given spin $\sigma$. 
As one can see from Figs. \ref{Fig:Wavefunction_gamma0_phi0p5pi_k1n0p4_l199}-\ref{Fig:Wavefunction_gamma0_phi0p5pi_k10p4_l200} $|\Psi_{B,\sigma,k_1}(x_2)|=|\Psi_{R,\sigma,k_1}(x_2)| \neq |\Psi_{G,\sigma,k_1}(x_2)|$. The reason for this is a symmetry of the lattice. At one edge we have sites of $R$ and $B$ sublattice and at the other edge we have sites of the $G$-sublattice.  

For finite $\gamma$ the spin up and spin down states are not decoupled any more. Each of the edge state $\Psi_{\alpha,k_1}(x_2)$ describing fermions in $R$, $B$, and $G$ sublattice are superpositions of wave functions $\Psi_{\alpha,\sigma,k_1}(x_2)$ for both spin projections. 
This can be observed by comparing the upper panel Figs. \ref{Fig:Wavefunction_gamma0_phi0p5pi_k1n0p4_l199}-\ref{Fig:Wavefunction_gamma0_phi0p5pi_k10p4_l200} with the lower panel Figs. \ref{Fig:Wavefunction_gamma0p05_phi0p5pi_k1n0p4_l199}-\ref{Fig:Wavefunction_gamma0p05_phi0p5pi_k10p4_l200}. For $\gamma=0$, the edge states exhibit finite amplitude only in one of the spin states, while for $\gamma=0.05$, both spin states have a finite amplitude in the edge state wavefunction, 
nevertheless contain dominantly either up spins or down spins.


One way of experimentally detecting the topological phase is measuring the  spin Hall conductivity $\sigma^H$. \cite{li.ch.12, ii.im.18, th.ko.82, ma.ku.16, ma.ma.19} But it is important to point out that the spin Hall conductivity $\sigma^H$ is only quantized if spin-orbit coupling $\gamma$ is vanishing.\cite{ma.ma.19} 
In this case for each spin component the Hall conductivity is proportional to the respective Chern number. The spin Hall conductivity is the difference of both Chern numbers, see discussion on Eq. \eqref{nuequaldifferenceC}.
For finite $\gamma$ there is no simple relation between $\mathbb{Z}_2$ number $\nu$ and $\sigma^H$ evaluated in the spin up and spin down spin basis. 
The spin Hall conductivity for zero temperature is given by the following expression
\begin{equation}
\label{Hall_conductivity}
\sigma^H(\mu)=\frac{1}{4N_1N_2}
\sum_{{\cal E}_{{\bf k},n} \leq \mu}\sum_{{\cal E}_{{\bf k},m} >\mu} 
\left(\Omega_{n{\bf k},m{\bf k}}-\Omega_{m{\bf k},n{\bf k}}\right) \,,
\end{equation}
where
\begin{equation}
\Omega_{n{\bf k},m{\bf k}}
=-\mbox{Im}\frac{\langle {\bf k}n|\mathcal{J}_1^s|{\bf k}m\rangle 
\langle {\bf k}m|\mathcal{J}_2|{\bf k}n\rangle}
{({\cal E}_{{\bf k},n}-{\cal E}_{{\bf k},m})^2} \,.
\end{equation}
Here ${\mathcal{J}_1^s=\frac{1}{2}[\mathcal{J}_1, \sigma_z \otimes \mathbb{1}_3]}$ is the spin current and ${\mathcal{J}_a=d{\cal H}({\bf k})/dk_{a}}$ is the charge current along direction ${a=1,2}$.  $\mathbb{1}_3$ is the $3 \times 3$ unit matrix and describes different unit cells, while the $\sigma_z$ Pauli matrix describes spin degrees of freedom. 
Here ${\cal H}({\bf k})$ is the $6\times 6$  Hamiltonian matrix from Eq. \eqref{Hk_matrix} (we again use the periodic boundary conditions). ${\cal E}_{{\bf k},n}$ and $|{\bf k}n\rangle$ are the eigenvalues and the eigenvectors of Hamiltonian ${\cal H}({\bf k})$. Finally $\mu$ is the chemical potential.  

In Fig. \ref{Fig:sigmaH} we plot $\sigma^H(\mu)$ as a function of $\mu$. We obtain that when the chemical potential is inside the gap, the spin Hall conductivity has a plateau. As it was mentioned above, for $\gamma=0$ spin Hall conductivity is quantized and proportional to $\mathbb{Z}_2$ number $\nu$: $\pi \sigma^H \cdot \hbar/e^2=\nu=1$.  For $\gamma=0.05$ despite the fact that we have plateau spin Hall conductivity is not quantized and $\pi \sigma^H \cdot \hbar/e^2 \simeq 1.013$ and deviates from $\mathbb{Z}_2$ number $\nu$.  A similar result was also obtained in Ref. \onlinecite{ma.ma.19}\,.

\begin{figure}[t!]
\includegraphics[width=8cm]{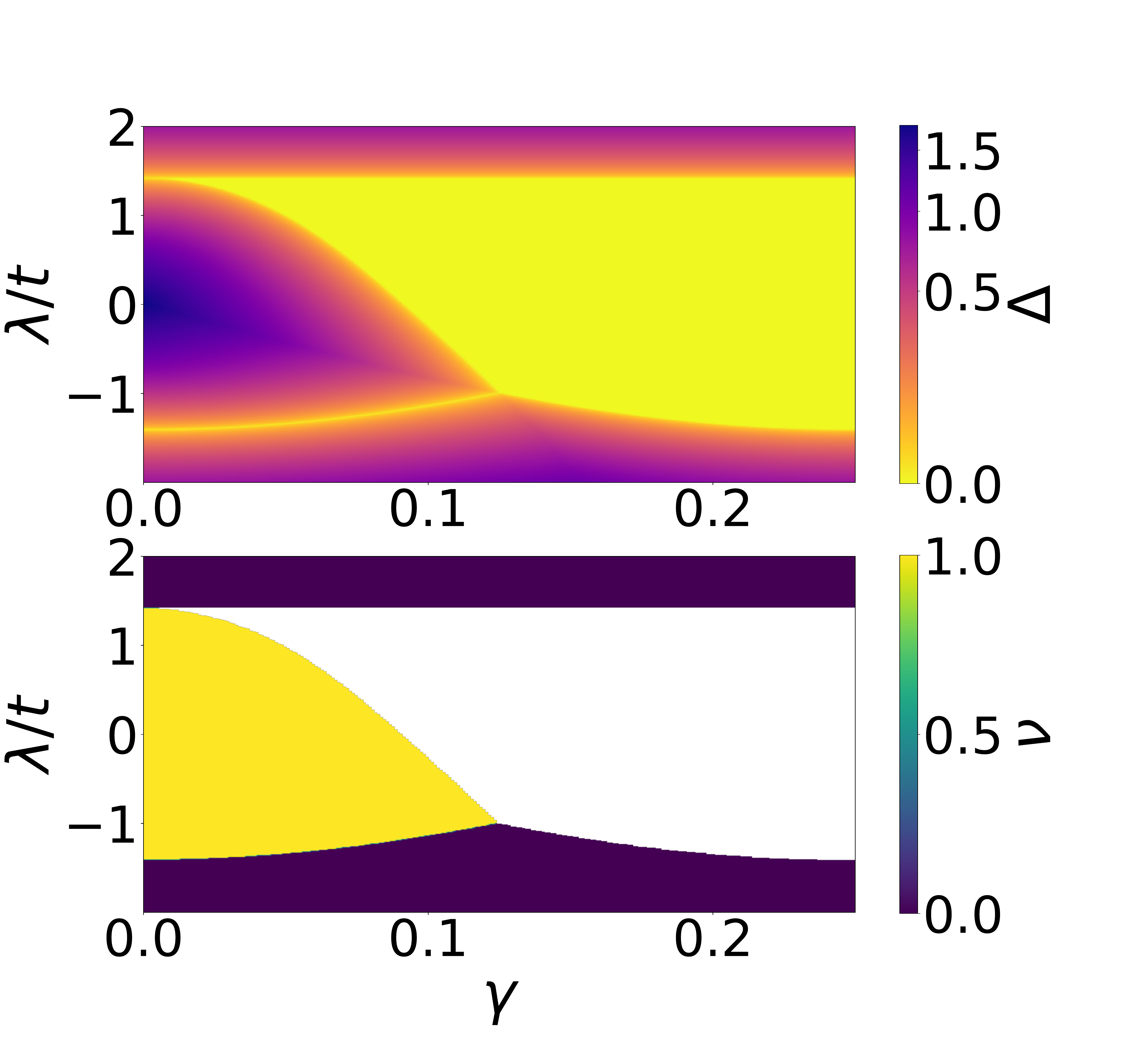}
\caption{The phase diagram for $n=2/3$ filling and $\phi=\pi/2$. $\lambda_B=-\lambda_R=\lambda$ and $\lambda_G =0$. The upper panel shows the size of the gap $\Delta$, while the lower panel shows the $\mathbb{Z}_2$ number $\nu$. White region in the lower panel corresponds to the parameter set when the gap is closed.}
\label{Fig:phi0p5pi_band1} 
\end{figure}

\subsection{Three different on-site energies:  $\lambda_B=-\lambda_R=\lambda$ and $\lambda_G=0$}
\label{tree-different}

\subsubsection{The numerical approach}

Now we consider the case where different on-site energies are applied at different sublattice sites. In particular $\lambda_B=-\lambda_R=\lambda$ and $\lambda_G=0$. We consider again a filling $n=2/3$ and a flux $\phi=\pi/2$. Note that for this flux, particularly for $\lambda=0$ and $\gamma=0$, the middle band is flat.

We obtain a metallic phase and three gapped phases (see upper panel of Fig.~\ref{Fig:phi0p5pi_band1}). According to our results, there is always a metallic phase between the middle gapped phase ($|\lambda|\lessapprox \sqrt{2}t$) and the gapped phase with $\lambda \gtrapprox \sqrt{2} t$, except for $\gamma=0$. For $\gamma=0$ the gap closes only at $\lambda = \pm\sqrt{2} t$. In contrast, we do not observe a finite metallic phase region between the middle gapped phase and the gapped phase with $\lambda\lessapprox -\sqrt{2}t$. We obtain that the gap closes only at a single value of $\lambda$ for each given $\gamma$.

To find out whether one of these gapped phases is topologically non-trivial, we calculate the $\mathbb{Z}_2$ number using twisted boundary conditions. Our results are shown in the lower part of Fig.~\ref{Fig:phi0p5pi_band1}. According to our results for the gapped phases $|\lambda| \gtrapprox \sqrt{2} t$ the $\mathbb{Z}_2$ number is equal to $0$. These two gapped phases thus correspond to the trivial band insulator. In contrast, for the gapped phase $|\lambda| \lessapprox \sqrt{2}t$ we obtain the $\mathbb{Z}_2$ number $\nu=1$. Thus, the middle gapped phase corresponds to the topological insulator.

\subsubsection{The analytical approach}
\label{Three_different_on-site_energy_analytical}

At $\gamma=0$, from the analytical computation method, we reproduce the numerical results for the $\mathbb{Z}_2$ number and we generalize it for all flux $\phi$ and arbitrary on-site energy $\lambda$. Both lowest spin species have the same energy, that we denote $E$. It is given by
\begin{equation}
E =- 2 \sqrt{\dfrac{\lambda^2 + 4t^2 \left[1 + f({\bf k})\right]}{3}} \cos \dfrac{\theta({\bf k})}{3},
\end{equation}
with $f({\bf k}) = 2 \prod_{\alpha=1}^3 \cos {\bf k}\cdot{\bf b}_\alpha $ and $0 \leq \theta({\bf k}) \leq \pi$ defined by 
\begin{equation}
\theta({\bf k}) 
=\arccos \dfrac{3^{3/2}\left( f({\bf k}) \cos\phi + \dfrac{\lambda\left(\varepsilon_2^2({\bf k}) -\varepsilon_3^2({\bf k})\right)}{8t^3} \right)}{2\left(1 + f({\bf k})+\left(\dfrac{\lambda}{2t}\right)^2\right)^{3/2}} \,.
\end{equation}
From this expression, we can show that the gap between the lowest band and the middle one only closes at $\lambda=\pm \sqrt{2}t$. It is valid for each spin species, to which associated middle band energy for both spin species is given by 
\begin{equation}
E_{\textrm{m}} =- 2 \sqrt{\dfrac{\lambda^2 + 4t^2 \left[1 + f({\bf k})\right]}{3}} \cos \dfrac{\theta({\bf k})-2 \pi}{3} 
\end{equation}

We introduce the 3 nonequivalent high symmetry $\bf M$ points, ${\bf M}_1=-\dfrac{1}{2}{\bf g}_1$, ${\bf M}_2=\dfrac{1}{2}{\bf g}_2$ and $ {\bf M}_3=\dfrac{1}{2} \left({\bf g}_1+ {\bf g}_2\right)$ (see Fig. \ref{domains}). At $\lambda=\sqrt{2}t$ the energy bands touch in reciprocal space at the ${\bf M}_3$ point, while at $\lambda=-\sqrt{2}t$ the energy bands touch at the ${\bf M}_1$ point, and we have
\begin{itemize}
	\item at $\lambda \leq -\sqrt{2}t $, we have $E \leq \lambda$, with the equality occurring at the ${\bf M}_1$ point,
 	\item at $-\sqrt{2}t<\lambda < \sqrt{2}t$, we have $E< - |\lambda|$,
 	\item at $ \lambda \geq \sqrt{2}t$, we have $E \leq - \lambda$, with the equality occurring at the ${\bf M}_3$ point.
\end{itemize}
In the following, we will only describe the computation of the lowest band Chern number for the case $\lambda < 0$ because the case $\lambda > 0$ can be studied following almost the same steps. 
We rewrite the state associated to the $\sigma$ spin species as
\begin{equation}
\ket{u_{\sigma,{\bf k}}} = \dfrac{\left(r_\sigma({\bf k}) c_{R,{\bf k},\sigma}^{\dagger} + b_\sigma({\bf k}) c_{B,{\bf k},\sigma}^{\dagger} + g_\sigma({\bf k}) c_{G,{\bf k},\sigma}^{\dagger}\right) \ket{0}}{\sqrt{\left|r_\sigma({\bf k})\right|^2+\left|b_\sigma({\bf k})\right|^2+\left|g_\sigma({\bf k})\right|^2}} \,.
\end{equation}
Here $\ket{0}$ is a vacuum state and by definition, we have ${\cal H}_\sigma({{\bf k}}) \ket{u_{\sigma,{\bf k}}} = E \ket{u_{\sigma,{\bf k}}}$. From this relation, we get the $r_\sigma({\bf k})$, the $b_\sigma({\bf k})$ and the $g_\sigma({\bf k})$ coefficients. We describe 3 gauge choices $G_I$, $G_{II}$ and $G_{III}$ for the eigenvectors that we respectively write $\ket{u_{\sigma,{\bf k},I}}$, $\ket{u_{\sigma,{\bf k},II}}$ and $\ket{u_{\sigma,{\bf k},III}}$. We will see later that these choices allow to correctly define the Bloch eigenvectors over the BZ.
\begin{itemize}
\item \underline{Gauge choice $G_I$}: the coefficient $g_\sigma({\bf k})$ is real. More precisely, we choose 
\begin{equation}
\label{gsigma_GI}
g_\sigma({\bf k}) = \rho_\sigma({\bf k})  \,,
\end{equation}
with
\begin{equation}
\rho_\sigma({\bf k}) e^{i\varphi_\sigma({\bf k})} = -\dfrac{(E-\lambda) \varepsilon_2({\bf k})}{2\varepsilon_1({\bf k})}  -\frac{1}{2} e^{-i s_z \phi} \varepsilon_3({\bf k}) \,,
\end{equation}
where $\rho_\sigma({\bf k})$ and $\varphi_\sigma({\bf k})$ are both real numbers.
Then we have
\begin{equation}
r_\sigma({\bf k}) = -\dfrac{E(E-\lambda)  - \varepsilon_3^2({\bf k})}{2 \varepsilon_1({\bf k})} e^{-i\varphi_\sigma({\bf k})} \,,
\end{equation}
and
\begin{equation}
b_\sigma({\bf k}) = \dfrac{ E(E+\lambda) - \varepsilon_2^2({\bf k})}{\rho_{1,\sigma}({\bf k})} \rho_\sigma({\bf k})e^{-i\varphi_{1,\sigma}({\bf k})} \,,
\end{equation} 
with
\begin{equation}
\rho_{1,\sigma}({\bf k}) e^{i\varphi_{1,\sigma}({\bf k})} = (E+\lambda) \varepsilon_3({\bf k}) e^{-is_z \phi} + \varepsilon_1({\bf k})\varepsilon_2({\bf k}) \,,
\end{equation}
where $\rho_{1,\sigma}({\bf k})$ and $\varphi_{1,\sigma}({\bf k})$  are both real numbers.
\item \underline{Gauge choice $G_{II}$}: the coefficient $r_\sigma({\bf k})$ is real. We choose \begin{equation}
g_\sigma({\bf k}) = \rho_\sigma({\bf k})e^{i\varphi_\sigma({\bf k})}.
\end{equation}
Then we have
\begin{equation}
r_\sigma({\bf k}) = -\dfrac{E(E-\lambda)  - \varepsilon_3^2({\bf k})}{ 2\varepsilon_1({\bf k})}  
\end{equation}
and
\begin{equation}
b_\sigma({\bf k}) = \dfrac{E(E+\lambda) - \varepsilon_2^2({\bf k})}{\rho_{1,\sigma}({\bf k})} \rho_\sigma({\bf k})e^{i\left[\varphi_\sigma({\bf k})-\varphi_{1,\sigma}({\bf k})\right]} \,.
\end{equation} 
\item \underline{Gauge choice $G_{III}$}: the coefficient $b_\sigma({\bf k})$ is real. We choose \begin{equation}
g_\sigma({\bf k}) = \rho_\sigma({\bf k})e^{i\varphi_{1,\sigma}({\bf k})}  \,.
\end{equation}
Then we have
\begin{equation} 
r_\sigma({\bf k}) = -\dfrac{E(E-\lambda)  - \varepsilon_3^2({\bf k})}{ 2\varepsilon_1({\bf k})} e^{-i\left[\varphi_\sigma({\bf k})-\varphi_{1,\sigma}({\bf k})\right]} 
\end{equation}
and
\begin{equation}
\label{bsigma_GIII}
b_\sigma({\bf k}) = \dfrac{E(E+\lambda) - \varepsilon_2^2({\bf k})}{\rho_{1,\sigma}({\bf k})} \rho_\sigma({\bf k}) \,.
\end{equation} 
\end{itemize}

Notice that $\varphi_\sigma({\bf k})$ is well-defined (modulo $2 \pi$) for all ${\bf k}$ except when $\rho_\sigma({\bf k}) = 0$ or when ${\varepsilon_1({\bf k})=-2t\cos {\bf k}\cdot{\bf b}_1 = 0}$ and $\varphi_{1,\sigma}({\bf k})$ is well-defined for all ${\bf k}$ except when ${\rho_{1,\sigma}({\bf k}) = 0}$. 

Now we have to distinguish two different cases ${\lambda<-\sqrt{2}t}$ and  $-\sqrt{2}t \leq \lambda \leq 0$. 
For the case $\lambda<-\sqrt{2}t$, then $E \leq \lambda$. The points for which $\rho_\sigma({\bf k}) = 0$ are the same than those for which $\rho_{1,\sigma}({\bf k}) = 0$. These are the $\boldsymbol{M}_2 = \dfrac{1}{2} {\bf g}_2$ point at which $ \cos {\bf k}\cdot{\bf b}_2 = \cos {\bf k}\cdot{\bf b}_3=0$, and the $\boldsymbol{M}_1$ point at which $ \cos {\bf k}\cdot{\bf b}_1 = \cos {\bf k}\cdot{\bf b}_3=0$ and $E=\lambda$. In this case, we apply the gauge choice $G_{III}$ for all the points of the BZ, and then we can show that the eigenvector $\ket{u_{\sigma,{\bf k},III}}$ is uniquely and smoothly defined, as is the Berry gauge field $\boldsymbol{A}_{\sigma,{\bf k},III} = \bra{u_{\sigma,{\bf k},III}}\boldsymbol{\nabla}_{{\bf k}} \ket{u_{\sigma,{\bf k},III}}$. Because the BZ is a torus, we find that the lowest band Chern numbers $C_\sigma$, are both vanishing. A very similar proof can be done for the $\lambda>\sqrt{2}t$ case. Therefore, at $n=2/3$ filling, for all flux $\phi \in \, ]0,\pi[$, the phase associated to $\lambda_B=-\lambda_R=\lambda$ and $\lambda_G=0$, $|\lambda|>\sqrt{2}t$ is a trivial insulator.

Now, let us study the $-\sqrt{2}t \leq \lambda\leq0$ case. In this situation, $E< \lambda$. We have $\rho_\sigma({\bf k}) = 0$ at the $\boldsymbol{M}_2$ point. We have $\rho_1({\bf k}) = 0$ at the $\boldsymbol{M}_2$ and the $\boldsymbol{M}_1$  points. Here, it is impossible to find a unique and smooth gauge everywhere in the BZ. 
For this purpose we split the BZ into two non-overlapping domains. One domain, which we call the $\mathcal{D}_{II}$ domain, contains the point where $\rho_\sigma({\bf k})$ vanishes. The other one, the $\mathcal{D}_I$ domain, contains all the points where 
$E(E-\lambda)  - \varepsilon_3^2({\bf k})$
vanishes or 
$\varepsilon_1({\bf k}) $
vanishes. The boundary between $\mathcal{D}_I$ and $\mathcal{D}_{II}$ does not contain any of the $\rho_\sigma({\bf k})=0$, $E(E-\lambda)  - \varepsilon_3^2({\bf k}) =0$ and $\varepsilon_1({\bf k})=0$ points. We also define $\Gamma$ a closed path along this boundary, surrounding once the $\boldsymbol{M}_2$ point. We refer the reader to Fig.~\ref{domains} for the notations. Now, we apply the $G_I$ gauge choice for the points contained in $\mathcal{D}_I$ and the $G_{II}$ gauge choice for the points contained in $\mathcal{D}_{II}$. Then we can show that the associated eigenvector $\ket{u_{\sigma,{\bf k},I}}$ and $\ket{u_{\sigma,{\bf k},II}}$ and the Berry gauge fields $\boldsymbol{A}_{\sigma,{\bf k},I} = \bra{u_{\sigma,{\bf k},I}}\boldsymbol{\nabla}_{{\bf k}} \ket{u_{\sigma,{\bf k},I}}$ and $\boldsymbol{A}_{\sigma,{\bf k},II} = \bra{u_{\sigma,{\bf k},II}}\boldsymbol{\nabla}_{{\bf k}} \ket{u_{\sigma,{\bf k},II}}$ are uniquely and smoothly defined respectively on $\mathcal{D}_I$ and $\mathcal{D}_{II}$. 
Along $\Gamma$, we have $\ket{u_{\sigma,{\bf k},I}} =\textrm{e}^{-i\varphi_\sigma({\bf k})}\ket{u_{\sigma,{\bf k},II}}$. We can define $\varphi_\sigma({\bf k})$ so that it is smooth along the whole $\Gamma$ path.
Therefore we have (see Eq.~\ref{chern_number})
\begin{equation} 
	C_\sigma = - \dfrac{1}{2 \pi} \oint_{\Gamma} d{\bf k} \cdot \boldsymbol{\nabla}_{{\bf k}} \varphi_\sigma ({\bf k}).
\end{equation}

Now, $C_\sigma$ is found by studying how $\varphi_\sigma({\bf k})$ evolves when moving along $\Gamma$. Generally speaking, when the $\Gamma$ path surrounds a $\varphi_\sigma ({\bf k})$'s divergence (here at the $\boldsymbol{M}_2$ point), the accumulated phase increases or decreases by $\pm 2 \pi z, \, z \in \mathbb{Z}$, which gives a quantized $C_\sigma$, as expected. 
We can check it explicitly, and we find within this analytical argument that
\begin{equation}
	C_\sigma =-s_z\textrm{sgn}\left( \sin \phi\right). 
\end{equation} 
A very similar proof can be done for the $0<\lambda<\sqrt{2}t$ case. Therefore, at $n=2/3$ filling, for all flux $\phi \in \, ]0,\pi[$, the phase associated to $\lambda_B=-\lambda_R=\lambda$ and $\lambda_G=0$, $|\lambda|<\sqrt{2}t$ is a topological insulator. 


This computation allows to understand the discontinuous changes in the  $\mathbb{Z}_2$ number. Let us denote $\max(E_l)$ the highest energy eigenvalue associated to the lowest band. At $\gamma=0$, when the $\lambda$ parameter is such that $|\lambda| < \sqrt{2} t$, we have $\max(E_l) <-|\lambda|$. From the eigenvector's coefficients (see Eqs. \eqref{gsigma_GI}-\eqref{bsigma_GIII}, we see that we can not define a continuous gauge choice in the whole BZ. When the $\lambda$ parameter is such that $|\lambda| > \sqrt{2} t$, $\max(E_l)=-|\lambda|$. It is now possible to define a continuous gauge choice in the whole BZ. When varying the $\lambda$ parameter and crossing the $|\lambda| = \sqrt{2} t$ point, this discontinuity in the value of $\max(E_l)$ (from $\max(E_l) <-|\lambda|$ to $\max(E_l)=-|\lambda|$) gives a discontinuity in the $\mathbb{Z}_2$ number.
It is important to notice that when $|\lambda| > \sqrt{2} t$, $\max(E_l)$ is reached at either the high symmetry ${\bf M}_1$ point or the high symmetry ${\bf M}_3$ point. In Appendix~\ref{appendixA.1}. the method for computing the $\mathbb{Z}_2$ number is different. But still, the discontinuity in the value of $\max(E_l)$ yields a discontinuity in parity eigenvalue (associated the lowest band) at the high symmetry points, which gives a discontinuity in the $\mathbb{Z}_2$ number.

\begin{figure}[t!]
\includegraphics[width=8cm]{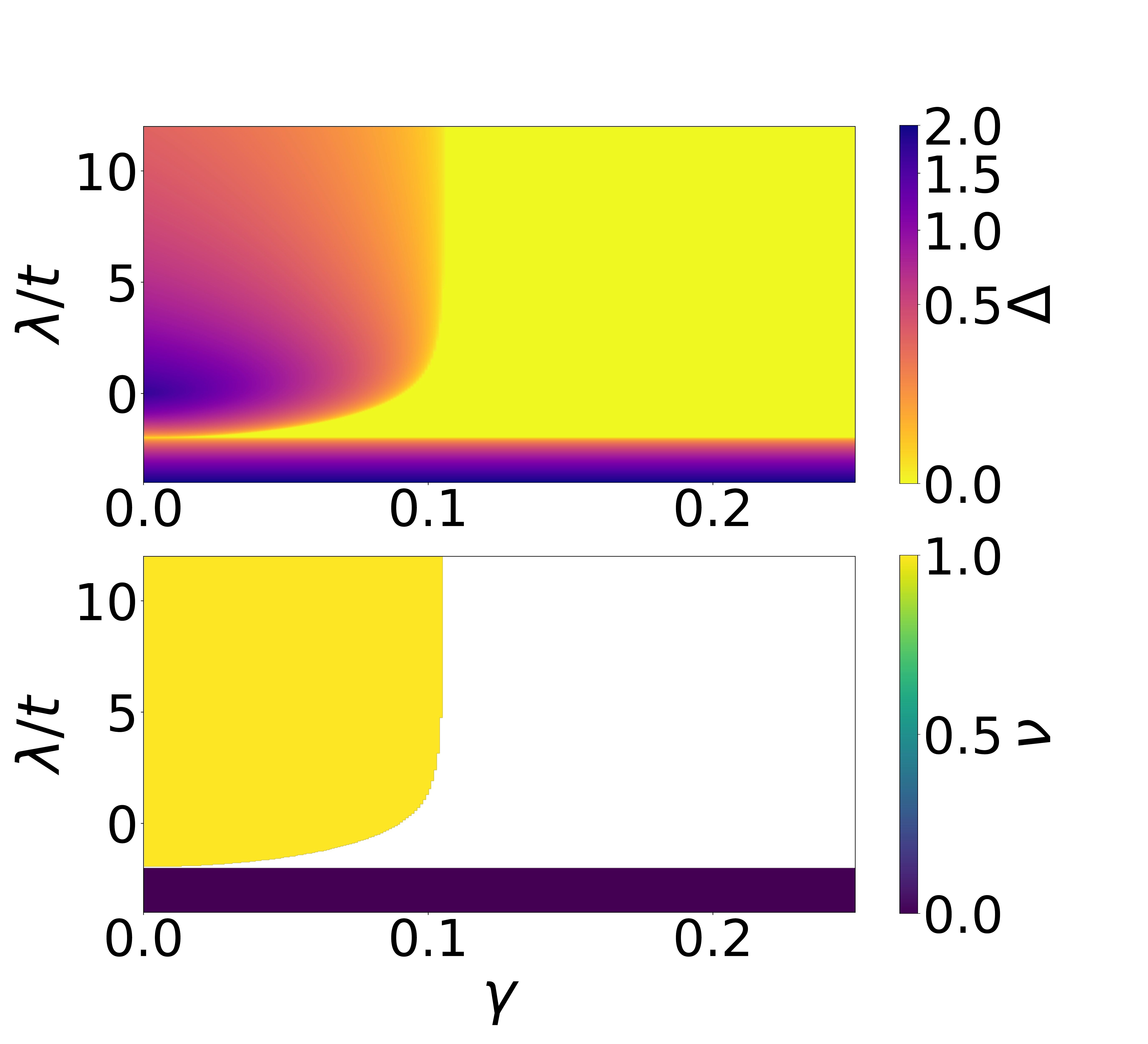}
\caption{The phase diagram for $n=2/3$ filling and $\phi=\pi/2$. $\lambda_R=\lambda$ and $\lambda_B=\lambda_G =0$. The upper panel shows the size of the gap $\Delta$, while the lower panel shows the $\mathbb{Z}_2$ number $\nu$. White region in the lower panel corresponds to the parameter set when the gap is closed.}
\label{Fig:Phase-Diagram_R_n2d3}
\end{figure}

\begin{figure}[t!]
\includegraphics[width=8cm]{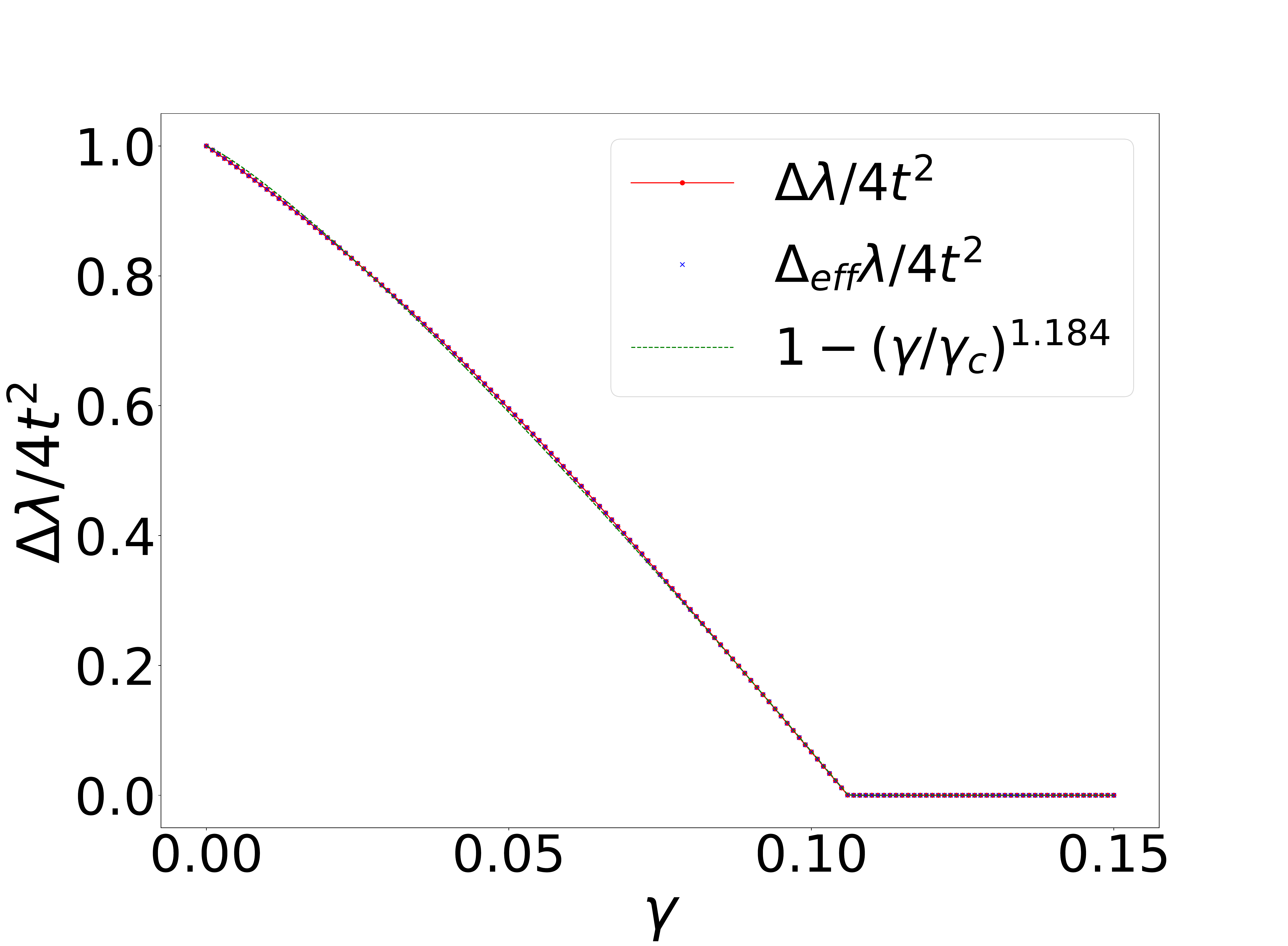}
\caption{Behavior of the gap $\Delta$ as a function of $\gamma$  for filling  $n=2/3$ and flux $\phi=\pi/2$, with $\lambda_R=\lambda=10^3t$ and $\lambda_B=\lambda_G=0$. 
The red curve with circles corresponds to our numerical calculations with the original Hamiltonian (see Eq. \ref{Hamiltonian}). Crosses correspond to numerical calculations obtained from the effective Hamiltonian  Eq.~\eqref{Effective_Hamiltonian_real_space} discussed in section \ref{sec_effective_hamiltonian} and in the appendix \ref{Details_Effective Hamiltonian}. Finally, the dashed curve is a fit function (see Eq.~\eqref{Gap_lambda_gamma}). 
}
\label{Fig:Gap_R_vs_gamma}
\end{figure}

\subsection{On-site energy applied  in $R$ sublattice sites: $\lambda_R=\lambda$ and $\lambda_B=\lambda_G=0$}
\label{on-site-energy-R}

\subsubsection{The numerical approach}

Now we consider the case where one of the sublattices has a finite on-site energy, while on the other two sublattices the on-site energy vanishes.  We start with  $\lambda_R=\lambda$ and $\lambda_B=\lambda_G=0$. We again consider $n=2/3$ filling and we choose the flux $\phi=\pi/2$. 

We observe two gapped phases for ${\lambda<-2t}$ and for ${\lambda>-2t}$ (see upper panel of Fig.~\ref{Fig:Phase-Diagram_R_n2d3}). Except for ${\gamma=0}$, there is always a finite metallic region between these two insulating phases. For ${\gamma=0}$ the gap is closing only for ${\lambda=-2t}$.

To determine if any of these gapped phases are topologically nontrivial we again calculate the  $\mathbb{Z}_2$ number using twisted boundary conditions. We observe that the gapped phase with ${\lambda>-2t}$ is a topological insulator with the $\mathbb{Z}_2$ number ${\nu=1}$, while the gapped phase with ${\lambda<-2t}$ is a trivial band insulator with ${\nu=0}$  (see lower panel of Fig.~\ref{Fig:Phase-Diagram_R_n2d3}). So we obtain three different phases: topological insulator, band insulator and metallic phase.

Our results presented in Fig.~\ref{Fig:Phase-Diagram_R_n2d3} suggest that the topological phase exists even in the limit $\lambda \rightarrow \infty$. To check this claim, we perform calculations for divergingly large $\lambda$ (see Fig.~\ref{Fig:Gap_R_vs_gamma}). We obtain that the gap $\Delta$ is inversely proportional to the on-site energy $\lambda$ and for large $\lambda$ the gap is closing for $\gamma=\gamma_c \simeq 1/3\pi$ (see Fig.~\ref{Fig:Gap_R_vs_gamma}). 
Based on the fit the behavior of the gap for large $\lambda$ is well described by  
\begin{equation}
\label{Gap_lambda_gamma}
\Delta(\lambda,\gamma)=
\frac{4t^2(1-(\gamma/\gamma_c)^{1.184})}{\lambda} \,.
\end{equation}
We thus obtain that for $\gamma < \gamma_c$ the gap is finite and never closed for $\lambda \geq 0$. So, since we know that the system is topological for $\lambda=0$, it should also be topological for any $\lambda>0$ for $\gamma < \gamma_c$.

\subsubsection{The analytical approach}
\label{On-site_energy_R_analytical}

At $\gamma=0$, we compute the $\mathbb{Z}_2$ number for all flux $\phi$ and arbitrary on-site energy $\lambda$. The computation is similar to the one we have done for the $\lambda_B=-\lambda_R=\lambda$ and $\lambda_G=0$ case. Both lowest spin species have the same energy $E$. We denote $\tilde{\lambda}=\lambda/3t$, and then we have
\begin{equation}
E =2t \sqrt{\tilde{\lambda}^2 + \dfrac{4}{3}\left[1 + f({\bf k})\right]} \cos \dfrac{\theta({\bf k})+2\pi}{3},
\end{equation}
with $f({\bf k}) = 2 \prod_{\alpha=1}^3 \cos {\bf k}\cdot{\bf b}_\alpha$ and $0 \leq \theta({\bf k}) \leq \pi$ defined by 
\begin{equation}
\theta({\bf k}) = \arccos \dfrac{\tilde{\lambda}^3+ 2\tilde{\lambda}\left( 1+f({\bf k})- \frac{3}{4t^2}\varepsilon_3^2({\bf k}) \right) -4f({\bf k}) \cos \phi}{\left[ \tilde{\lambda}^2 + \dfrac{4}{3}\left[1 + f({\bf k})\right] \right]^{3/2}} \,.
\end{equation}
From this expression, we can show that the gap between the lowest band and the middle one only closes for the parameter value $\lambda=- 2t$, at the reciprocal space ${\bf M}_3$ point. It is valid for each spin species, to which associated middle band energy is given by 
\begin{equation}
{E_\textrm{m} =2t \sqrt{\tilde{\lambda}^2 + \dfrac{4}{3}\left[1 + f({\bf k})\right]} \cos \dfrac{\theta({\bf k})-2 \pi}{3}} \,.
\end{equation}
We have
\begin{itemize}
\item for $\lambda \leq -2t  $, $E \leq \lambda$, with the equality occurring at the ${\bf M}_3$ point,
\item for $ \lambda > -2t$, $E< \lambda$.
\end{itemize}
Now we can proceed in the same way as in Sec.~\ref{Three_different_on-site_energy_analytical}. We keep the same notations for the coefficients of the wave function and we introduce three new gauge choices.
\begin{itemize}
\item \underline{Gauge choice $G_{IV}$}: the coefficient $g_\sigma({\bf k})$ is real. More precisely, we choose 
\begin{equation}
g_\sigma({\bf k}) = \rho_{2,\sigma}({\bf k}),
\end{equation}
with
\begin{equation}
\rho_{2,\sigma}({\bf k}) e^{i\varphi_{2,\sigma}({\bf k})} = \dfrac{(-E+\lambda) \varepsilon_3({\bf k})}{2\varepsilon_1({\bf k})}  e^{-i s_z \phi} - \frac{1}{2} \varepsilon_2({\bf k}) \,,
\end{equation}
where $\rho_{2,\sigma}({\bf k})$ and $\varphi_{2,\sigma}({\bf k})$ are both real numbers. Then we have
\begin{equation}
r_\sigma({\bf k}) = \dfrac{E^2 - \varepsilon_3^2({\bf k})}{\rho_{3,\sigma}} \rho_{2,\sigma} e^{-i\varphi_{3,\sigma}({\bf k})}
\end{equation}
and
\begin{equation}
b_\sigma({\bf k}) = -\dfrac{E(E-\lambda) - \varepsilon_2^2({\bf k})}{2\varepsilon_1( {\bf k})} e^{-i\varphi_{2,\sigma}({\bf k})} \,,
\end{equation} 
with
\begin{equation}
\rho_{3,\sigma}({\bf k}) e^{i\varphi_{3,\sigma}({\bf k})} =  E \varepsilon_2({\bf k}) + e^{-is_z \phi} \varepsilon_1({\bf k}) \varepsilon_2({\bf k}) \,,
\end{equation}
where $\rho_{3,\sigma}({\bf k})$ and $\varphi_{3,\sigma}({\bf k})$  are both real numbers.
\item \underline{Gauge choice $G_{V}$}: the coefficient $b_\sigma({\bf k})$ is real. We choose 
\begin{equation}
g_\sigma({\bf k}) = \rho_{2,\sigma}({\bf k})e^{i\varphi_{2,\sigma}({\bf k})}.
\end{equation}
Then we have
\begin{equation}
r_\sigma({\bf k}) = \dfrac{E^2 - \varepsilon_3^2({\bf k})}{\rho_{3,\sigma}} \rho_{2,\sigma}  e^{i\left[\varphi_{2,\sigma}({\bf k})-\varphi_{3,\sigma}({\bf k})\right]}
\end{equation}
and
\begin{equation}
b_\sigma({\bf k}) = -\dfrac{E(E-\lambda) - \varepsilon_2^2({\bf k})}{2\varepsilon_1({\bf k})} \,.
\end{equation} 
\item \underline{Gauge choice $G_{VI}$}: the coefficient $r_\sigma({\bf k})$ is real. We choose 
\begin{equation}
g_\sigma({\bf k}) = \rho_{2,\sigma}({\bf k})e^{i\varphi_{3,\sigma}({\bf k})}.
\end{equation}
Then we have
\begin{equation} 
r_\sigma({\bf k}) = \dfrac{E^2 - \varepsilon_3^2({\bf k})}{ \rho_{3,\sigma}} \rho_{2,\sigma} 
\end{equation}
and
\begin{equation}
b_\sigma({\bf k}) =- \dfrac{E(E-\lambda) - \varepsilon_2^2({\bf k})}{2\varepsilon_1({\bf k})} e^{-i\left[\varphi_{2,\sigma}({\bf k})-\varphi_{3,\sigma}({\bf k})\right]} \,.
\end{equation} 
\end{itemize}

When $\lambda < -2t $, we can show that the gauge choice $G_{VI}$ is applicable to the whole BZ. This indicates that at $n=2/3$ filling, for all flux $\phi \in \, ]0,\pi[$, our model with $\lambda_R=\lambda$ and $\lambda_B=\lambda_G=0$, $\lambda<-2t$ is characterized by a trivial insulating phase. 

When {$ \lambda > -2t$}, we split the BZ into two non-overlapping domains referring here to  $\mathcal{D}_{IV}$ and $\mathcal{D}_{V}$ domains. The $\mathcal{D}_{V}$  domain contains the point where $\rho_2({\bf k})$ vanishes and the $\mathcal{D}_{IV}$ domain contains all the points where $E(E-\lambda)  - \varepsilon_2^2({\bf k})$ vanishes or $\varepsilon_1({\bf k})$ vanishes. The application of gauge choice $G_{IV}$ and $G_{V}$ in respectively $\mathcal{D}_{IV}$ and $\mathcal{D}_{V}$ indicates us that at $n=2/3$ filling, for all flux $\phi \in \, ]0,\pi[$, our model with $\lambda_R=\lambda$ and $\lambda_B=\lambda_G=0$, $\lambda>-2t$ is characterized by a topological insulating phase.

\begin{figure}[t!]
\includegraphics[width=8cm]{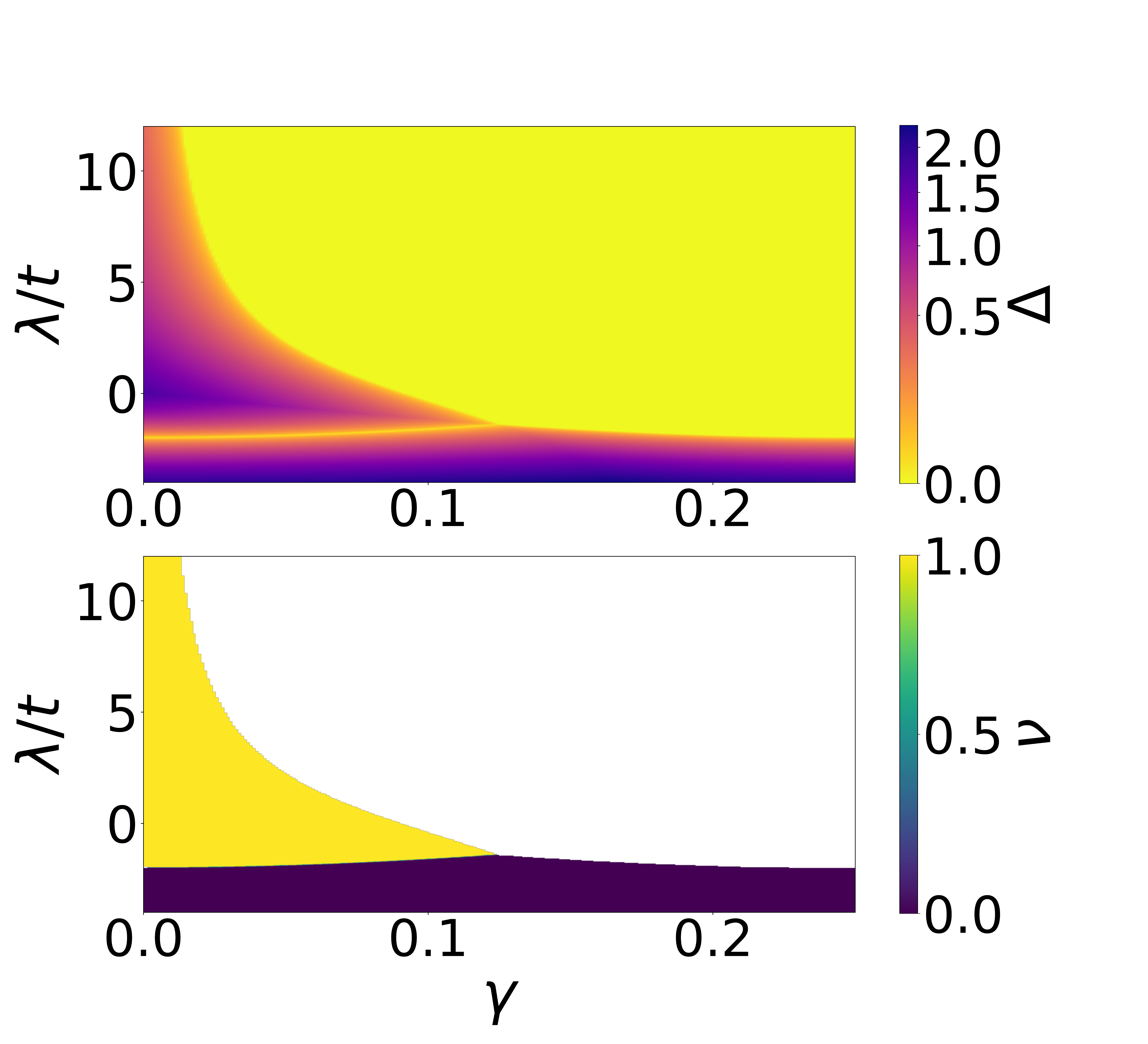}
\caption{The phase diagram for $n=2/3$ filling and $\phi=\pi/2$. $\lambda_B=\lambda$ and $\lambda_R=\lambda_G =0$. The upper panel shows the size of the gap $\Delta$, while the lower panel shows the $\mathbb{Z}_2$ number $\nu$. White region in the lower panel corresponds to the parameter set when the gap is closed.}
\label{Fig:Phase-Diagram_B_n2d3}
\end{figure}

\begin{figure}[t!]
\includegraphics[width=8cm]{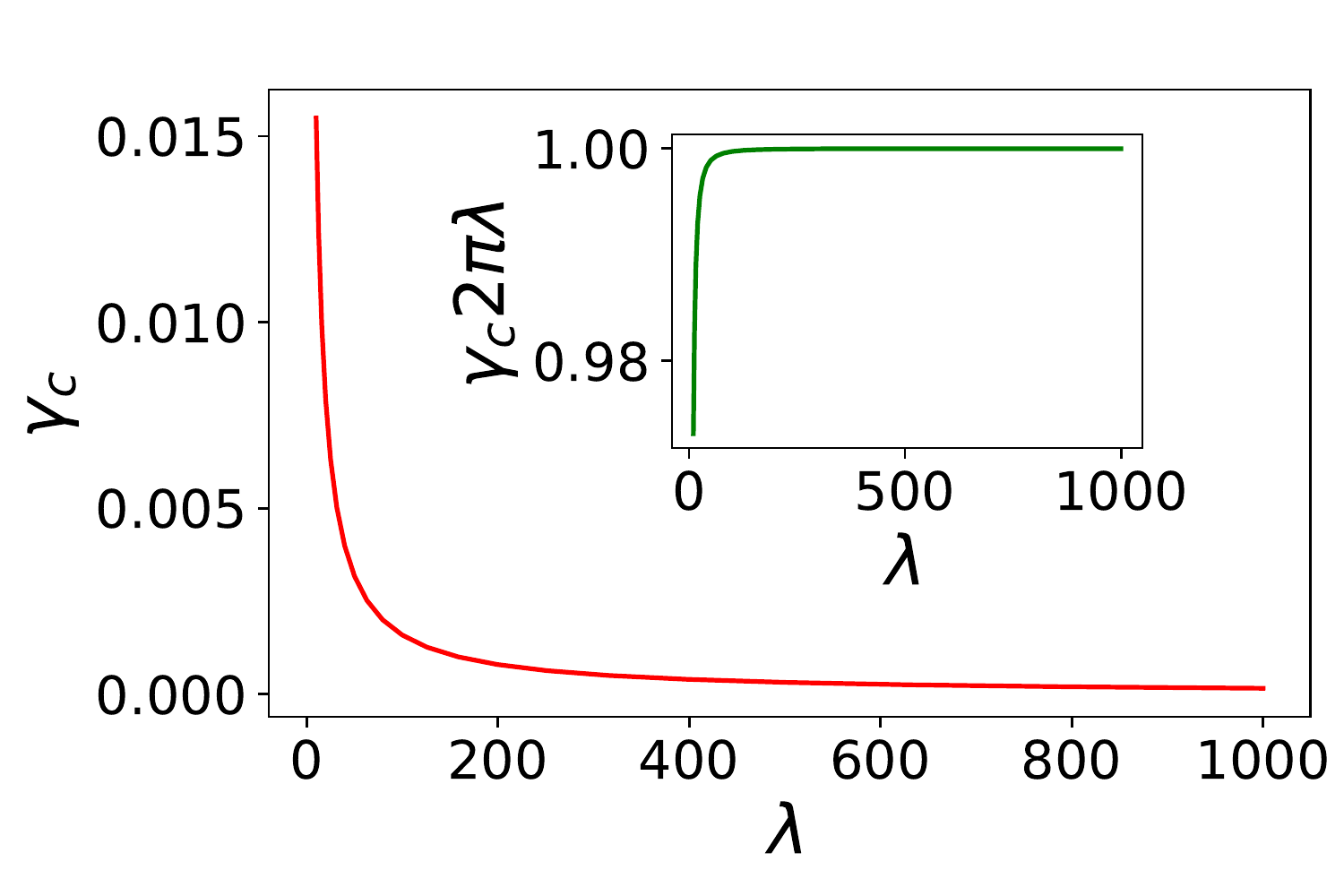}
\caption{The critical value $\gamma_c$ as a function of $\lambda$ for 
$n=2/3$ filling and $\phi=\pi/2$.  $\lambda_B=\lambda$ and $\lambda_R=\lambda_G=0$. The inset shows $2\pi\lambda \gamma_c$ as a function of $\lambda$.}
\label{Fig:gamma_c_B_n2d3}
\end{figure}

\subsection{On-site energy applied  in $B$ sublattice sites: 
${\lambda_B=\lambda}$ and ${\lambda_R=\lambda_G=0}$}
\label{on-site-energy-B}

\subsubsection{The numerical approach}

Now we consider the setup where ${\lambda_B=\lambda}$ and ${\lambda_R=\lambda_G=0}$. The filling is again $n=2/3$ and we fix the flux $\phi=\pi/2$. Similar to the case ${(\lambda_R,\lambda_B,\lambda_G)=(\lambda,0,0)}$, we observe two gapped phases: for $\lambda<-2t$ and  for $\lambda>-2t$, but in contrast to the case when the on-site energy was applied in $R$ sublattice sites, the critical value $\gamma_c$ where the gap is closing, is decreasing with increasing on-site energy $\lambda$ (see upper panel of Fig.~\ref{Fig:Phase-Diagram_B_n2d3}). We analyze this decrease for large $\lambda$ and obtain from the fitting the critical value (see also Fig.~\ref{Fig:gamma_c_B_n2d3})
\begin{equation}
\label{gammaclambda_vs_lambda}
\gamma_c=\frac{t^2}{2\pi\lambda}
\end{equation}

Another difference between the cases when the on-site energy is applied in $R$ sublattice sites and in $B$ sublattice sites is that for the latter, we do not obtain a finite metallic region between two insulator phases in contrast to the former. We observe that when a finite on-site energy is applied in $B$ sublattice sites for fixed $\gamma <1/4$ the gap is closing only at one value of $\lambda$.

We calculate the $\mathbb{Z}_2$ number using twisted boundary conditions. We obtain for the gapped phase with $\lambda<-2t$  $\nu=0$, which is topologically trivial, while for $\lambda>-2$t we again obtain that the $\mathbb{Z}_2$ number $\nu=1$ and the system is in the topological insulator phase (see lower panel of Fig.~\ref{Fig:Phase-Diagram_B_n2d3}). So we again obtain three different phases: topological insulator, band insulator and metallic phase. According to our results, the topological phase exists for $\lambda \rightarrow \infty$, but in this configuration only for $\gamma \to 0$.

\subsubsection{The analytical approach}

At $\gamma=0$, we compute the $\mathbb{Z}_2$ number for all flux $\phi$ and arbitrary on-site energy $\lambda$. The computation is very similar to the one done at $\lambda_R=\lambda$ and $\lambda_B=\lambda_G=0$. We can show that, for all flux $\phi \in \, ]0,\pi[$, the $\lambda<-2t$ parameter space is characterized by a trivial insulating phase while the $\lambda>-2t$ parameter space is characterized by a topological insulating phase.

\begin{figure}[t]
\includegraphics[width=8cm]{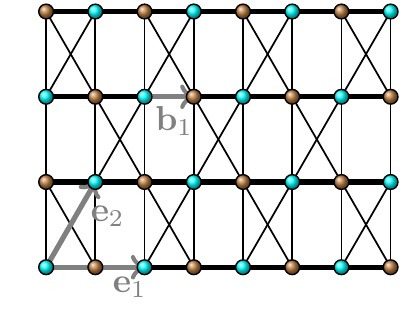}
\caption{
The schematic representation of the lattice corresponding to the effective Hamiltonian in Eq. \eqref{Effective_Hamiltonian_real_space}. The lattice contains two sites per unit cell, which we depict by aqua ($a$) and brown ($b$). Unit cells are arranged on a triangular lattice. Thick lines correspond to the hoppings existing also in the original Hamiltonian in Eq. \eqref{Hamiltonian},  while thin lines correspond to the hoppings which arise in the effective Hamiltonian. }
\label{Fig:effective_schematicp}
\end{figure}

\subsection{Effective Hamiltonian}
\label{sec_effective_hamiltonian}

To understand the behavior of the system for large $\lambda$, when the finite  on-site energy is applied in one of the sublattice sites, we derive the effective Hamiltonian (details of the derivation are given in Appendix \ref{Details_Effective Hamiltonian}). We obtain that the effective model is defined on a rectangular lattice with alternating diagonal hoppings. The resulting structure has two sites in the unit cell ($a$ and $b$), while the unit cells are arranged in a triangular lattice  (see Fig.~\ref{Fig:effective_schematicp}).
The effective Hamiltonian reads
\begin{eqnarray}
\label{Effective_Hamiltonian_real_space}
{\cal H}_{\rm eff}&=&\sum_{n,m}\Biggl[a_{n,m}^{\dagger} \hat t_{h,+} b_{n,m}^{\phantom\dagger}
+a_{n,m}^{\dagger} \hat t_{h,-} b_{n-1,m}^{\phantom\dagger}
\nonumber\\
&+&a_{n,m}^{\dagger} \hat t_{v,+} b_{n-1,m+1}^{\phantom\dagger}
+a_{n,m}^{\dagger} \hat t_{v,-} b_{n,m-1}^{\phantom\dagger}
\nonumber\\
&+&a_{n,m}^{\dagger} \hat t_{d,a} a_{n,m+1}^{\phantom\dagger}
+b_{n,m}^{\dagger} \hat t_{d,b} b_{n-1,m+1}^{\phantom\dagger}
+h.c.\Biggl] \nonumber\\
&&\hspace{-0.2cm}+\sum_{n,m}\left[a_{n,m}^{\dagger} \hat \varepsilon_a a_{n,m}^{\phantom\dagger} + b_{n,m}^{\dagger} \hat \varepsilon_b b_{n,m}^{\phantom\dagger}  \right] \,.
\end{eqnarray}
Here, $a_{n,m}^\dagger$ and $b_{n,m}^\dagger$ create fermions in the unit cell $(n,m)$ on the $a$ and $b$ sublattices. $\hat t_{h,\pm}$, $\hat t_{v,\pm}$, $\hat t_{d,a}$, $\hat t_{d,b}$ are hopping matrices and $\hat \varepsilon_a$ and $\hat \varepsilon_b$ describe matrices of on-site energies.

For $\lambda_R=\lambda$ and $\lambda_B=\lambda_G=0$ we are left with $B$ and $G$ sites. We obtain
\begin{eqnarray}
&&\hat t_{h,\pm}  =-t e^{i \phi\sigma^z}
-\frac{t^2}{\lambda } e^{\pm i 2\pi\gamma\sigma^x}\\
&&\hat t_{v,+}=\hat t_{v,-}^\dagger=-\frac{t^2}{\lambda } e^{i 2\pi\gamma\sigma^x} \\
&&\hat t_{d,a}=\hat \varepsilon_a=\hat \varepsilon_b
=-\frac{t^2}{\lambda } \mathbb{1}\\
&&\hat t_{d,b}=-\frac{t^2}{\lambda } e^{i 4\pi\gamma\sigma^x}  \,.
\end{eqnarray}

For $\lambda_B=\lambda$ and $\lambda_R=\lambda_G=0$ we are left with $G$ and $R$ sites. We obtain
\begin{eqnarray}
&&\hat t_{h,\pm} =-t e^{\pm i2\pi\gamma\sigma^x} 
-\frac{t^2}{\lambda } e^{-i \phi\sigma^z}\\
&&\hat t_{v,+}=\hat t_{v,-}=-\frac{t^2}{\lambda } e^{-i \phi\sigma^z} \\
&&\hat t_{d,a}=\hat t_{d,b}=\hat \varepsilon_a=\hat \varepsilon_b
=-\frac{2t^2}{\lambda}\mathbb{1} \,.
\end{eqnarray}

Based on the effective Hamiltonian in Eq. \eqref{Effective_Hamiltonian_real_space}, we calculate the size of the gap for large values of $\lambda_R=\lambda$ (and $\lambda_B=\lambda_G=0$). We present our results for $\lambda=1000t$ in Fig.~\ref{Fig:Gap_R_vs_gamma}. As one can expect we get perfect agreement with the results obtained by the ``original'' Hamiltonian.

We also analytically investigate the effective Hamiltonian for large values of  $\lambda_R=\lambda$ and $\lambda_B=\lambda_G=0$ (more details see Appendix \ref{Details_Effective Hamiltonian}). For $\gamma=0$ the spin up and spin down fermions are decoupled from each other. Therefore instead of finding eigenvalues of a $4 \times 4$ matrix (the unit cell contains 2 sites and a factor 2 arises due to the spin) one needs to perform calculations for two equivalent $2 \times 2$ matrices. We obtain
\begin{eqnarray}
E_{\sigma,\mp}&=&-\frac{2t^2}{\lambda}\Bigl(1+\cos k_1 \cos(k_1-2k_2)\Bigl)
\\
&\pm&\sqrt{4t^2\cos^2 k_1 + \frac{4t^4}{\lambda^2}\Bigl(1 + \cos k_1\cos  (k_1 - 2k_2) }
\,. \nonumber
\end{eqnarray}
As a result, we obtain (more details see Appendix \ref{Details_Effective Hamiltonian})  
\begin{eqnarray}
\Delta =\min\left[E_{\sigma,+}\right]-\max\left[E_{\sigma,-}\right] \simeq \frac{4t^2}{\lambda} \,,
\end{eqnarray}
in agreement with the result obtained from fitting the numerical data (see Eq.~\eqref{Gap_lambda_gamma})

\section{Results: staggered potential}
\label{section_staggered_potential}

In this Section, we consider the results for a finite staggered potential. As it was mentioned above $V_{\alpha,{\bf r}}=\lambda_{\alpha,1}$
for ${\bf r}=2n_1{\bf e}_1+n_2{\bf e}_2$ and  
$V_{\alpha,{\bf r}}=\lambda_{\alpha,2}$ for ${{\bf r}=(2n_1+1){\bf e}_1+n_2{\bf e}_2}$.  We consider: (i)~${\lambda_{R,1}=\lambda_{B,1}=\lambda_{G,1}=-\lambda_{R,2}=-\lambda_{B,2}=-\lambda_{G,2}=\lambda}$ and (ii)~${\lambda_{R,1}=-\lambda_{R,2}=\lambda}$ and $\lambda_{B,s=1,2}=\lambda_{G,s=1,2}=0$.

\begin{figure}[t!]
\subfigure[]{
\label{Fig:Staggered_4d3_Gap}
{\centering
\includegraphics[width=8cm]{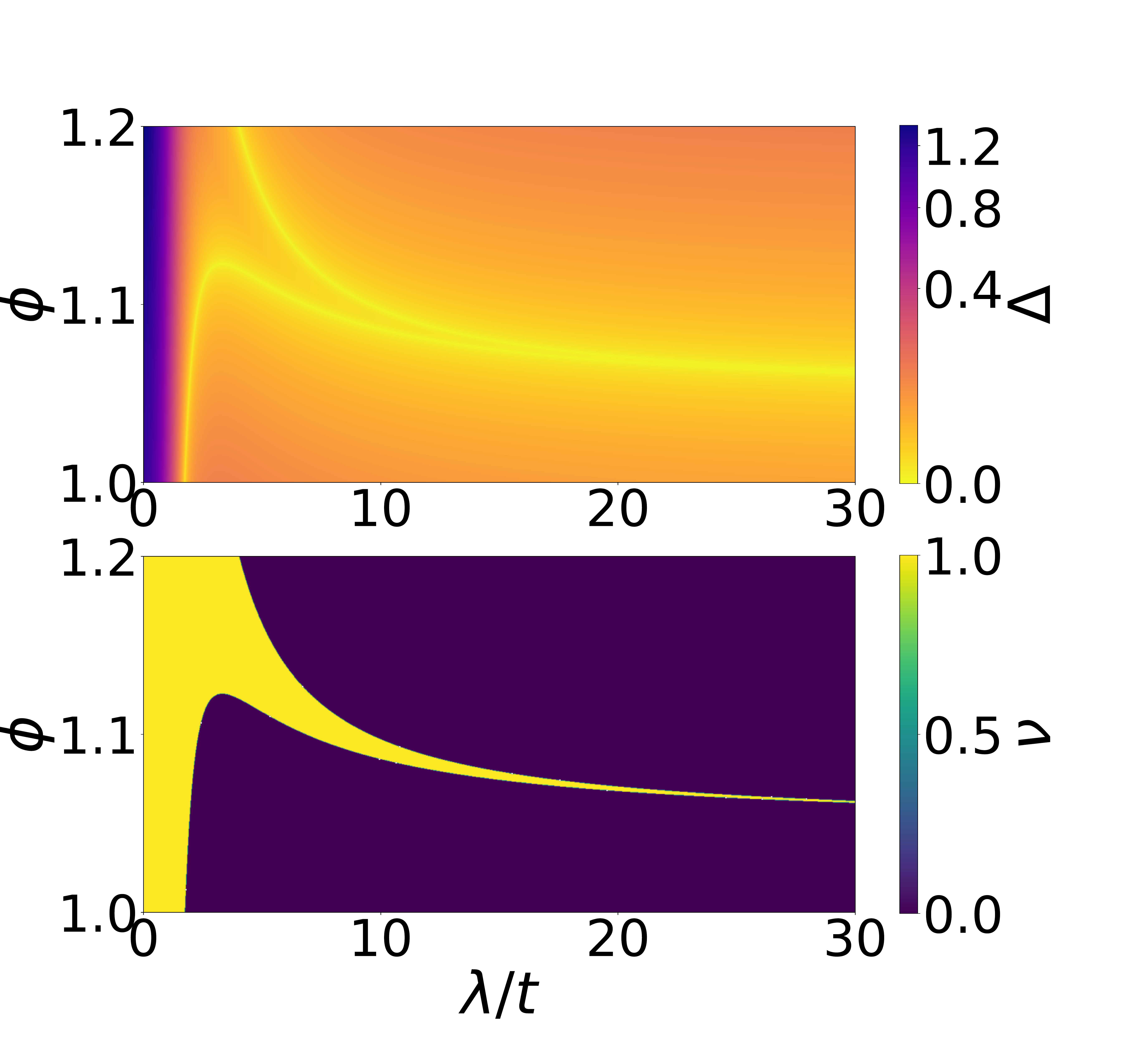}} 
}
\\
\subfigure[]{
\label{Fig:Staggered_Largelambda_fitting}
{\centering 
\includegraphics[width=6cm]{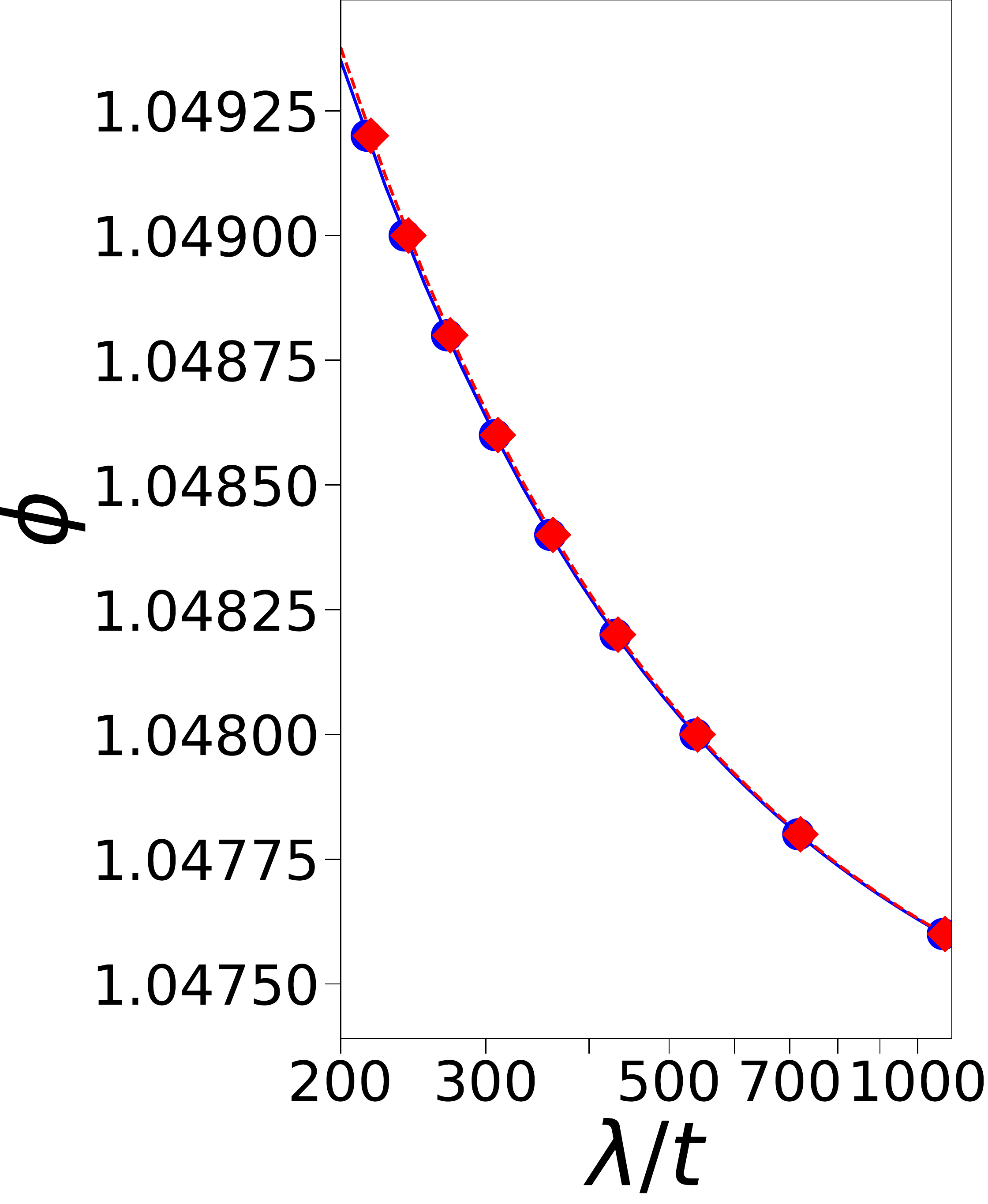}}
}
\caption{
The phase diagram for filling $n=2/3$, ${\lambda_{\alpha,1}=-\lambda_{\alpha,2}=\lambda}$, and $\gamma=0$ and $\phi=\pi/2$. (a) The upper panel shows the size of the gap $\Delta$, while the lower panel shows the $\mathbb{Z}_2$ number $\nu$.  
(b) Topological phase for large values of $\lambda$ between red and blue curves. Red and blue points are obtained by our numerical calculations, while blue and red curves are obtained by fitting (see Eq.~\eqref{critical_phic}). Fitting parameters are $a^{u}=0.4335 \pm 0.0001$, $\lambda^{u}/t= 1.12 \pm 0.09$, $a^{l}=0.4325 \pm 0.0001$, and $\lambda^{l}/t=-0.79 \pm 0.08$. 
}
\label{Fig:4d3_Staggered}
\end{figure}

\begin{figure}[t!]
\centering \includegraphics[width=8cm]{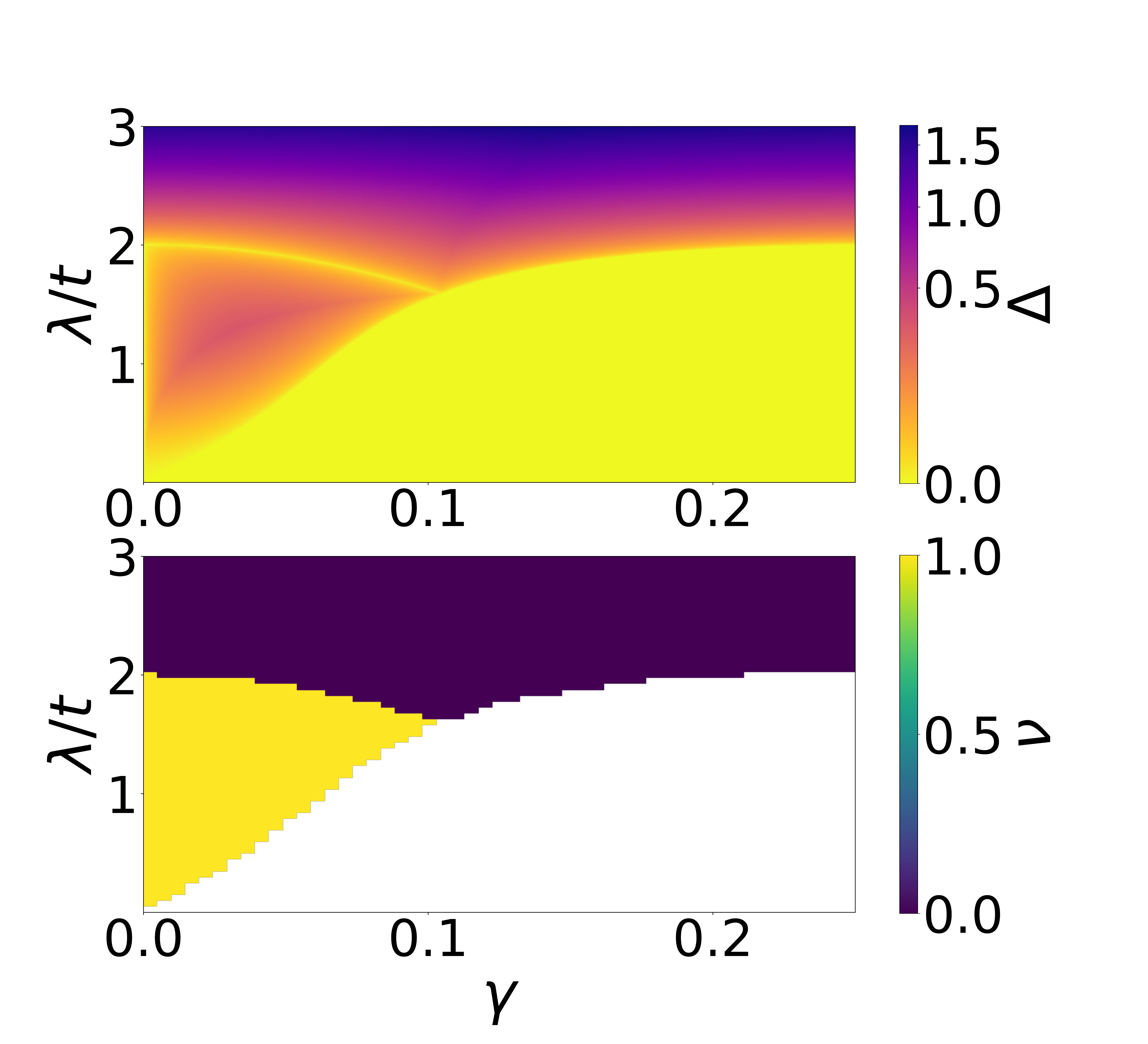}
\caption{
The phase diagram for half-filling ($n =1$), $\phi=\pi/2$, and ${\lambda_{\alpha,1}=-\lambda_{\alpha,2}=\lambda}$. The upper panel shows the size of the gap $\Delta$, while the lower panel shows the $\mathbb{Z}_2$ number $\nu$. White region in the lower panel corresponds to the parameter set when the gap is closed.
}
\label{Fig:HF_Staggered}
\end{figure}

\subsection{Uniform potential inside the unit cell of the Kagome lattice: 
${\lambda_{\alpha,1}=-\lambda_{\alpha,2}=\lambda}$}
\label{Uniform_inside_unitcell}

First, we consider $n=2/3$ filling and $\gamma=0$. We obtained three gapped phases. There is no finite metallic region between them (see Fig.~\ref{Fig:Staggered_4d3_Gap}).  We observe that the gapped phase which  also exists for $\lambda=0$ persists for arbitrarily large $\lambda$, which has been numerically confirmed up to $\lambda = 1000t$. However, the gap size shrinks with increasing $\lambda$. Both boundaries of the topological phase can be conveniently fitted to an inverse proportionality 
\begin{equation}
\label{critical_phic}
\phi_{c}^{b}=\frac{a^{b}t}{\lambda-\lambda_0^{b}}+\frac{\pi}{3} 
\end{equation}
for large $\lambda$. The fit parameters $a^{b}$ and $\lambda_0^{b}$ for the upper ($b=u$) and the lower ($b=l$) boundaries are obtained using 10 points for each boundary. We obtain $a^{u}=0.4335 \pm 0.0001$, $\lambda^{u}/t= 1.12 \pm 0.09$, $a^{l}=0.4325 \pm 0.0001$, and $\lambda^{l}/t=-0.79 \pm 0.08$ 
(see also Fig.~\ref{Fig:Staggered_Largelambda_fitting}).

Based on our results for the system without on-site energies (see Sec.~\ref{Without_on-site_energies}), we can predict that the gapped phase, which also exists for $\lambda=0$ (no on-site energies), must be a topological insulator. We calculate the $\mathbb{Z}_2$ number using twisted boundary conditions and for the gapped phase mentioned above we indeed obtain $\nu=1$ in agreement with our prediction. Our calculations show that the other two gapped phases are topologically trivial band insulators. To conclude, we again obtain a topological phase in the limit $\lambda \rightarrow \infty$ for $\phi =\pi/3$.  
We also perform calculations for the finite $\gamma$ and we obtain a similar phase diagram (not shown).

For the half-filled case (n = 1), in contrast to the case without staggered potential ($V_{\alpha,{\bf r}}=\lambda_\alpha$) which is always metallic for half filling, we recognize two distinct insulating phases for $\phi \neq 0$. For $\phi \lesssim 0.2$, we additionally observe a third insulating phase for small $\gamma$ and $\lambda  \approx 1.5$ (not shown). We do not observe a finite metallic region between these three insulating phases for $\phi \neq 0$, hence the gap is closing at a specific value of $\lambda$ for each given $\gamma$. 
For $\gamma=0$, the closing of the gap takes place at $\lambda=2t$. Our results for the flux $\phi=\pi/2$ are presented in the upper panel of Fig.~\ref{Fig:HF_Staggered}. 
For $\phi = \pi/2$ a gapped phase does exist for all $\gamma \neq 0$. Here, for $\gamma \to 0$, there exists a gapped phase for all values of $\lambda$ except $0$ and $2$. If $\phi \neq \pi/2$, the gap closes for finite values of $\lambda$, and remains open for $\gamma = 0$ (not shown).

To find out if one of these gapped phases are topologically non-trivial, we calculate the $\mathbb{Z}_2$ number using twisted boundary conditions. Our results are presented in the lower panel of Fig.~\ref{Fig:HF_Staggered}. We obtain that the gapped phase with $\lambda<2t$ corresponds to the topological insulator, while the gapped phase with $\lambda>2t$ and the third gapped phase with $\lambda \approx 1.5$ are trivial band insulators.

\begin{figure}[t!]
%
\label{Fig:Staggered_2d3_R_Gap}
\centering \includegraphics[width=8cm]{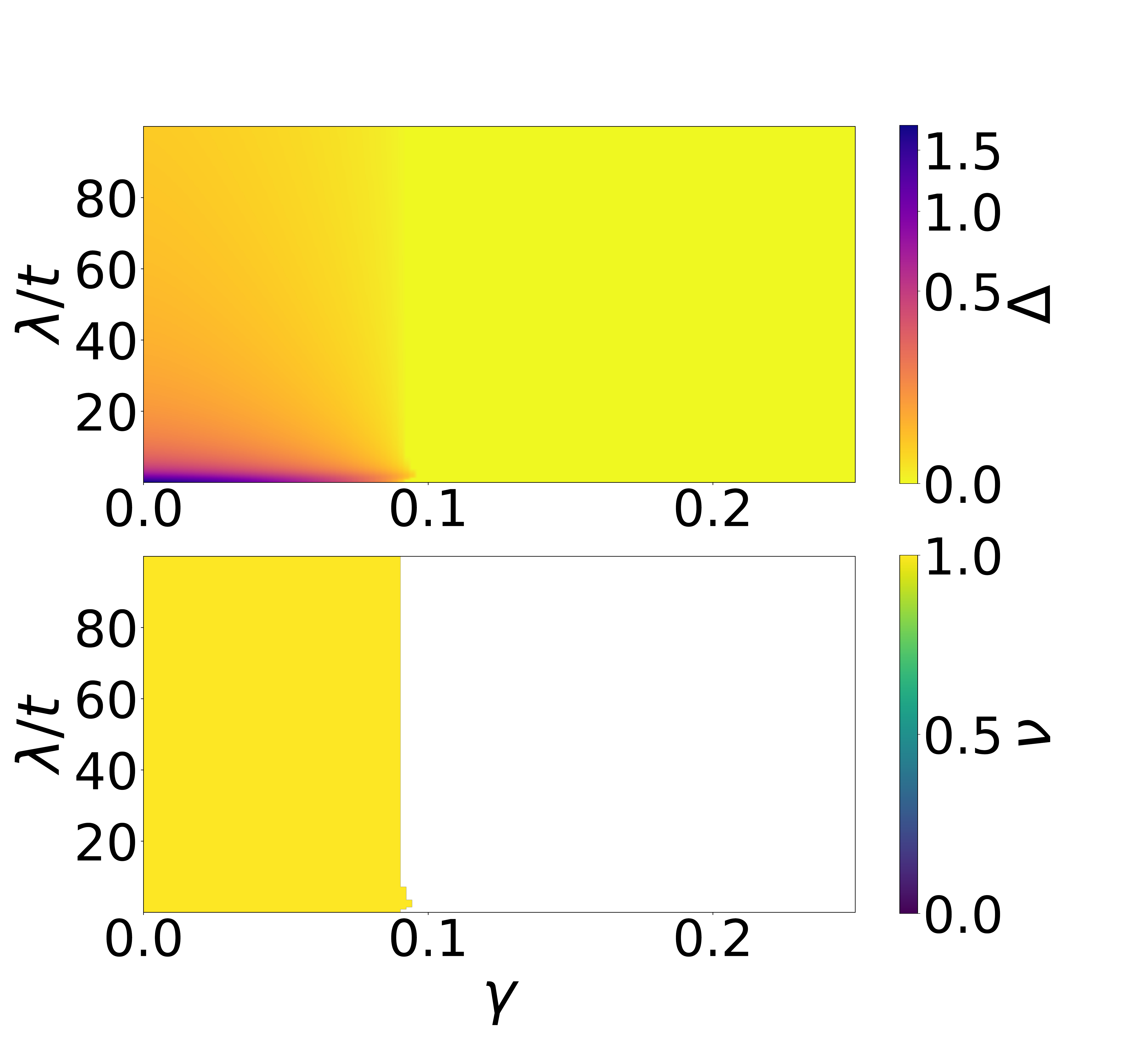}
\\
\caption{
The phase diagram for filling $n=2/3$, $\phi=\pi/2$, and ${\lambda_{R,1}=-\lambda_{R,2}=\lambda}$ and $\lambda_{B,s=1,2}=\lambda_{G,s=1,2}=0$. The upper panel shows the size of the gap $\Delta$, while the lower panel shows the $\mathbb{Z}_2$ number $\nu$. White region in the lower panel corresponds to the parameter set when the gap is closed. 
}
\label{Fig:2d3_R_Staggered}
\end{figure}

\subsection{On-site energies applied in $R$-sublattice sites: ${\lambda_{R,1}=-\lambda_{R,2}=\lambda}$ and $\lambda_{B,s=1,2}=\lambda_{G,s=1,2}=0$}
\label{Staggered_R_sublattice}

Here, we study the behavior of the system when a finite on-site energy is applied in one of the sublattice sites only. We present our results for filling $n = 2/3$. We consider the flux  $\phi=\pi/2$  and on-site energies ${\lambda_{R,1}=-\lambda_{R,2}=\lambda}$ and $\lambda_{B,s=1,2}=\lambda_{G,s=1,2}=0$. We observe only one gapped phase which also exists for very large values of $\lambda$. We check this numerically up to $\lambda=100t$. Our results are presented in Fig.~\ref{Fig:2d3_R_Staggered}. By calculating the $\mathbb{Z}_2$ number we obtain that the gapped phase is a topological insulator. Again for this setup, we obtain a topological phase in the limit $\lambda \rightarrow\infty$.

We also perform calculations for other values of $\phi$ (not shown) and we obtain a similar phase diagram.

\section{Conclusions}
\label{Conclusions}
In this work, we have studied topological properties of the spin-orbit coupled time-reversal 
tight-binding model  with flux on the Kagome lattice. In addition to the spin-orbit coupling $\gamma$ and the flux $\phi$, we also considered the effect of the on-site energies $V_{\alpha,{\bf r}}$. First, we considered the case where the on-site energies are independent of the spatial coordinate $\bf r$ but differ within the unit cell, i.e., $V_{\alpha,{\bf r}}=\lambda_\alpha$. In this case, we obtain that the system has three bands that are potentially non-overlapping for any value of $\bf k$.  It follows that for filling situations $n=2/3$ and $n=4/3$ a gapped phase can occur. Because of the symmetries of our model, the results for these two fillings are related by Eq. \eqref{symmetry}. Here we have presented results for the the filling $n=2/3$. 
We also performed calculations for a staggered potential. For the latter we have $V_{\alpha,{\bf r}}=\lambda_{\alpha,1}$ for ${\bf r}=2n_1{\bf e}_1+n_2{\bf e}_2$ and  
$V_{\alpha,{\bf r}}=\lambda_{\alpha,2}$ for ${\bf r}=(2n_1+1){\bf e}_1+n_2{\bf e}_2$. In this case, we obtain that the system has six potentially non-overlapping bands for all values of $\bf k$. It follows that a gapped phase can occur for fillings $n=1/3$, $n=2/3$, $n=1$, $n=4/3$, and $n=5/3$. In this work we have presented results for $n=2/3$ and for half-filling $n=1$.

The model has time-reversal symmetry, and to determine its topological nature we calculated the $\mathbb{Z}_2$ invariant. For this purpose, we used three methods. One of them is numerical and based on the twisted boundary conditions\cite{fu.ha.05, fu.ha.07, ku.me.16, ir.zh.20}, while the other two are analytical. 
One  of them introduces different smooth fields for several domains in the reciprocal space and was developed by us. The phase accumulated at the boundary between the smooth fields' domains is related to the topological properties.
It is important to remind that such a smooth fields approach has recently been shown to be useful on the honeycomb lattice, e.g., 
to indicate that transport and light-matter properties can be revealed from the Dirac points only\cite{kl.gr.02, hu.hu.02u}.

The other approach described in Appendix \ref{appendixA.1} is along the line of the method introduced on the honeycomb lattice for $\mathbb{Z}_2$ topological insulators\cite{ka.me.05a, ka.me.05b, fu.ka.07}. We also aimed at developing the "counting points" method in an energy band, as discussed previously in Ref. \onlinecite{pe.ho.12}. Here, we observe gap closing effects when $\gamma \neq 0$ such that the results must be taken with care in that situation; see Appendix~\ref{appendixA.2}.

Depending on the model parameters, we obtain the topological insulator, the trivial band insulator, and the metallic phase. We show that the obtained topological phases are stable after applying on-site energies. 
Even more, if the flat band for half-filling is split due to the the staggered potential in combination with flux and spin-orbit coupling, stable topological phases appear for arbitrarily small gap sizes.

One of the most interesting results we obtain is the existence of the topological phase for infinitely large on-site energies.  We show that for selected sets of parameters, in particular when on-site energies were applied only on one of the sublattice sides for $n=2/3$, we obtain a topological insulator for infinitely large on-site energies.  

The situations with finite on-site energies were studied previously related to topological insulators. In the original paper by Kane and Mele, the effect of a finite staggered potential was already considered\cite{ka.me.05a, ka.me.05b}.
Topological phases were obtained for finite on-site energies, while for large on-site energies, the system was shown to be topologically trivial. Recently, a topological phase for infinitely large on-site energies was observed for a mixture of three component fermions, on a triangular lattice at $1/3$ filling, in the presence of a gauge potential stabilizing a quantum Hall insulator\cite{ha.zh.20}.

Experimental progress in ultracold atomic gases allows to realize and study such systems.  Topological phases have already been realized by loading ultracold atoms in optical lattices\cite{ai.at.13, ai.lo.15, mi.hi.13, jo.me.14, fl.re.16, ma.pa.15, st.lu.15}. For this purpose, artificial gauge fields\cite{ge.da.10, da.ge.11, go.da.14, go.ju.14, go.bu.16, ga.ju.19, ga.sp.13} were used.  In particular, ultracold atomic gases also allow the realization of SOC\cite{li.ji.11, wa.yu.12, ch.so.12,hu.me.16}. As mentioned above, the Kagome lattice can be realized experimentally by superimposing two optical triangular lattices with different wavelengths.\cite{jo.gu.12}  
The experimental realization of periodically oscillating on-site energies, i.e. superlattices, is also well established\cite{gu.ve.98, ri.ge.06, se.an.06, fo.tr.07, ch.tr.08, ge.da.10, la.ca.12, na.ch.12}.
The physics obtained in this paper can be observed for systems cooled down to temperatures $T<\Delta/k_B$, as the gap $\Delta$ is of the order of $t$ and the typical hopping amplitude for ultracold atomic gases is of the order $0.1$kHz\cite{bl.da.08}, the system must reach temperatures of the order of $1$nK. In real materials $\phi \leq \pi/4$ and spin-orbit coupling is always smaller compared to the hopping amplitude between corresponding nearest neighbors, which in our notations means that $\gamma \leq 1/4$.\cite{ma.ko.15, wa.xu.18, wa.su.16}  Ultracold atomic gases allow to consider larger intervals of Peierls phase  $\phi$. The maximal value of the Peierls phase which can be obtained is $\phi=\pi/2$ while $\gamma \leq 1/4$.\cite{st.ga.10, go.sa.10} We conclude that the results obtained in our work may be very relevant for upcoming experiments.

\begin{acknowledgments}
This work was supported by the Deutsche Forschungsgemeinschaft (DFG, German Research Foundation) under Project No.~277974659 via Research Unit FOR 2414. This work was also supported by the DFG via the high performance computing center Goethe-HLR. 
The research on the topological Kagome lattice is also funded by ANR BOCA (KLH) for which JL is also supported for his PhD.
\end{acknowledgments}

\appendix
\section{Two other ways of computing the $\mathbb{Z}_2$ number at $\gamma =0$.} \label{appendixA}

\begin{table*}
\begin{tabular}{|c|c|c|c|c|c|} 
\hline
$\lambda$ & $(n_1,n_2)$& Eigenvalues $E_l$ &Eigenvector's coefficient $\left(r_\sigma({\bf k}), b_\sigma({\bf k}), g_\sigma({\bf k})\right)$ & Parity operator & $(-1)^\nu$ \\
\hline
\multirow{3}*{$\lambda<-\sqrt{2}t$} &  $(1,0)$ 
& $\lambda$  &$(0,t,0)$ & diag$(1,-1,1)$ & \multirow{3}*{$+1$} \\ 
 \cline{2-5}
&  $(0,1)$&$-\sqrt{4t^2+\lambda^2}$  & $(-2t^2 /\left(\lambda+E_l\right),t,0)$ &diag$(1,1,-1)$ & \\
\cline{2-5}
& $(1,1)$ &$\left(\lambda-\sqrt{16t^2+\lambda^2}\right)/2$ &  $(0,2t^2 \textrm{e}^{i s_z \phi}/\left(\lambda-E_l\right),t)$ &diag$
 (1,-1,-1)$ &  \\
\hline
\multirow{3}*{$-\sqrt{2}t<\lambda<\sqrt{2}t$} & $(1,0)$
& $-\left(\lambda+\sqrt{16t^2+\lambda^2}\right)/2$   & $(-2t^2 /\left(\lambda+E_l\right),0,t)$ & diag$(1,-1,1)$ & \multirow{3}*{$-1$} \\
\cline{2-5}
& $(0,1)$ &$-\sqrt{4t^2+\lambda^2}$   & $(-2t^2 /\left(\lambda+E_l\right),t,0)$ &diag$(1,1,-1)$ &  \\
\cline{2-5}
&  $(1,1)$ &$\left(\lambda-\sqrt{16t^2+\lambda^2}\right)/2$   & $(0,2t^2 \textrm{e}^{i s_z \phi}/\left(\lambda-E_l\right),t)$ & diag$
 (1,-1,-1)$ & \\
\hline
\multirow{3}*{$\lambda>\sqrt{2}t$} & $(1,0)$ 
&$-\left(\lambda+\sqrt{16t^2+\lambda^2}\right)/2$  & $(-2t^2 /\left(\lambda+E_l\right),0,t)$ & diag$(1,-1,1)$ & \multirow{3}*{$+1$} \\
\cline{2-5}
& $(0,1)$ &$-\sqrt{4t^2+\lambda^2}$  & $(-2t^2 /\left(\lambda+E_l\right),t,0)$ &diag$(1,1,-1)$ &  \\
\cline{2-5}
&  $(1,1)$ &$-\lambda$ & $(t,0,0)$ & diag$
 (1,-1,-1)$ & \\
\hline
\end{tabular}
\caption{\label{table0}Table resuming the computation of the $Z_2$ topological invariant $\nu$ for an inversion and time-reversal symmetric system. In this example, we took the chemical potentials to be $\lambda_B=-\lambda_R=\lambda$ and $\lambda_G=0$. For definition $(n_1,n_2)$ see Eq. \eqref{App_Gamma}, while for definitions $\left(r_\sigma({\bf k}), b_\sigma({\bf k}), g_\sigma({\bf k})\right)$ see Sec.~\ref{Three_different_on-site_energy_analytical}}
\end{table*} 

\begin{table*}
\begin{tabular}{|c|c|c|c|c|c|} 
\hline
$\lambda$ & $(n_1,n_2)$ &Eigenvalues $E_l$ & Eigenvector's coefficient $\left(r_\sigma({\bf k}), b_\sigma({\bf k}), g_\sigma({\bf k})\right)$ & Parity operator & $(-1)^\nu$ \\
\hline
\multirow{3}*{$\lambda<-2t$} &  $(1,0)$ 
&  $\left(\lambda-\sqrt{16t^2+\lambda^2}\right)/2$  &$(-E_l/2,0,t)$ & diag$(1,-1,1)$ & \multirow{3}*{$+1$} \\ 
 \cline{2-5}
 &  $(0,1)$ &  $\left(\lambda-\sqrt{16t^2+\lambda^2}\right)/2$  & $(-E_l/2,t,0)$ &diag$(1,1,-1)$ & \\
\cline{2-5}
 & $(1,1)$ &  $\lambda$ & $(t,0,0)$ &diag$
 (1,-1,-1)$ &  \\
\hline
\multirow{3}*{$\lambda>-2t$} & $(1,0)$ 
&  $\left(\lambda-\sqrt{16t^2+\lambda^2}\right)/2$  & $(-E_l/2,0,t)$ & diag$(1,-1,1)$ & \multirow{3}*{$-1$} \\
\cline{2-5}
& $(0,1)$ &  $\left(\lambda-\sqrt{16t^2+\lambda^2}\right)/2$ & $(-E_l/2,t,0)$ &diag$(1,1,-1)$ &  \\
\cline{2-5}
&  $(1,1)$ &  $-2t$ & $(0,-t\textrm{e}^{i s_z \phi},1)$ & diag$
 (1,-1,-1)$ & \\
\hline
\end{tabular}
\caption{\label{table1}Table resuming the computation of the $Z_2$ topological invariant $\nu$ for an inversion and time-reversal symmetric system. In this example, we took the chemical potentials to be $\lambda_R=\lambda$ and $\lambda_B=\lambda_G=0$. For definition $(n_1,n_2)$ see Eq. \eqref{App_Gamma}, while for definitions $\left(r_\sigma({\bf k}), b_\sigma({\bf k}), g_\sigma({\bf k})\right)$ see Sec.~\ref{On-site_energy_R_analytical}}
\end{table*}

\subsection{Using inversion symmetry.} \label{appendixA.1}

When our system is inversion symmetric (\textit{i.e.} $\gamma = 0$), the $\mathbb{Z}_2$ topological invariant $\nu$ (defined for time-reversal invariant Hamiltonian, \textit{i.e.} $B=0$) is given by\cite{fu.ka.07}
\begin{equation} \label{invsym}
(-1)^\nu = \prod_{\substack{n_1=0,1\\n_2=0,1 }}\delta_{(n_1,n_2)}
\end{equation}
with $\delta_{(n_1,n_2)} = \prod_{m=1}^N p_{2m}\left( \Gamma_{(n_1,n_2)} \right)$ is the product of the parity eigenvalues $p_{2m}$, at $\Gamma_{(n_1,n_2)}$, associated to each occupied Kramers' degenerate pair ($p_{2m}$ and $p_{2m-1}$ are identical). The $\Gamma_{(n_1,n_2)}$ points are the time-reversal invariant points, \textit{i.e.} the points such that $H(\Gamma_{(n_1,n_2)}) = \Theta H(\Gamma_{(n_1,n_2)}) \Theta^{-1}$. These points can be written
\begin{equation}
\Gamma_{(n_1,n_2)} = \dfrac{1}{2} \left( n_1 \boldsymbol{g}_1 + n_2 \boldsymbol{g}_2 \right),
\label{App_Gamma}	
\end{equation}
with $\boldsymbol{g}_1$ and $\boldsymbol{g}_2$ both reciprocal lattice basis vectors. These points correspond to the so-called $\bf \Gamma$ point and to the three $\bf M$ points. 

We find the parity eigenvalues $p_{2m}$ associated to each of the four $\Gamma_{(n_1,n_2)}$ points. We need to express the Hamiltonian and the momentum space parity operator at these points. When $\gamma =0$, the Hamiltonian decouples into two independent parts, and it gives 3 energy bands associated to the spin up states, which are degenerated with the 3 energy bands associated to the spin down states. At $n=2/3$ filling, a Kramers' degenerate pair is occupied. This pair contains the lowest energy band associated to each spin species. Formula~\ref{invsym} only implies the parity eigenvalues associated to one band, because the ones associated to the other band are identical. 
Therefore, in the following, we consider separately each of both diagonal parts of the Hamiltonian~\ref{Hk_matrix}.

The parity operator can be defined in real space by 
\begin{equation}
P \left(c_{R,\boldsymbol{r}}, c_{B,\boldsymbol{r}}, c_{G,\boldsymbol{r}} \right) = \left(c_{R,-\boldsymbol{r}}, c_{B,-\boldsymbol{r}-\boldsymbol{e}_1}, c_{G,-\boldsymbol{r}-\boldsymbol{e}_2} \right).
\end{equation}
A Fourier transformation of the fermionic operators enables to see that, in momentum space, the parity operator reads 
\begin{equation}
P_{{\bf k}} = \textrm{diag}\left( 1, \textrm{e}^{- i \boldsymbol{e}_1 {\bf k}}, \textrm{e}^{- i \boldsymbol{e}_2 {\bf k}}\right).
\end{equation}
Therefore, at the ${\bf \Gamma}_{(0,0)}$ point, $P_{\boldsymbol{\Gamma}_{(0,0)}}$ is diagonal so $\delta_{{(0,0)}}=1$ and at the other points we have ${P_{\boldsymbol{\Gamma}_{(1,0)}} = \textrm{diag}\left( 1,-1, 1\right)}$, ${P_{\boldsymbol{\Gamma}_{(0,1)}} = \textrm{diag}\left( 1,1,-1\right)}$, and ${P_{\boldsymbol{\Gamma}_{(1,1)}} = \textrm{diag}\left( 1,-1,-1\right)}$. 

Now we need to evaluate the lowest energy associated eigenvectors of ${\cal H}_\uparrow({\bf k})$ at $\boldsymbol{\Gamma}_{(1,0)}$, $\boldsymbol{\Gamma}_{(0,1)}$, and $\boldsymbol{\Gamma}_{(1,1)}$. 
%
We consider two different setups $\lambda_B=-\lambda_R=\lambda$ and $\lambda_G=0$ or $\lambda_R=\lambda$ and $\lambda_B=\lambda_G=0$. 
We start with $\lambda_B=-\lambda_R=\lambda$ and $\lambda_G=0$. 
We denote the lowest energy eigenvalue $E_l({\bf k})$ and we write $\ket{u_{\sigma,{\bf k}}}$ the $\sigma$ spin species associated eigenvector 
\begin{equation}
\ket{u_{\sigma,{\bf k}}} = \dfrac{\left(r_\sigma({\bf k}) c_{R,{\bf k},\sigma}^{\dagger} + b_\sigma({\bf k}) c_{B,{\bf k},\sigma}^{\dagger} + g_\sigma({\bf k}) c_{G,{\bf k},\sigma}^{\dagger}\right) \ket{0}}{\sqrt{\left|r_\sigma({\bf k})\right|^2+\left|b_\sigma({\bf k})\right|^2+\left|g_\sigma({\bf k})\right|^2}} \,.
\end{equation}
We notice that
\begin{itemize}
\item at the $\boldsymbol{\Gamma}_{(1,0)}$ point, for $\lambda<-\sqrt{2}t$, the smallest eigenvalue is ${E_l=\lambda}$  and the associated eigenvector is proportional to $(0,1,0)$, while for ${\lambda>-\sqrt{2}t}$, the smallest eigenvalue is ${E_l=-\left(\lambda+\sqrt{16t^2+\lambda^2}\right)/2}$, with associated eigenvector proportional to $(-2t /\left(\lambda+E_l\right),0,1)$.
\item at the $\boldsymbol{\Gamma}_{(0,1)}$ point, the smallest eigenvalue is ${E_l=-\sqrt{4t^2+\lambda^2}}$ and the eigenvector associated to it is proportional to $(-2t /\left(\lambda+E_l\right),1,0)$. 
\item at the $\boldsymbol{\Gamma}_{(1,1)}$ point, if $\lambda>\sqrt{2}t$, then ${E_l=-\lambda}$ is the smallest eigenvalue and the associated eigenvector is proportional to $(1,0,0)$, while if $\lambda<\sqrt{2}t$, then the smallest eigenvalue is ${E_l=\left(\lambda-\sqrt{16t^2+\lambda^2}\right)/2}$, with associated eigenvector proportional to $(0,2t \textrm{e}^{i s_z \phi}/\left(\lambda-E_l\right),1)$.
\end{itemize}
Applying the parity operator and the eigenvectors, we find that the phases $\lambda>\sqrt{2}t$ and $\lambda<-\sqrt{2}t$ are trivial insulators while the phase $-\sqrt{2}t<\lambda<\sqrt{2}t$ is a topological insulator. These results are sum up in table~\ref{table0}.

Now we consider
$\lambda_R=\lambda$ and $\lambda_B=\lambda_G=0$.
We notice that 
\begin{itemize}
\item at the $\boldsymbol{\Gamma}_{(1,0)}$ point, the smallest eigenvalue is ${E_l=\left(\lambda-\sqrt{16t^2+\lambda^2}\right)/2}$, with associated eigenvector proportional to $(-E_l/2t,0,1)$.
\item at the $\boldsymbol{\Gamma}_{(0,1)}$ point, the smallest eigenvalue is ${E_l=\left(\lambda-\sqrt{16t^2+\lambda^2}\right)/2}$, with associated eigenvector proportional to $(-E_l/2t,1,0)$.
\item at the $\boldsymbol{\Gamma}_{(1,1)}$ point, if $\lambda<-2t$, then $E_l=\lambda$ is the smallest eigenvalue and the associated eigenvector is proportional to $(1,0,0)$, while if $\lambda>-2t$, then the smallest eigenvalue is $E_l=-2t$, with associated eigenvector proportional to $(0,-\textrm{e}^{i s_z \phi},1)$.
\end{itemize}

Applying the parity operator on the eigenvectors, we find that the phase $\lambda<-2t$ is a trivial insulator while the phase $\lambda>-2t$ is a topological insulator. These results are sum up in Table~\ref{table1}.

\begin{figure}[t!]
\begin{center}
\includegraphics[width=8cm]{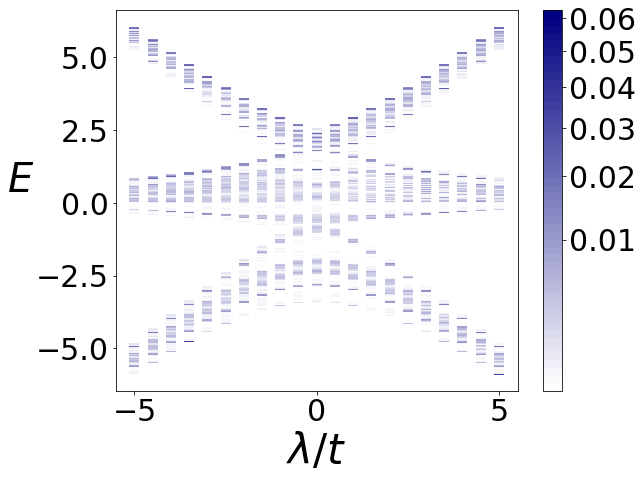}

\includegraphics[width=8cm]{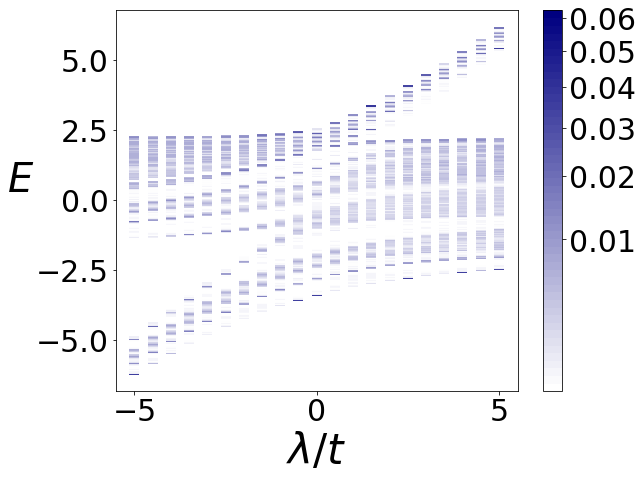}
 
\end{center}
\caption{Density of states for the spin up part of the Hamiltonian with $\gamma=0$, $1/f =5$, and at $-\lambda_R=\lambda_B=\lambda$, $\lambda_G=0$ (upper panel) and $\lambda_R=\lambda$, $\lambda_B=\lambda_G=0$ (lower panel).}
\label{DoS}
\end{figure}

\subsection{Counting the subband number under a weak magnetic field.}
\label{appendixA.2}

Here we consider the case $V_{\alpha,{\bf r}}=\lambda_{\alpha}$, $\gamma=0$. We review another method to determine the Chern number associated to the energy bands of our model. It is based on the numerical determination of the energy spectrum when we add a weak magnetic field $\boldsymbol{\cal B} = {\cal B} \hat{\boldsymbol{z}}$ orthogonal to the Kagome lattice plane. That is we add the term
\begin{equation}
-{\cal B}\sum_{\bf r}\sum_{\alpha=R,B,G}\left(n_{\alpha,{\bf r},\uparrow} - n_{\alpha,{\bf r},\downarrow}\right)
\end{equation}
in the Hamiltonian and also an extra coordinate dependent flux $\phi_1({\bf r})$ for the  hopping between nearest neighbor sites.

As we argue before, the Hamiltonian decouples into two spin independent parts. Spin up and spin down part are associated to opposite energy bands' Chern number. Here for simplicity, we only consider the spin up part of the Hamiltonian. We call the energy bands when ${{\cal B}=0}$ the "parents bands". Under a weak magnetic field, each parent band $i$ is spitted into a certain number of subband that we denote $D_i$. We can link this number to the amplitude of the magnetic field and to the Chern number $\nu_i$ of the $i^{th}$ parent band\cite{ch.ni.95,pe.ho.12}. We introduce $f =q n_\phi$ where $ n_\phi = \dfrac{{\cal B} a^2}{h/e}$ is the number of flux quanta in the system and $q$ is the area of the first magnetic Brillouin zone. In the following, we choose $\cal B$ such that $1/f$ is an integer. Then we have: 
\begin{equation} \label{nd}
	\nu_i = \dfrac{1}{f} - D_i.
\end{equation}
First we show that this method is useful at $\lambda_{\alpha} =0$, $\alpha = {R,B,G}$, or $\lambda_{\alpha} \ll t$. We implement the effect of the $\cal B$ field in the Hamiltonian. We compute the energy spectrum and the energy density of states (DoS) so that we can count the number of subbands arising from each parent band. We consider for instance $1/f=5$. Figure~\ref{DoS} show the density of states (DoS) at different values $\lambda_\alpha$. At $\lambda_\alpha=0, \, \alpha =R,B,G$, we see that we have $D_1 = 4$ subbands, $D_2=5$ subbands and $D_3=6$ subbands, which gives $\nu_1=-\nu_3=1$, and $\nu_2=0$. We see that when $|\lambda_\alpha|$ start increasing, counting the subbands is becoming more and more difficult; when $\lambda_{\alpha}$ reaches a value of the order $t$ or greater, the determination of the Chern number relying on this method becomes impracticable. 

When $\gamma \neq 0$, several issues prevent us from extending this method to compute the energy bands' Chern number. First, the Hamiltonian does not decouple into two spin independent parts. Formula~\ref{nd}, relies on a semiclassical dynamics description of the Bloch electrons in a magnetic field with no interband tunneling. At $\gamma \neq 0$, this description is much more tricky to draw because the bands are crossing. Moreover, we can not identify the subbands arising from each parent band, so counting the former is impracticable.
\newline

\section{Conservation of the average current along ${\cal E}_2$ direction}
\label{Conservation_along_E2}

Here we compute the time derivative of the average value of the current operator for one directional line, that we denote ${\cal E}_2$, along the ${\bf e}_2$ direction.
We remind that we consider an inversion symmetric lattice with line-shape boundaries.
The ${\cal E}_2$ line crosses a certain number of unit cells and may have 2 extremities if the system is open in the ${\bf e}_2$ direction. Both atoms at the extremity of the line have the same color that we denote $\alpha_e$, $\alpha_e = R$ or $G$. For each unit cell at position ${\bf r}$, we need to consider the currents along ${\cal E}_2$, which are ${\bf j}_{(G,{\bf r}),(R,{\bf r})}(\tau)$ and ${\bf j}_{(R,{\bf r}+{\bf e}_2),(G,{\bf r})}(\tau)$. We have 
\widetext
\begin{eqnarray} \label{com1}
\left[{\cal H}(\tau),{\bf j}_{(G,{\bf r}),(R,{\bf r})}(\tau)\right] &=&
-{\bf e}_2 \left[ i t^2 \left( \sin^2 \tau -\cos^2 \tau\right) \sum_{\sigma= \{\uparrow,\downarrow\}} \left[ \left( c_{R,{\bf r}+{\bf e}_2,\sigma}^{\dagger}c_{R,{\bf r},\sigma}^{\phantom\dagger} - c_{G,{\bf r}-{\bf e}_2,\sigma}^{\dagger}c_{G,{\bf r},\sigma}^{\phantom\dagger} \right) + H.c. \right] \right.\nonumber\\
&+&i t^2 \cos \tau \sum_{\sigma= \{\uparrow,\downarrow\}} \left[ \left( c_{G,{\bf r},\sigma}^{\dagger}c_{B,{\bf r},\sigma}^{\phantom\dagger} + c_{G,{\bf r},\sigma}^{\dagger}c_{B,{\bf r}-{\bf e}_1,\sigma}^{\phantom\dagger} \right) + H.c. \right] \nonumber\\
&-&i t^2 \cos \tau \sum_{\sigma= \{\uparrow,\downarrow\}} \left[ \left( e^{is_z \phi} c_{B,{\bf r},\sigma}^{\dagger}c_{R,{\bf r},\sigma}^{\phantom\dagger} + e^{is_z \phi} c_{B,{\bf r}+{\bf e}_3,\sigma}^{\dagger}c_{R,{\bf r},\sigma}^{\phantom\dagger} \right) + H.c. \right] \nonumber\\
&+& 2 i t^2 \cos^2 \tau \sum_{\sigma= \{\uparrow,\downarrow\}} \left( c_{G,{\bf r},\sigma}^{\dagger}c_{G,{\bf r},\sigma}^{\phantom\dagger} - c_{R,{\bf r},\sigma}^{\dagger}c_{R,{\bf r},\sigma}^{\phantom\dagger} \right) \nonumber \\
&+& 2 t^2 \cos \tau \sin \tau \sum_{\sigma= \{\uparrow,\downarrow\}} \left( c_{R,{\bf r}+{\bf e}_2,\sigma}^{\dagger}c_{R,{\bf r}, \overline{\sigma}}^{\phantom\dagger} + c_{G,{\bf r}-{\bf e}_2,\sigma}^{\dagger}c_{G,{\bf r},\sigma}^{\phantom\dagger} \right) - H.c. \nonumber \\
&-& t^2 \sin \tau \sum_{\sigma= \{\uparrow,\downarrow\}} \left( c_{G,{\bf r},\sigma}^{\dagger}c_{B,{\bf r},\overline{\sigma}}^{\phantom\dagger} + c_{G,{\bf r},\sigma}^{\dagger}c_{B,{\bf r}-{\bf e}_1,\overline{\sigma}}^{\phantom\dagger} \right) - H.c.  \nonumber\\
&+&\left. t^2 \sin \tau \sum_{\sigma= \{\uparrow,\downarrow\}} \left( e^{i s_z(\sigma) \phi} c_{B,{\bf r},\sigma}^{\dagger}c_{R,{\bf r},\overline{\sigma}}^{\phantom\dagger} + e^{is_z(\sigma) \phi} c_{B,{\bf r}+{\bf e}_3,\sigma}^{\dagger}c_{R,{\bf r},\overline{\sigma}}^{\phantom\dagger} \right) - H.c.\right]  
\end{eqnarray}
\begin{equation} \label{dtau1}
\dfrac{d}{d\tau} {\bf j}_{(G,{\bf r}),(R,{\bf r})}(\tau) = {\bf e}_2\,\, t \sum_{\sigma= \{\uparrow,\downarrow\}} \left( i \sin \tau \, c_{G,{\bf r},\sigma}^{\dagger}c_{R,{\bf r},\sigma}^{\phantom\dagger} + \cos \tau \, c_{G,{\bf r},\sigma}^{\dagger}c_{R,{\bf r},\overline{\sigma}}^{\phantom\dagger} \right) + H.c. \, ,
\end{equation}
\begin{eqnarray} \label{com2}
\left[{\cal H}(\tau),{\bf j}_{(R,{\bf r}+{\bf e}_2),(G,{\bf r})}(\tau)\right] &=& -{\bf e}_2
\left[i t^2 \left(\cos^2 \tau-\sin^2 \tau \right) \sum_{\sigma= \{\uparrow,\downarrow\}} \left[ \left( c_{R,{\bf r}+{\bf e}_2,\sigma}^{\dagger}c_{R,{\bf r},\sigma}^{\phantom\dagger} - c_{G,{\bf r}+{\bf e}_2,\sigma}^{\dagger}c_{G,{\bf r},\sigma}^{\phantom\dagger} \right) + H.c. \right] \right.\nonumber\\
&-&i t^2 \cos \tau \sum_{\sigma= \{\uparrow,\downarrow\}} \left[ \left( c_{G,{\bf r},\sigma}^{\dagger}c_{B,{\bf r}+{\bf e}_3,\sigma}^{\phantom\dagger} + c_{G,{\bf r},\sigma}^{\dagger}c_{B,{\bf r}+{\bf e}_2,\sigma}^{\phantom\dagger} \right) + H.c. \right] \nonumber\\
&+&i t^2 \cos \tau \sum_{\sigma= \{\uparrow,\downarrow\}} \left[ \left( e^{is_z \phi} c_{B,{\bf r},\sigma}^{\dagger}c_{R,{\bf r}+{\bf e}_2,\sigma}^{\phantom\dagger} + e^{is_z \phi} c_{B,{\bf r}+{\bf e}_3,\sigma}^{\dagger}c_{R,{\bf r}+{\bf e}_2,\sigma}^{\phantom\dagger} \right) + H.c. \right] \nonumber \\
&+& 2 i t^2 \cos^2 \tau \sum_{\sigma= \{\uparrow,\downarrow\}} \left( c_{R,{\bf r}+{\bf e}_2,\sigma}^{\dagger}c_{R,{\bf r}+{\bf e}_2,\sigma}^{\phantom\dagger} - c_{G,{\bf r},\sigma}^{\dagger}c_{G,{\bf r},\sigma}^{\phantom\dagger} \right) \nonumber \\
&-& 2 t^2 \cos \tau \sin \tau \sum_{\sigma= \{\uparrow,\downarrow\}} \left( c_{R,{\bf r}+{\bf e}_2,\sigma}^{\dagger}c_{R,{\bf r}, \overline{\sigma}}^{\phantom\dagger} - c_{G,{\bf r}+{\bf e}_2,\sigma}^{\dagger}c_{G,{\bf r},\sigma}^{\phantom\dagger} \right) - H.c. \nonumber \\
&-& t^2 \sin \tau \sum_{\sigma= \{\uparrow,\downarrow\}} \left( c_{G,{\bf r},\sigma}^{\dagger}c_{B,{\bf r}+{\bf e}_3,\overline{\sigma}}^{\phantom\dagger} + c_{G,{\bf r},\sigma}^{\dagger}c_{B,{\bf r}+{\bf e}_2,\overline{\sigma}}^{\phantom\dagger} \right) - H.c. \nonumber\\
&+&\left. t^2 \sin \tau \sum_{\sigma= \{\uparrow,\downarrow\}}  \left( e^{i s_z(\sigma) \phi} c_{B,{\bf r},\sigma}^{\dagger}c_{R,{\bf r}+{\bf e}_2,\overline{\sigma}}^{\phantom\dagger} + e^{is_z(\sigma) \phi} c_{B,{\bf r}+{\bf e}_3,\sigma}^{\dagger}c_{R,{\bf r}+{\bf e}_2,\overline{\sigma}}^{\phantom\dagger} \right) - H.c. \right] \,, 
\end{eqnarray}
and
\begin{equation} \label{dtau2}
\dfrac{d}{d\tau} {\bf j}_{(R,{\bf r}+{\bf e}_2),(G,{\bf r})}(\tau) = 
{\bf e}_2 \,\,t \sum_{\sigma= \{\uparrow,\downarrow\}} \left( i \sin \tau \, c_{G,{\bf r},\sigma}^{\dagger}c_{R,{\bf r}+{\bf e}_2,\sigma}^{\phantom\dagger} + \cos \tau \, c_{G,{\bf r},\sigma}^{\dagger}c_{R,{\bf r}+{\bf e}_2,\overline{\sigma}}^{\phantom\dagger} \right) + H.c.,
\end{equation}
\endwidetext
We can notice that, integrated along the whole line ${\cal E}_2$, the $1^{\textrm{st}}$ and the $5^{\textrm{th}}$ terms of the commutator~\ref{com1} are canceled by the ones of the commutator~\ref{com2}, and the $4^{\textrm{th}}$ term of both commutators give a resulting contribution $ 2 i t^2 \cos^2 \tau \sum_{\sigma= \{\uparrow,\downarrow\}} \left( c_{\alpha_e,{\bf r}_1,\sigma}^{\dagger}c_{\alpha_e,{\bf r}_1,\sigma}^{\phantom\dagger} - c_{\alpha_e,{\bf r}_2,\sigma}^{\dagger}c_{\alpha_e,{\bf r}_2,\sigma}^{\phantom\dagger} \right)$, with ${\bf r}_1$ and ${\bf r}_2$ both positions at the extremity of ${\cal E}_2$. 
Moreover, we notice that the Hamiltonian is invariant under inversion symmetry added to the transformation $\gamma \rightarrow - \gamma$. This transforms the $2^{\textrm{nd}}$, $3^{\textrm{rd}}$, $6^{\textrm{th}}$ and $7^{\textrm{th}}$ terms in the commutator~\ref{com1} (taken at some position ${\bf r}$) respectively into the opposite of the same terms in the commutator~\ref{com2} (taken at the inversion symmetric position ${\bf r}_R$) and the term $ 2 i t^2 \cos^2 \tau \sum_{\sigma= \{\uparrow,\downarrow\}} \left( c_{\alpha_e,{\bf r}_1,\sigma}^{\dagger}c_{\alpha_e,{\bf r}_1,\sigma}^{\phantom\dagger} - c_{\alpha_e,{\bf r}_2,\sigma}^{\dagger}c_{\alpha_e,{\bf r}_2,\sigma}^{\phantom\dagger} \right)$ into its opposite, while it leaves the terms~\ref{dtau1} and \ref{dtau2} unchanged. It means that we have 
\widetext
\begin{eqnarray}
&& \sum_{{\bf r} \in \mathcal{E}_2} \bra{\Psi(\tau)} \left(  \dfrac{i}{\hbar} \left[{\cal H}(\tau),{\bf j}_{(G,{\bf r}),(R,{\bf r})}(\tau)\right] + \dfrac{i}{\hbar} \left[{\cal H}(\tau),{\bf j}_{(R,{\bf r}+{\bf e}_2),(G,{\bf r})}(\tau)\right] \right)\ket{\Psi(\tau)} \nonumber \\ 
&&= \sum_{{\bf r} \in \mathcal{E}_2} \bra{\Psi(\tau)} \left( - \dfrac{i}{\hbar} \left[{\cal H}(\tau),{\bf j}_{(R,{\bf r}+{\bf e}_2),(G,{\bf r})}(\tau)\right] - \dfrac{i}{\hbar} \left[{\cal H}(\tau),{\bf j}_{(G,{\bf r}),(R,{\bf r})}(\tau)\right]  \right)\ket{\Psi(\tau)}, \nonumber \\
\end{eqnarray}
\endwidetext
\noindent
where $\mathcal{E}_2$ is the ensemble containing the positions of all the unit cells belonging to the ${\cal E}_2$ line. From this we deduce that the term 
\begin{eqnarray*}
&&\hspace{-1.75cm}\sum_{{\bf r} \in \mathcal{E}_2} \bra{\Psi(\tau)} \Biggl(  \dfrac{i}{\hbar} \left[{\cal H}(\tau),{\bf j}_{(G,{\bf r}),(R,{\bf r})}(\tau)\right] \\
&&+ \dfrac{i}{\hbar} \left[{\cal H}(\tau),{\bf j}_{(R,{\bf r}+{\bf e}_2),(G,{\bf r})}(\tau)\right] \Biggl)\ket{\Psi(\tau)} 
\end{eqnarray*}
amounts to a vanishing contribution. 

The Hamiltonian is also invariant under ${c_{\alpha,{\bf r}}^{\dagger}=\left(c_{\alpha,{\bf r},\uparrow}^{\dagger},c_{\alpha, {\bf r},\downarrow}^{\dagger}\right) \rightarrow c_{\alpha,{\bf r}}^{\dagger \, \prime}= \left(-c_{\alpha,{\bf r},\uparrow}^{\dagger},c_{\alpha, {\bf r},\downarrow}^{\dagger}\right),}$ ${\alpha=R\,,\,B\,,\,G}$, added to the transformation $\gamma \rightarrow - \gamma$. From this we deduce that each of the operators $\bra{\Psi(\tau)} \left(\dfrac{d}{d\tau} {\bf j}_{(G,{\bf r}),(R,{\bf r})}(\tau)\right)\ket{\Psi(\tau)}$ and 
$\bra{\Psi(\tau)} \left(\dfrac{d}{d\tau} {\bf j}_{(R,{\bf r}+{\bf e}_2),(G,{\bf r})}(\tau)\right)\ket{\Psi(\tau)}$ gives a vanishing contribution. Eventually we conclude that we have 
\begin{equation}
\sum_{{\bf r} \in {\cal E}_2} \dfrac{d }{d\tau} \bra{\Psi(\tau)}  {\bf j}_{(G,{\bf r}),(R,{\bf r})}(\tau) + {\bf j}_{(R,{\bf r}+{\bf e}_2),(G,{\bf r})}(\tau) \ket{\Psi(\tau)} = 0.
\end{equation}

\section{Effective Hamiltonian}
\label{Details_Effective Hamiltonian}

\subsection{Real space}

Here, we derive the effective Hamiltonian for the case for which the large on-site energy (either positive or negative) is applied only on one site per unit cell, while for the other two sites, the on-site energies are equal to zero or smaller than the hopping amplitude $t$.

To derive the effective Hamiltonian, we split our original Hamiltonian in three parts
\begin{equation}
\label{App_Hamiltonian}
{\cal H}= {\cal H}_o + {\cal H}_\lambda + {\cal H}_\Delta \,.
\end{equation}
Here, ${\cal H}_\lambda$ denotes the on-site potential for the sites with large on-site energies, while ${\cal H}_o$ denotes the Hamiltonian describing the reduced lattice, without these sites. ${\cal H}_\Delta$ represents the coupling of these two subsystems.

In matrix form we have
\begin{equation}
{\cal H}=\left(
\begin{array}{cc}
{\cal H}_o & {\cal H'}_\Delta \\
{\cal H'}_\Delta^\dagger & {\cal H}_\lambda
\end{array}
\right) \,.
\end{equation}
Here ${\cal H}_\Delta={\cal H'}_\Delta + {\cal H'}_\Delta^\dagger $. The dimension of the matrix ${\cal H}$ is $6N_1N_2 \times 6N_1N_2$ since the system contains $3N_1N_2$ sites and therefore there are $6N_1N_2$  single particle energy levels (additional factor $2$ comes from the spin degrees of freedom). The dimension of the matrix ${\cal H}_o$ is $4N_1N_2 \times 4N_1N_2$, because it contains $2N_1N_2$ sites.
The dimension of the matrix ${\cal H}_\lambda$ is $2N_1N_2 \times 2N_1N_2$, as the latter contains only $N_1N_2$ sites. Finally the dimension of the matrix ${\cal H'}_\Delta$  (${\cal H'}_\Delta^\dagger$) is $4N_1N_2 \times 2N_1N_2$ ($2N_1N_2 \times 4N_1N_2$).

So we have
\begin{eqnarray*}
&&\hspace{-0.5cm}{\cal H}|\psi \rangle =\mathbb{1}  E|\psi \rangle
\quad \Longrightarrow \quad  \det\left[{\cal H} - \mathbb{1} E\right]=0 
\nonumber\\
&&\hspace{-0.5cm} 
\left|\begin{array}{cc}
{\cal H}_o - \mathbb{1} E& {\cal H'}_\Delta \\
{\cal H'}_\Delta^\dagger & {\cal H}_\lambda - \mathbb{1} E
\end{array}
\right|=\det\left[{\cal H}_\lambda - \mathbb{1} E\right] \times
\nonumber\\
&&\hspace{0.5cm}\times
\det\left[{\cal H}_o - \mathbb{1} E 
- {\cal H'}_\Delta \left({\cal H}_\lambda - \mathbb{1} E\right)^{-1} 
{\cal H'}_\Delta^\dagger\right]=0 \,.
\end{eqnarray*}
Here, $E$ corresponds to the eigenvalues of  Hamiltonian \eqref{App_Hamiltonian}. 
For further evaluation, we will restrict ourselves to the case that $\det\left[{\cal H}_\lambda - \mathbb{1} E\right] \neq 0$. So we have $\det\left[{\cal H}_{\rm eff}(E)-\mathbb{1}E\right]=0$. Here 
\begin{eqnarray}
\label{App_1Effective_H}
{\cal H}_{\rm eff}(E)={\cal H}_o 
- {\cal H'}_\Delta \left({\cal H}_\lambda - \mathbb{1} E\right)^{-1}  
{\cal H'}_\Delta^\dagger \,.
\end{eqnarray}
Here, ${\cal H}_{\rm eff}(E)$ is a $4N_1N_2 \times 4N_1N_2$ matrix.
Assuming that the sites with large on-site energies are decoupled from one another, we get ${\cal H}_\lambda=\rm{diag}\{\lambda_{{\bf r}_\lambda}\}$. Thus we have
\begin{equation}
\label{App_2Effective_H} 
{\cal H}_{\rm eff}(E)= {\cal H}_o - \hspace{-0.15cm} \sum_{{\bf r}_1,{\bf r}_2 \in \Omega_o}
\sum_{{\bf r}_\lambda \in \Omega_\lambda} \hspace{-0.25cm}
\frac{|{\bf r}_1\rangle\langle {\bf r}_1|{\cal H'}_\Delta
|{\bf r}_\lambda\rangle \langle {\bf r}_\lambda|{\cal H'}_\Delta^\dagger|{\bf r}_2\rangle \langle {\bf r}_2|}{\lambda_{{\bf r}_\lambda}-E} \,.
\end{equation}
where 
$$
|{\bf r}\rangle=\left(
\begin{array}{c}
|{\bf r},\uparrow\rangle\\
|{\bf r},\downarrow\rangle
\end{array}
\right) \,,
$$
with ${\bf r}_1$ and ${\bf r}_2$ running over all lattice sites except of those with large on-site energies ($\Omega_o$), while ${\bf r}_\lambda$ runs over all lattice sites with large on-site energies ($\Omega_\lambda$).

We note that up to this point no approximation was made and Eqs. \eqref{App_1Effective_H} and \eqref{App_2Effective_H} are exact expressions, but they are nonlinear equations.

We are interested in the low energy sector, i,e. $|E| \ll  |\lambda_{{\bf r}_\lambda}|$. So we can neglect $E$ compared to $\lambda_{{\bf r}_\lambda}$ and obtain the following effective Hamiltonian: 
\begin{equation}
\label{App_3Effective_H} 
{\cal H}_{\rm eff}= {\cal H}_o - \hspace{-0.15cm} \sum_{{\bf r}_1,{\bf r}_2 \in \Omega_o}
\sum_{{\bf r}_\lambda \in \Omega_\lambda} \hspace{-0.25cm}
\frac{|{\bf r}_1\rangle\langle {\bf r}_1|{\cal H'}_\Delta
|{\bf r}_\lambda\rangle \langle {\bf r}_\lambda|{\cal H}_\Delta^\dagger|{\bf r}_2\rangle \langle {\bf r}_2|}{\lambda_{{\bf r}_\lambda}} \,.
\end{equation}

Now we calculate $\langle {\bf r}_\alpha|{\cal H}_\Delta|{\bf r}_{\alpha'} \rangle$:
\begin{eqnarray*}
\begin{array}{|c|c|c|}
\hline
\mbox{Matrix elements}&\mbox{Same unit cell}&\mbox{Different unit cell}\\ 
\hline
&&\\[-2.2ex]
\langle {\bf r}_R|{\cal H}_\Delta|{\bf r}_B\rangle 
&- t \mathbb{1} 
&- t \mathbb{1}\\
\hline 
&&\\[-2.2ex]
\langle {\bf r}_B|{\cal H}_\Delta|{\bf r}_R\rangle 
&- t \mathbb{1}
&- t \mathbb{1}\\
\hline 
&&\\[-2.2ex]
\langle {\bf r}_R|{\cal H}_\Delta|{\bf r}_G\rangle 
&-t e^{- i2\pi\gamma\sigma^x} 
&-t e^{ i2\pi\gamma\sigma^x} \\
\hline 
&&\\[-2.2ex]
\langle {\bf r}_G|{\cal H}_\Delta|{\bf r}_R\rangle 
&-t e^{i2\pi\gamma\sigma^x}
&-t e^{-i2\pi\gamma\sigma^x}\\
\hline 
&&\\[-2.2ex]
\langle {\bf r}_B|{\cal H}_\Delta|{\bf r}_G\rangle 
&-t e^{i \phi\sigma^z}
&-t e^{i \phi\sigma^z}\\
\hline
&&\\[-2.2ex]
\langle {\bf r}_G|{\cal H}_\Delta|{\bf r}_B\rangle 
&-t e^{-i \phi\sigma^z}
&-t e^{-i \phi\sigma^z}\\
\hline 
\end{array}
\end{eqnarray*}

For the case where we apply a large on-site energy $\lambda$ to either $R$, $B$ or $G$ for each unit cell, we obtain
\begin{eqnarray}
\label{App_Effective_Hamiltonian_real_space}
{\cal H}_{\rm eff}&=&\sum_{n,m}\Biggl[a_{n,m}^{\dagger} \hat t_{h,+} b_{n,m}^{\phantom\dagger}
+a_{n,m}^{\dagger} \hat t_{h,-} b_{n-1,m}^{\phantom\dagger}
\nonumber\\
&+&a_{n,m}^{\dagger} \hat t_{v,+} b_{n-1,m+1}^{\phantom\dagger}
+a_{n,m}^{\dagger} \hat t_{v,-} b_{n,m-1}^{\phantom\dagger}
\nonumber\\
&+&a_{n,m}^{\dagger} \hat t_{d,a} a_{n,m+1}^{\phantom\dagger}
+b_{n,m}^{\dagger} \hat t_{d,b} b_{n-1,m+1}^{\phantom\dagger}
+h.c.\Biggl] \nonumber\\
&&\hspace{-0.2cm}+\sum_{n,m}\left[a_{n,m}^{\dagger} \hat \varepsilon_a a_{n,m}^{\phantom\dagger} + b_{n,m}^{\dagger} \hat \varepsilon_b b_{n,m}^{\phantom\dagger}  \right] \,,
\end{eqnarray}
where, $a_{n,m}^\dagger$ and $b_{n,m}^\dagger$ create fermions in the unit cell $(n,n)$ on $a$ and $b$ sublattices. $\hat t_{\pm}$, $\hat t_{v,\pm}$, $\hat t_{d,a}$, $\hat t_{d,b}$ are hopping matrices and $\hat \varepsilon_a$ and $\hat \varepsilon_b$ describe on-site energies.

For $\lambda_R=\lambda$ and $\lambda_B=\lambda_G=0$, we are left with $B$ and $G$ sites. We obtain
\begin{eqnarray}
&&\hat t_{h,\pm}  =-t e^{i \phi\sigma^z}
-\frac{t^2}{\lambda } e^{\pm i 2\pi\gamma\sigma^x}\\
&&\hat t_{v,+}=\hat t_{v,-}^\dagger=-\frac{t^2}{\lambda } e^{i 2\pi\gamma\sigma^x} \\
&&\hat t_{d,a}=\hat \varepsilon_a=\hat \varepsilon_b
=-\frac{t^2}{\lambda } \mathbb{1}\\
&&\hat t_{d,b}=-\frac{t^2}{\lambda } e^{i 4\pi\gamma\sigma^x} \,.
\end{eqnarray}

For $\lambda_B=\lambda$ and $\lambda_R=\lambda_G=0$, we are left with $G$ and $R$ sites. We obtain
\begin{eqnarray}
&&\hat t_{h,\pm} =-t_2 e^{\pm i2\pi\gamma\sigma^x} 
-\frac{t_1 t_3}{\lambda } e^{-i \phi\sigma^z}\\
&&\hat t_{v,+}=\hat t_{v,-}=-\frac{t_1 t_3}{\lambda } e^{-i \phi\sigma^z} \\
&&\hat t_{d,a}=\hat t_{d,b}=\hat \varepsilon_a=\hat \varepsilon_b
=-\frac{2t^2}{\lambda}\mathbb{1} \,.
\end{eqnarray}

\subsection{Momentum space}

In this Section, we rewrite the effective Hamiltonian Eq.~\eqref{App_Effective_Hamiltonian_real_space} in momentum space. Similar to the original lattice, also for the effective system, unit cells are arranged on a  triangular lattice (see Fig.~\ref{Fig:effective_schematicp}).

We perform a Fourier transformation 
\begin{eqnarray*}
&&a_{n,m}=\frac{1}{\sqrt{N_1N_2}}\sum_{\bf k}e^{i {\bf k} \cdot {\bf r}}a_{\bf k}\\
&&b_{n,m}=\frac{1}{\sqrt{N_1N_2}}\sum_{\bf k}e^{i {\bf k} \cdot ({\bf r}+{\bf b}_1)}b_{\bf k} \,.
\end{eqnarray*}
Here
\begin{equation}
{\bf r}=n{\bf e}_1+m{\bf e}_2 \,. 
\end{equation}

We obtain
\begin{eqnarray}
{\cal H}_{\rm eff}&=&\sum_{\bf k}\Biggl[
a_{\bf k}^{\dagger}\left(
e^{i {\bf k} \cdot {\bf b}_1}\hat t_{h,+}
+e^{-i {\bf k} \cdot {\bf b}_1}\hat t_{h,-}
\right. \nonumber\\
&+&\left. e^{i {\bf k} \cdot({\bf e}_2-{\bf b}_1)} \hat t_{v,+}
+e^{-i {\bf k} \cdot({\bf e}_2-{\bf b}_1)} \hat t_{v,-}
\right)b_{\bf k}^{\phantom\dagger}
\nonumber\\
&+&a_{\bf k}^{\dagger} e^{i {\bf k} \cdot {\bf e}_2} \hat t_{d,a} a_{\bf k}^{\phantom\dagger}
+b_{\bf k}^{\dagger} e^{i {\bf k} \cdot({\bf e}_2-{\bf e}_1)} \hat t_{d,b}
b_{\bf k}^{\phantom\dagger}
+h.c.\Biggl] \nonumber\\
&&\hspace{-0.2cm}+\sum_{\bf k}\left[a_{\bf k}^{\dagger}\hat\varepsilon_a a_{\bf k}^{\phantom\dagger} + b_{\bf k}^{\dagger}\hat\varepsilon_b b_{\bf k}^{\phantom\dagger}  \right]  \,.
\end{eqnarray}
So we have
\begin{eqnarray}
&&{\cal H}_{\rm eff}=\sum_{\bf k}\Biggl[
a_{\bf k}^{\dagger}\left(
e^{i k_1}\hat t_{h,+} +e^{-i k_1}\hat t_{h,-}
\right. \\
&&\hspace{1.4cm}+\left.e^{i (2k_2-k_1)} \hat t_{v,+} +e^{-i (2k_2-k_1)} \hat t_{v,-}\right)b_{\bf k}^{\phantom\dagger}
\nonumber\\
&&\hspace{1.4cm}+b_{\bf k}^{\dagger}\left(
e^{-i k_1}\hat t_{h,+}^\dagger +e^{i k_1}\hat t_{h,-}^\dagger
\right.\nonumber\\
&&\hspace{1.4cm}\left.+e^{-i (2k_2-k_1)} \hat t_{v,+}^\dagger +e^{i (2k_2-k_1)} \hat t_{v,-}^\dagger \right)a_{\bf k}^{\phantom\dagger}
\nonumber\\
&&\hspace{1.4cm}+a_{\bf k}^{\dagger} \left(e^{i 2k_2} \hat t_{d,a} 
+ e^{-i 2k_2} \hat t_{d,a}^\dagger  +\hat\varepsilon_a\right)a_{\bf k}^{\phantom\dagger}
\nonumber\\
&&\hspace{1.4cm}+b_{\bf k}^{\dagger}\left( e^{i 2(k_2-k_1)} \hat t_{d,b} + e^{-i 2(k_2-k_1)} \hat t_{d,b}^\dagger +\hat\varepsilon_b\right)b_{\bf k}^{\phantom\dagger}\Biggl] \,.
\nonumber
\end{eqnarray}

In the matrix form, the Hamiltonian can be written as follows
\begin{eqnarray}
\label{App_Effective_Hamiltonian_momentum_space}
&&{\cal H}_{\rm eff} =\sum_{\bf k}
\phi_{\bf k}^{\dagger} {\cal H}_{\rm eff}({\bf k})\phi_{\bf k}^{\phantom\dagger}\\
\label{App_Effective_Hamiltonian_k_momentum_space}
&&{\cal H}_{\rm eff}({\bf k})=\left(
\begin{array}{cccc}
{\cal H}_{11} & {\cal H}_{12} & {\cal H}_{13} & {\cal H}_{14} \\
{\cal H}_{21} & {\cal H}_{22} & {\cal H}_{23} & {\cal H}_{24} \\
{\cal H}_{31} & {\cal H}_{32} & {\cal H}_{33} & {\cal H}_{34} \\
{\cal H}_{41} & {\cal H}_{42} & {\cal H}_{43} & {\cal H}_{44} 
\end{array}
\right)\\
&&\phi_{\bf k}^{\dagger}=
\left(
a_{{\bf k},\uparrow}^{\dagger},
a_{{\bf k},\downarrow}^{\dagger},
b_{{\bf k},\uparrow}^{\dagger},
b_{{\bf k},\downarrow}^{\dagger}
\right) \,.
\end{eqnarray}

For $\lambda_R=\lambda$ and $\lambda_G=\lambda_B=0$ we have
\begin{eqnarray*}
&&{\cal H}_{11}={\cal H}_{22}=-\frac{2t^2}{\lambda }\cos(2k_2) -\frac{2t^2}{\lambda }\\
&&{\cal H}_{33}={\cal H}_{44}=-\frac{2t^2 \cos(4\pi\gamma)}{\lambda } \cos(2(k_2-k_1))-\frac{2t^2}{\lambda }\\
&&{\cal H}_{12}={\cal H}_{21}=0\\
&&{\cal H}_{34}={\cal H}_{43}=\frac{2t^2 \sin(4\pi\gamma)}{\lambda } \sin(2(k_2-k_1))\\
&&{\cal H}_{13}={\cal H}_{31}^\dagger=-2t\cos k_1 e^{i \phi} 
\nonumber\\
&&\hspace{1.5cm}-\frac{2t^2 \cos(2\pi\gamma)}{\lambda } (\cos(2k_2-k_1) +\cos k_1) \\
&&{\cal H}_{24}={\cal H}_{42}^\dagger=-2t\cos k_1 e^{-i \phi}  
\nonumber\\
&&\hspace{1.5cm}-\frac{2t^2 \cos(2\pi\gamma)}{\lambda } (\cos(2k_2-k_1) + \cos k_1) \\
&&{\cal H}_{14}={\cal H}_{41}=\frac{2t^2 \sin(2\pi\gamma)}{\lambda } (\sin(2k_2-k_1)+\sin k_1)\\
&&{\cal H}_{23}={\cal H}_{32}=\frac{2t^2 \sin(2\pi\gamma)}{\lambda } (\sin(2k_2-k_1) +\sin k_1) \,.
\end{eqnarray*}

While for $\lambda_B=\lambda$ and $\lambda_R=\lambda_B=0$ we have
\begin{eqnarray*}
&&{\cal H}_{11}={\cal H}_{22}=-\frac{2t^2}{\lambda }\cos(2k_2) -\frac{2t^2}{\lambda }\\
&&{\cal H}_{33}={\cal H}_{44}=-\frac{2t^2}{\lambda } \cos(2(k_2-k_1))-\frac{2t^2}{\lambda }\\
&&{\cal H}_{12}={\cal H}_{21}=0\\
&&{\cal H}_{34}={\cal H}_{43}=0\\
&&{\cal H}_{13}={\cal H}_{31}^\dagger=-2t\cos(2\pi\gamma)\cos k_1 
\nonumber\\
&&\hspace{1.5cm}-\frac{2t^2}{\lambda} (\cos(2k_2-k_1)+\cos k_1) e^{-i \phi}\\
&&{\cal H}_{24}={\cal H}_{42}^\dagger=-2t\cos(2\pi\gamma)\cos k_1 
\nonumber\\
&&\hspace{1.5cm}-\frac{2t^2}{\lambda} (\cos(2k_2-k_1)+\cos k_1) e^{i \phi}\\
&&{\cal H}_{14}={\cal H}_{41}=2t\sin(2\pi\gamma)\sin k_1\\
&&{\cal H}_{23}={\cal H}_{32}=2t\sin(2\pi\gamma)\sin k_1  \,.
\end{eqnarray*}

\subsection{Analytical solution for $\lambda_R=\lambda$, $\lambda_B=\lambda_G=0$  and $\gamma=0$}

We now calculate eigenvalues of the effective Hamiltonian for $\lambda_R=\lambda$, $\lambda_B=\lambda_G=0$. For $\gamma=0$, spin up and spin down fermions are decoupled from each other. Therefore instead of finding eigenvalues of a $4 \times 4$ matrix (the unit cell contains 2 sites and a factor 2 comes due to the spin) one needs to perform calculations for two equivalent $2 \times 2$ matrices. We obtain
\begin{eqnarray}
E_{\sigma,-}&=&-\frac{2t^2}{\lambda}\Bigl(1+\cos k_1 \cos(k_1-2k_2)\Bigl)
\\
&-&\sqrt{4t^2\cos^2 k_1 + \frac{4t^4}{\lambda^2}\Bigl(1 + \cos k_1 \cos(k_1 - 2k_2) \Bigl)^2 }
\nonumber\\
E_{\sigma,+}&=&-\frac{2t^2}{\lambda}\Bigl(1+\cos k_1 \cos(k_1-2k_2)\Bigl)
\\
&+&\sqrt{4t^2\cos^2 k_1 + \frac{4t^4}{\lambda^2}\Bigl(1 + \cos k_1 \cos  (k_1 -2 k_2) \Bigl)^2} \,.
\nonumber
\end{eqnarray}

The size of the gap is the energy difference between the lowest energy of the higher band and the highest energy of the lower band. So we have to find the minimal value of $E_{\sigma,+}$ and maximal value of $E_{\sigma,-}$. One can easily notice that $E_{\sigma,+} \geq 0$. From here directly follows that $\min[E_{\sigma,+}]=0$. This minimum occurs for $k_1=\pm \pi/2$. What concerns $E_{\sigma,-}$, it is maximal when $k_1-2k_2=\pi$. So we have to find the maximum for 
\begin{equation}
 f=-\frac{4t^2}{\lambda}\sin^2\frac{k_1}{2}
-\sqrt{4t^2\cos^2 k_1+\frac{16t^4}{\lambda^2}\sin^4\frac{k_1}{2}} \,.
\end{equation}
We have to solve $\frac{df}{dk_1}=0$, leading to
\begin{equation}
-\frac{2t^2}{\lambda}\sin k_1 -
\frac{-8t^2\cos k_1 \sin k_1 
+\frac{16t^4}{\lambda^2}\sin^2\frac{k_1}{2}\sin k_1}{2\sqrt{4t^2\cos^2 k_1+\frac{16t^4}{\lambda^2}\sin^4\frac{k_1}{2}}}=0 \,.
\end{equation}
The solution $k_1=0$ corresponds to the local minimum, therefore the maximum is given by
\begin{equation}
\cos k_1  
-\frac{2t^2}{\lambda^2}\sin^2\frac{k_1}{2}
=\frac{t}{\lambda}\sqrt{\cos^2 k_1+\frac{4t^2}{\lambda^2}\sin^4\frac{k_1}{2}} \,. 
\end{equation}
We take the square of both sides of the equation and obtain
\begin{equation}
\cos^2 k_1-\frac{4t^2}{\lambda^2}\cos k_1\sin^2\frac{k_1}{2}
=\frac{t^2}{\lambda^2}\cos^2k_1  \,.
\end{equation}
Here the solution $\cos k_1=0$ also corresponds to a local minimum. So we have for the maximum
\begin{equation}
\cos k_1=\frac{t^2}{t^2+\lambda^2}  \,.
\end{equation}
Based on that we obtain 
\begin{equation}
k_1
=\frac{\pi}{2}-\arcsin\frac{t^2}{t^2+\lambda^2} \simeq
\frac{\pi}{2}-\frac{t^2}{\lambda^2} \,,
\end{equation}
and therefore
\begin{equation}
\max[E_{\sigma,-}]\simeq -\frac{4t^2}{\lambda}+\frac{3t^4}{\lambda^3} \simeq 
-\frac{4t^2}{\lambda} \,.
\end{equation}
So we get 
\begin{eqnarray}
\Delta =\min\left[E_{\sigma,+}\right]-\max\left[E_{\sigma,-}\right] \simeq  \frac{4t^2}{\lambda} \,.
\end{eqnarray}

\end{document}